\documentclass[aps, rmp, letterpaper, reprint, floatfix, amsmath, amssymb, notitlepage, longbibliography]{revtex4-1}

\usepackage{enumerate}
\usepackage{xcolor}
\usepackage{amsfonts}
\usepackage{mathrsfs}
\usepackage{bm}
\usepackage{graphicx}
\usepackage[colorlinks=true,citecolor=blue]{hyperref}
\hypersetup{bookmarksnumbered, pdfpagemode=UseOutlines, pdfdisplaydoctitle, 
colorlinks=true, citecolor=blue, filecolor=blue, linkcolor=blue, urlcolor=blue}

\usepackage{verbatim}
\usepackage{textcomp}
\usepackage[utf8]{inputenc}
\usepackage{amssymb}

\DeclareUnicodeCharacter{2009}{\,} 

\begin{document}

\title{Cavity Magnonics}
\author{Babak {Zare Rameshti}}
\email[Corresponding author's email: ]{bzarer@iust.ac.ir}
\affiliation{Department of Physics, Iran University of Science and Technology, Narmak, Tehran 16844, Iran}

\author{Silvia Viola Kusminskiy}
\affiliation{Max Planck Institute for the Science of Light, Staudtstra\ss e 2, 91058 Erlangen, Germany}
\affiliation{Institute for Theoretical Physics, University Erlangen-N\"{u}rnberg, Staudtstra\ss e 7, 91058 Erlangen, Germany}

\author{James A. Haigh}
\affiliation{Hitachi Cambridge Laboratory, Cambridge, CB3 0HE, UK}

\author{Koji Usami}
\affiliation{Research Center for Advanced Science and Technology (RCAST), The University of Tokyo, Meguro-ku, Tokyo 153-8904, Japan}

\author{Dany Lachance-Quirion}
\thanks{Present address: Nord Quantique, Sherbrooke, Qu\'{e}bec, J1K 0A5, Canada}
\affiliation{Research Center for Advanced Science and Technology (RCAST), The University of Tokyo, Meguro-ku, Tokyo 153-8904, Japan}

\author{Yasunobu Nakamura}
\affiliation{Research Center for Advanced Science and Technology (RCAST), The University of Tokyo, Meguro-ku, Tokyo 153-8904, Japan}
\affiliation{RIKEN Center for Quantum Computing (RQC), RIKEN, Wako-shi, Saitama 351-0198, Japan}

\author{Can-Ming Hu}
\affiliation{Department of Physics and Astronomy, University of Manitoba, Winnipeg R3T 2N2, Canada}

\author{Hong X. Tang}
\affiliation{Department of Electrical Engineering, Yale University, New Haven, Connecticut 06511, USA}

\author{Gerrit E. W. Bauer}
\affiliation{Institute for Materials Research and WPI-AIMR, Tohoku University, Sendai 980-8577, Japan}
\affiliation{Kavli Institute of NanoScience, Delft University of Technology, Lorentzweg 1, 2628 CJ Delft, The Netherlands}

\author{Yaroslav M. Blanter}
\email[Corresponding author's email: ]{Y.M.Blanter@tudelft.nl}
\affiliation{Kavli Institute of NanoScience, Delft University of Technology, Lorentzweg 1, 2628 CJ Delft, The Netherlands}

\date{\today{}}

\begin{abstract}
Cavity magnonics deals with the interaction of magnons --- elementary excitations in magnetic materials --- and confined electromagnetic fields. We introduce the basic physics and review the experimental and theoretical progress of this young field that is gearing up for integration in future quantum technologies. Much of its appeal is derived from the strong magnon-photon coupling and the easily-reached nonlinear regime in microwave cavities. The interaction of magnons with light as detected by Brillouin light scattering is enhanced in magnetic optical resonators, which can be employed to manipulate magnon distributions. The cavity photon-mediated coupling of a magnon mode to a superconducting qubit enables measurements in the single magnon limit.
\end{abstract}

\maketitle
\tableofcontents


\section{Introduction}\label{SecI}
Spectroscopy, the study of the reflection and  transmission of radiation (or its quanta, the photons) by a given sample as a function of frequency, relies on the interaction between electromagnetic (EM) fields and matter. In condensed matter physics, the electric and magnetic field components of an EM wave dominantly interact with the charge and spin of the electrons by the Coulomb and Zeeman interactions. Spectroscopy relies on the weakness of these interactions that allows treating the scattering process by perturbation theory. The observed amplitudes and intensities then give direct information about the electronic and magnetic structure of the scattering object.  

The EM \textit{cavities} trap photons in a finite spatial region in which they interfere to form standing waves. According to Fermi's Golden Rule the modulation of the photon density of states affects the scattering amplitudes. \cite{Purcell1946} pointed out that the light emission of excited matter can be strongly enhanced or suppressed in a cavity via the available photon states for the emitted radiation.
When the confinement is efficient, the cavity modes develop a discrete spectrum with a nearly singular density of states. At the cavity mode frequencies an intrinsically weak interaction may become so strong that perturbation theory breaks down. In this \textit{strong coupling regime} hybrid \(polariton\) states arise in which matter and radiation cannot be distinguished anymore.

Cavities and resonators differ in size and nature depending on the frequency of the photons they are designed to trap and many forms of matter can be inserted. 
 {\em Cavity quantum electrodynamics (cavity QED)} studies Rydberg atoms and trapped ions in optical and microwave (MW) cavities~\cite{Haroche1989,Walther2006}. Micro and nanostructured devices such as superconducting qubits or quantum dots behave in the MW regime like two-level systems or tunable ``artificial atoms'' \cite{Girvin2014}. Their study in MW cavities or {\em circuit QED} \cite{Blais2004,Wallraff2004,Blais2020}, has paved the way for quantum information processing. {\em Cavity optomechanics} studies the forces exerted by radiation pressure \cite{Braginski1967}  on  devices such as  mechanical resonators, i.e. the photon-phonon coupling. An important breakthrough has been the cavity-assisted cooling of the vibration of a macroscopic object to its (zero-phonon) quantum ground state \cite{Teufel2011cooling,Aspelmeyer2014}.

The present review addresses the electrodynamics of cavities that are filled by a magnetic material and tuned to the interaction of the cavity photons with {\em magnons}, the elementary excitations of the magnetic order.

Soykal and Flatt\'e \cite{Soykal2010} predicted strong coupling  of photons in a MW cavity to the quantum dynamics of a small ferromagnetic sphere. Subsequently, \cite{Huebl2013} reported the observation of strong coupling in the form of an anticrossing of the collective magnetic precession of the magnetization with MW cavity modes. These studies kick-started an international research activity on the coupling of magnons to photons, predominantly at MW and infrared frequencies. We call this field {\em cavity magnonics} but the terms cavity optomagnonics, cavity spintronics, and spin cavitronics are in use as well.

We review here the considerable progress achieved to understand cavity magnonics in terms of semiclassical physics. The field is presently in a watershed situation in which low temperature experiments dedicated to identify quantum effects on the level of cavity/circuit QED or cavity/circuit optomechanics are on their way. We therefore believe that a review of the concepts and main results will consolidate the present understanding and help with the challenges ahead.   



We organized this review as follows. Sec. \ref{secII} summarizes the concepts of an EM cavity, Sec. \ref{secIIm} the physics of ferromagnets and their low energy excitations, and Sec. \ref{SecIII}  the coupling between them. The remaining sections summarize and explain selected experiments, in MW cavities (Sec. \ref{SecIV}) and optical resonators (Sec. \ref{SecV}). We address a hybrid system of a magnet and a superconducting qubit in Sec. \ref{SecVI}. In Sec. \ref{SecVIII} we anticipate the developments in the near future.


\section{Electromagnetic cavities} \label{secII}

Classical and quantum waves that are trapped in a limited space or ``cavity'' where multiple scattering leads to interference have the photon density of states strongly modulated by this interference. Here we focus on EM cavities, i.e. structures that serve to confine EM fields. The cavity modes are the solutions of Maxwell's equations with appropriate boundary conditions at the confining potentials and contacts to the environment.

Wave guides confine the EM waves in one or two directions but are open in another direction. Fabry-Perot interferometers are one-dimensional wave guides that are partially open at the endpoints. Full confinement of the EM field in all directions with a discrete spectrum can be achieved when photons have a long lifetime, i.e. when they are not absorbed and cannot escape except through non-invasive ports. 

The functionality and quality of a cavity depends on the design, size, and material. The modulation of the photon density of states is optimized when the size of the cavity in the confining direction is comparable to the wavelength. MW cavities are made from metals (that may be superconducting) with dimensions in the centimeter
range. Confined MW modes also exist on top of metallic (superconducting) strips such as co-planar waveguides fabricated on insulating substrates. 
An interface between materials with a large dielectric constant mismatch can reflect light efficiently, so solid objects of 10--1000 microns size and a large dielectric constant trap optical (infrared to visible light) fields. Absorption and quality factor of such resonators is high when the material is an electric insulator with a fundamental energy gap higher than the light frequency. 
In the following we briefly discuss the main concepts of EM
cavities as open quantum systems, see also e.g. 
\citet{Meystre2007,Walls2008,Heebner2008,Aspelmeyer2014}. 

\begin{figure}[!t]
\centering
\includegraphics[width=0.8\columnwidth]{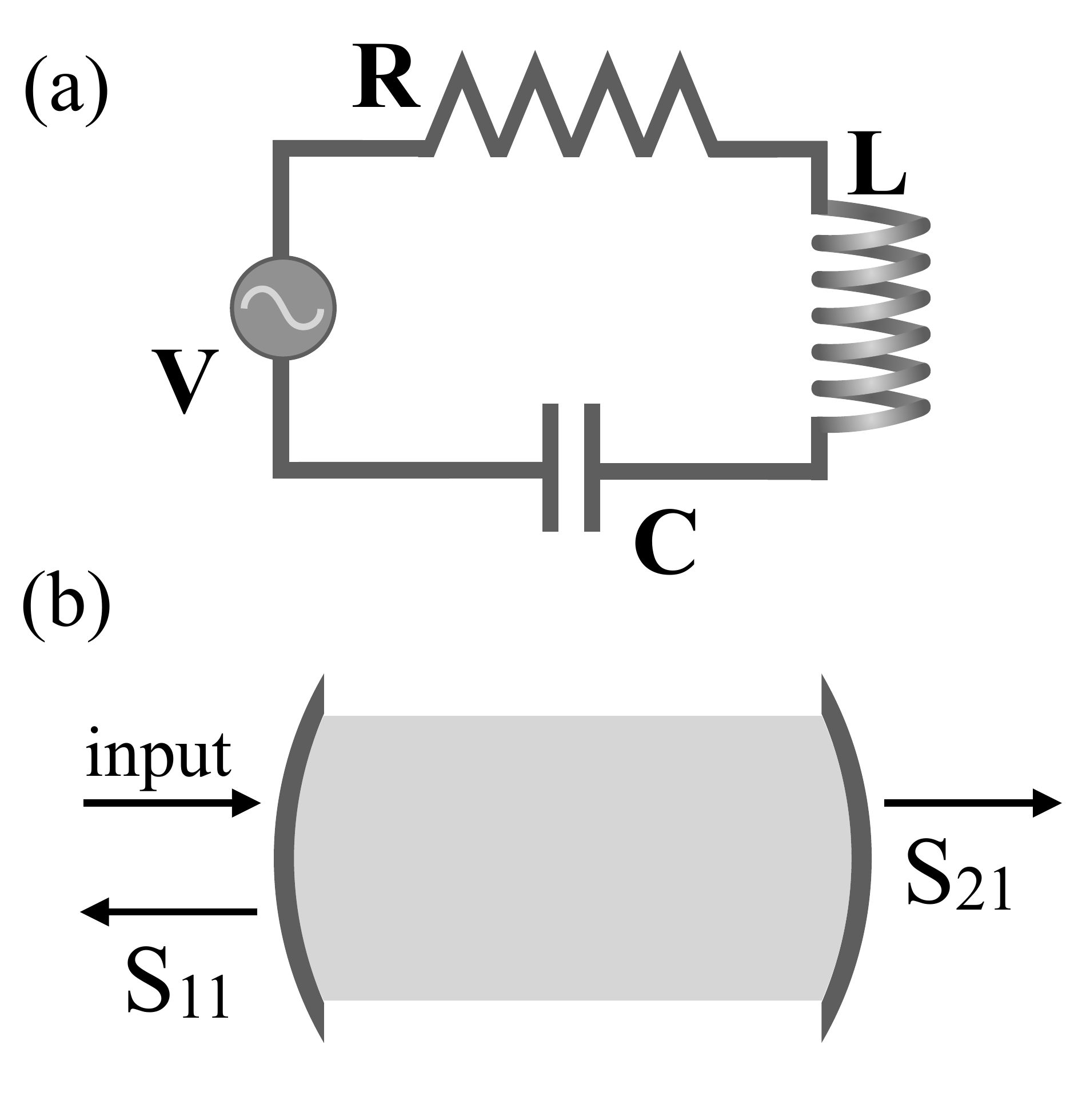}
\caption{\footnotesize{(Color online) (a) Classical $RLC$ circuit in series driven by a time-dependent potential $V$. (b) Fabry-Perot type of cavity defined by two semi-transparent mirrors $1$ (left) and $2$ (right). The ratio of the signal amplitude transmitted through $2$ and the input amplitude entering through $1$ is $S_{21}$ and analogous for reflection $S_{11}$. When the second mirror is totally reflecting or not monitored we call the cavity ``one-sided".}}
\label{Fig:RLC_FPC}
\end{figure}

\subsection{Free LC circuit}

We first illustrate the basic physics of a MW cavity by considering an $LC$ resonator. i.e. an electrically connected inductance $L$ and capacitance $C$. A voltage $V$ charges the capacitor as $Q_{\rm C}=CV$, while the the current $I$ generates a flux $\Phi=LI$ in the inductor. With $\dot{Q}_{\rm C}=I$ and $\dot{\Phi}=-V$, where the overdot indicates the time derivative, we arrive at the equation for a harmonic oscillator,
\begin{equation}
LC\ddot{I}+I=0,\label{MW_cavity_classical_oscillator}
\end{equation}
with frequency $\omega_{\rm c}=1/\sqrt{LC}$. For MWs typically $\omega_{\rm c}/2\pi \sim 5$ GHz. The $LC$ circuit stores energy
\begin{equation}
U=\frac{CV^{2}}{2}+\frac{LI^{2}}{2}.\label{LC_oscillator_energy}
\end{equation}

In reality, a cavity loses energy at a rate $\kappa_{\rm c}$ that is the sum of internal Ohmic dissipation $\kappa_{\rm 0}$ and radiation leakage $\kappa_{{\rm ex}}$ loss rates,
\begin{equation}
\kappa_{\rm c} =\kappa_{\rm 0}+\kappa_{{\rm ex}}.\label{eq:kappa}
\end{equation}
An important parameter is the cavity {\em quality factor},
\begin{equation}
Q=\omega_{\rm c}/\kappa_{\rm c}.\label{quality_factor}
\end{equation}
Including a dissipative element --- a resistor $R$ --- into the (RLC) circuit, see Fig. \ref{Fig:RLC_FPC}(a), introduces a viscous term into the equation of motion,
\begin{equation}
\ddot{I}+\frac{R}{L}\dot{I}+\omega_c^2 I=0\,,\label{RLC}
\end{equation}
and we have $Q=(1/R)\sqrt{L/C}$.

We may quantize a classical $LC$-oscillator by replacing the amplitudes $I$ and $V$
by {\em operators} \cite{Devoret1997,Girvin2014},
\begin{equation}
\hat{V}=\sqrt{\frac{\hbar\omega_{\rm c}}{2C}}\left(\hat{a}+\hat{a}^{\dagger}\right),\ \hat{I}=i\sqrt{\frac{\hbar\omega_{\rm c}}{2L}}\left(\hat{a}^{\dagger}-\hat{a}\right),\label{LC quantization}
\end{equation}
expressed in terms of photon creation $\hat{a}^{\dagger}$ and annihilation $\hat{a}$ operators that obey the boson commutation relation $[\hat{a},  \hat{a}^{\dagger}]=1$. The photon number operator is $\hat{n}=\hat{a}^{\dagger}\hat{a}$, and the energy Eq.~(\ref{LC_oscillator_energy}) becomes the {\em Hamilton operator} or {\em Hamiltonian},
\begin{equation}
\hat{H}_{\rm c}=\hbar\omega_{\rm c}\left(\hat{a}^{\dagger}\hat{a}+1/2\right),\label{energy_quantized}
\end{equation}
where zero-point energy $\hbar\omega_{\rm c}/2$ contributes a constant shift that we often simply disregard.

In the Heisenberg picture, an operator 
$\hat{A}$ obeys the equation of motion $(d/dt)\hat{A}=(i/\hbar)[\hat{H}_{\rm c},\hat{A}]$ that for the voltage operator,
\begin{equation}
\frac{d}{dt}\hat{V}=\frac{i}{\hbar}\left[\hat{H}_{\rm c},\hat{V}\right]=\frac{1}{C}\hat{I},\label{current_eq_motion}
\end{equation}
agrees with the classical equation $C\dot{V}=I$. The time dependence of the annihilation operator, found from
\begin{equation}
\frac{d}{dt}\hat{a}=\frac{i}{\hbar}\left[\hat{H}_{\rm c},\hat{a}\right]=-i\omega_{\rm c}\hat{a},\label{equation_motion_annihilation_operator}
\end{equation}
is $\hat{a}\propto\exp(-i\omega_{\rm c}t)$, while the creation operator  $\hat{a}^{\dagger}\propto\exp(i\omega_{\rm c}t)$. It is convenient to work in a {\em rotating frame} by introducing the operators $\hat{\tilde{a}}$ and $\hat{\tilde{a}}^{\dagger}$,
\begin{equation}
\hat{a}=\hat{\tilde{a}}e^{-i\omega_{\rm c}t},\quad \hat{a}^{\dagger}=\hat{\tilde{a}}^{\dagger}e^{i\omega_{\rm c}t},\label{rotating_frame}
\end{equation}
that are time-independent. The voltage operator in the rotating frame becomes 
\begin{equation}
\hat{\tilde{V}}=\sqrt{\frac{\hbar\omega_{0}}{2C}}\left(\hat{\tilde{a}}e^{-i\omega_{\rm c}t}\ +\hat{\tilde{a}}^{\dagger}e^{i\omega_{\rm c}t}\ \right).\ \label{voltage_rotating_frame}
\end{equation}
The number of photons $n_i=\langle \Psi_i |\hat{a}^{\dagger}\hat{a}|\Psi_i\rangle$ in an eigenstate \(|\Psi_i \rangle\) vanishes in the ground state. At a finite temperature  $T$, the photon number fluctuates with an average given by the Planck (or Bose-Einstein with zero chemical potential) distribution function,
\begin{equation}
n_B =\left[\exp\left(\frac{\hbar\omega_{\rm c}}{k_{B}T}\right) -  1 \right]^{-1},\label{Bose-Einstein}
\end{equation}
where $k_{{\rm B}}$ is the Boltzmann constant. Thermal photons are called ``incoherent" since their phases are uncorrelated and the thermal average vanishes, $\langle\hat{a}\rangle_B=0$. 

\subsection{Driven LC circuit} \label{DLC}

A time-dependent perturbation at or close to the resonance frequency \(\omega_{\mathrm{c}}\) ``drives" an LC circuit  into excited states.  When adding a time-dependent voltage $V_{\rm D}\cos\omega_{\rm D} t$ to Eq.~(\ref{LC_oscillator_energy}) with driving frequency
$\omega_{\rm D}$ the classical instantaneous energy becomes 
\begin{equation}
U(t)=\frac{CV^{2}}{2}+CVV_{\rm D}\cos\omega_{\rm D} t+\frac{CV_{\rm D}^{2}}{2}\cos^{2}\omega_{\rm D} t+\frac{LI^{2}}{2}.\label{LC_oscillator_energy_driven}
\end{equation}
The term proportional to $V_{\rm D}^{2}$ is the energy of the external drive. The interaction term in Eq. (\ref{LC_oscillator_energy_driven}) is
linear in both $V$ and $V_{\rm D}\cos\omega_{\rm D} t$ and contributes a drive $CV_{\rm D}\sin\omega_{\rm D} t$ to Eq. (\ref{MW_cavity_classical_oscillator})
that enhances the undamped oscillator amplitude by \(\propto 1/\Delta \), where $\Delta =\omega_{\rm D} - \omega_{\rm c}$ is the  {\em detuning}. Damping removes the divergence at a resonance ($\Delta = 0$) with a response proportional to the quality factor $Q$.

The first term in the Hamiltonian of the driven quantum cavity $\hat{H}=\hat{H}_{\rm c}+\hat{H}_{\rm D}$ is  Eq. (\ref{energy_quantized}). The four time-dependent terms in the drive,
\begin{align}
\hat{H}_{\rm D}&=\frac{V_{\rm D}}{2}\sqrt{\frac{\hbar\omega_{\rm c}C}{2}}
\left(\hat{a}^{\dagger}+\hat{a}\right)\left(e^{i\omega_{\rm D} t}+e^{-i\omega_{\rm D} t}\right) \nonumber\\
&=\frac{V_{\rm D}}{2}\sqrt{\frac{\hbar\omega_{\rm c}C}{2}}
\left(\hat{a}^{\dagger}e^{-i\omega_{\rm D} t}
+\hat{a}e^{i\omega_{\rm D} t}\right)
+\mathrm{h.c.},\label{driving_term_quantum}
\end{align}
are not equivalent since a harmonic oscillator
appreciably responds to a time-dependent external force only close to its resonance. In the Heisenberg representation,
 $\hat{a}\propto\exp(-i\omega_{\rm c}t)$, so $\hat{a}\exp(i\omega_{\rm D} t)$ oscillates with frequency $\vert\omega_{\rm D}-\omega_{\rm c}\vert$, while $\hat{a}\exp(-i\omega_{\rm D} t)$
oscillates with frequency $\omega_{\rm D} +\omega_{\rm c}$. When $\Delta$
is of the order or less than the damping rate
of the cavity $\kappa_{\rm c}$, the term $\hat{a}\exp(i\omega_{\rm D} t)$ becomes nearly constant, while the cavity cannot react to the rapidly oscillating  $\hat{a}\exp(-i\omega_{\rm D} t)$.
The amplitude amplification under resonant drive conditions corresponds to the generation of a large photon number that in contrast to  the thermal one are coherent, i.e. phase-locked to the drive with $\langle\hat{a}\rangle \ne 0$.

The {\em rotating wave approximation} (RWA), commonly used for driven systems, is equivalent to disregarding the Hermitian conjugate in Eq. (\ref{driving_term_quantum}),
\begin{equation}
\hat{H}_{\rm D}\approx\frac{V_{\rm D}}{2}\sqrt{\frac{\hbar\omega_{\rm c}C}{2}}\left(\hat{a}^{\dagger}e^{-i\omega_{\rm D} t}+\hat{a}e^{i\omega_{\rm D} t}\right).\label{driving_term_quantum_RWA}
\end{equation}
It holds for sufficiently small detunings or drive amplitudes, i.e., when the resonant response at $\Delta \le\kappa_{\rm c}$ is much larger than the non-resonant one ($\Delta \gg\kappa_{\rm c}$). If this is not the case, we enter the ultra-strong coupling regime, at which the RWA breaks down, see Sec.~\ref{SecIII}.

\subsection{Microwave and optical cavities} \label{SecII-MWOcavities}

MW resonators in the GHz regime come in various designs, see Table \ref{tbl:cavity} in Sec. \ref{SecIV}.  Conducting metal films on an insulating substrate, such as co-planar waveguides or notch filters, confine MW modes in their vicinity and populate them by applied ac currents.
Lumped-elements LC resonators are electric circuits consisting of inductors and capacitors.
Traditional cavities are boxes made from a metal with high conductivity with small holes (ports) for the input and output that confine MWs by screening electric fields and expelling magnetic ones. Cavities have in general more than one resonant frequency. Usually the line broadening governed by the quality factor $Q$ in Eq. \eqref{quality_factor} is much smaller than the mode separation that scales roughly with the square of the inverse cavity size. The single-mode approximation and the simple $RLC$ circuit picture are then appropriate.

Optical cavities operate typically for infrared light at frequencies of hundreds of THz. They consist of insulators with high dielectric constants and \(\rm{\mu}\)m sizes to match the corresponding wave lengths. The can be filled with photons by proximity optical fibers or prism that are illuminated by external lasers.
While the magnetic field component of the radiation dominates the interaction with spins in the MW regime, the direct Zeeman interaction is suppressed at high frequencies until the second order interaction of the spin with the electric field as mediated by spin-orbit coupling \cite{Fleury1968} takes over at optical frequencies. In the intermediate ~THz regime, the spin-photon interactions with both electric and magnetic fields are significant \cite{Kampfrath2013}. 

The textbook example of a cavity is a Fabry-Perot interferometer  (Fig. \ref{Fig:RLC_FPC}b). The solution of the Maxwell equations with reflecting boundary conditions at the two mirrors at a distance $\ell$  may be labeled by an positive integer $p$ with mode frequencies $\omega_{p}=\pi pc/\ell$ and amplitudes $u_{p}(\mbox{r})$, where $c$ is the speed of light. When the mirrors are slightly transparent or contain small holes, a cavity mode with frequency $\omega_{\rm D}$ can be populated by photons from a source on the left, leading to observable transmission $S_{21}(\omega_{\rm D})$ and reflection $S_{11}(\omega_{\rm D})$ amplitude spectra peaked at the mode frequencies $\omega_p$. 

The cavity fields can be quantized analogously to an \textit{LC}
resonator. By expanding the Cartesian components of the electric field into the cavity eigen modes $u_{p}(\mbox{r})$ 
\begin{equation}
\mbox{E}_{x}(\mbox{r},t)=\sum_{p}E_{p}u_{p}(\mbox{r})\left(\hat{a}_{p}e^{-i\omega_{p}t}+\hat{a}_{p}^{\dagger}e^{i\omega_{p}t}\right),\label{quantization_cavity_field}
\end{equation}
where $E_{p}\propto\sqrt{\omega_{p}}$ and  $\hat{a}_{p}$ is
the creation operator for a photon in the mode $p$ with bosonic commutators
$\left[\hat{a}_{p},\hat{a}_{p'}^{\dagger}\right]=\delta_{pp'}$,
$\left[\hat{a}_{p}^{\dagger},\hat{a}_{p'}^{\dagger}\right]=\left[\hat{a}_{p},\hat{a}_{p'}\right]=0$.
The EM Hamiltonian is then a sum of harmonic oscillators,
\begin{equation}
\hat{H}_{{\mathrm{} c}}=\hbar\sum_{p}\omega_{p}\hat{a}_{p}^{\dagger}\hat{a}_{p}\,,\label{eq:H_free_cavity}
\end{equation}
in which we disregarded the zero-point energy $\hbar \sum_{p} \omega_p/2$, even though it can affect quantum noise correlations \cite{Clerk2010}. 

More generally, we can quantize the vector potential $\mathbf{A}(\mathbf{r},t)$
(with $\mathbf{B}=\nabla\times\mathbf{A}(\mathbf{r},t)$ and $\mathbf{E}=-\partial\mathbf{A}(\mathbf{r},t)/\partial t$),
which is convenient in the Coulomb gauge $\nabla\cdot\mathbf{A}(\mathbf{r},t)=0$. The
 Maxwell equations in the absence of sources read
\begin{align}
\nabla\cdot{\bf D}= & 0, & \nabla\times{\bf E}=-\partial{\bf B}/\partial t, \label{eq:Maxwell_free}\\
\nabla\cdot{\bf B}= & 0, & \nabla\times{\bf H}=\partial{\bf D}/\partial t.\nonumber 
\end{align}
The magnetic induction ${\bf B}$ and the displacement field ${\bf D}$
depend on frequency and material dependent response functions. In linear response,
\begin{align}
{\bf B}=\overleftrightarrow{\mu}{\bf H},\qquad{\bf D}=\overleftrightarrow{\varepsilon}{\bf E},
\end{align}
where $\overleftrightarrow{\mu}$ and $\overleftrightarrow{\varepsilon}$ are the magnetic permeability and electric permittivity tensors, respectively. The magnetic induction then satisfies the wave equation,
\begin{align}
\nabla\times\left[\varepsilon \varepsilon_{0}\overleftrightarrow{\varepsilon}^{-1}\cdot\left(\nabla\times\mu \mu_{0}\overleftrightarrow{\mu}^{-1}\cdot{\bf B}\right)\right]-k^{2}{\bf B}=0,\label{WaveEquation}
\end{align}
where $k^{2}=\omega^{2}\varepsilon\varepsilon_{0}\mu\mu_{0} = (n \omega/c)^2$ with $\varepsilon_{0}$ ($\varepsilon$) and $\mu_{0}$ ($\mu$) are the scalar vacuum (relative) permittivity and permeability of the medium, respectively. Here $c = (\mu_0 \varepsilon_0)^{-1/2}$ is the speed of light in vacuum and $n=\sqrt{\varepsilon}$ is the refractive index of the cavity medium. A similar equation is satisfied by ${\bf D}$.

In air or non-magnetic dielectrics, $\overleftrightarrow{\mu}=\mu_{0}\mathbf{1}$, where \(\mathbf{1}\) is the unity tensor, is an excellent approximation at optical frequencies at which the magnetic response is negligibly small. In an isotropic medium ${\bf D}=\varepsilon_{0}\varepsilon{\bf E}$.  At interfaces, the fields inside and outside of a body obey boundary conditions at the surface such as
\begin{align}
{\bf n}\times\left({\bf E}_\mathrm{out}-{\bf E}_\mathrm{in}\right)=0,\quad{\bf n}\cdot \left({\bf B}_\mathrm{out}-{\bf B}_\mathrm{in}\right)=0,\label{BCs}
\end{align}
where the unit vector ${\bf n}$ is the outward normal. 

Equations (\ref{eq:Maxwell_free}) reduce to the wave equation for the vector potential $\mathbf{A}(\mathbf{r},t)$. In a homogeneous material,
\begin{equation}
\nabla^{2}\mathbf{A}=\mu_{0}\varepsilon\partial^{2}\mathbf{A}/\partial t^{2}\,.\label{eq:wave_eq_A}
\end{equation}
Working with complex phase factors implies working with positive and negative frequencies with time dependence $\mathbf{A}(\mathbf{r},t)=\mathbf{A}^{+}(\mathbf{r},t)+\mathbf{A}^{-}(\mathbf{r},t)$,
and $\mathbf{A}^{+}(\mathbf{r},t)=\left[\mathbf{A}^{-}(\mathbf{r},t)\right]^{*}$. The function $\mathbf{A}^{+}(\mathbf{r},t)=\sum_{k}a_{k}\mathbf{u}_{k}(\mathbf{r})e^{-i\omega_{k}t}$, with $\omega_k = ck/n$,
solves Eq. (\ref{eq:wave_eq_A}). Quantization proceeds
by promoting the amplitudes $a_{k}$ and $a_{k}^{*}$ to bosonic annihilation and
creation operators $\hat{a}_{k}$ and $\hat{a}_{k}^{\dagger}$, respectively. The Hamiltonian reduces again to collection of harmonic oscillators, as in Eq. \eqref{eq:H_free_cavity}. The solutions of the Helmholtz equation,
\begin{equation}
\left(\nabla^{2}+k^{2}\right)\mathbf{u}_{k}(\mathbf{r})=0, \label{eq:Helmholtz}
\end{equation}
form an orthogonal complete set that that can be normalized, for example, to the volume of the cavity $\int \mathbf{u}_{k}\cdot\mathbf{u}^*_{k'} {\rm{d}}^3 r=V\delta_{k,k'}$. The eigenstates $\mathbf{u}_k$ are two-dimensional vectors in a given polarization basis. They are subject to boundary conditions as Eq. (\ref{BCs}), which in turn depend on the specific cavity, e.g. geometry and material.
Dissipation can be taken into account by imaginary component of $\omega_{k}$ that is proportional to the the loss rate $\kappa_{\rm c}$ that to leading order does not modify the mode functions $\mathbf{u}_{k}(\mathbf{r})$ of the ideal cavity. 

We thus arrive at the operators for electric and magnetic field
\begin{align} 
  \hat{\mathbf{E}}^{+}(\mathbf{r},t) & =i\sum_{k}\sqrt{\frac{\hbar\omega_{k}}{2V\varepsilon_{0}\varepsilon}}\hat{a}_{k}\mathbf{u}_{k}(\mathbf{r})e^{-i\omega_{k}t},
  \label{eq:E op}\\
\hat{\mathbf{B}}^{+}(\mathbf{r},t) & =i\sum_{k} \sqrt{\frac{\hbar}{2V\varepsilon_{0}\varepsilon\omega_{k}}}\hat{a}_{k} \mathbf{k}\times \mathbf{u}_{k}(\mathbf{r})e^{-i\omega_{k}t}, \label{eq:M op}
\end{align}with  $\hat{\mathbf{E}}^{-}(\mathbf{r},t)=\left(\hat{\mathbf{E}}^{+}(\mathbf{r},t)\right)^{\dagger}$, $\hat{\mathbf{B}}^{-}(\mathbf{r},t)=\left(\hat{\mathbf{B}}^{+}(\mathbf{r},t)\right)^{\dagger}$.

For dielectric cavities it can be convenient to replace the volume $V$ in Eqs.~\eqref{eq:E op} and \eqref{eq:M op} by an effective mode volume $V_{k}$, defined as  
\begin{equation} \label{eq:modevolume}
V_{k} =\frac{\int{ |\mathbf{E}_{k}(\mathbf{r})|^2 {\rm{d}}^3 r}}{\max{|\mathbf{E}_{k}(\mathbf{r})|^2}},
\end{equation}
where $\mathbf{E}_{k}$ is the mode function for mode $k$ with arbitrary normalization. 
When the amplitude $\mathbf{E}_{k}$ is chosen such that the energy stored in the mode is that of a single photon $\hbar \omega_k$ as in the Hamiltonian \eqref{eq:H_free_cavity}, we obtain the maximum amplitude of the electric field per photon $\max{|\mathbf{E}_{k}(\mathbf{r})|}= \sqrt{\hbar \omega_k/(2\varepsilon_0\varepsilon V_k)}$ and the modified normalization condition $\int \mathbf{u}_{k}\cdot\mathbf{u}^*_{k'} {\rm{d}}^3 r=V_k\delta_{k,k'}$ \cite{Safavi-Naeini2014}. The effective mode volume is a measure of the spatial extension of the light field which can useful when dealing e.g. with optical surface states, see Sec. \ref{SecV}. 

The polarization degeneracy of photons in a continuum is broken at interfaces. The polarization states can often be classified as transverse electric (TE) and transverse magnetic (TM) modes, in which there are no magnetic and electric field components along the propagation direction, respectively, also at curved interfaces \cite{Joannopoulos2010}. The electric field components of quasi-TE and TM modes at dielectric resonators as in Sec. \ref{sec:VC} are polarized normal and parallel to the interface, respectively.  In the the following, we return to labeling the modes by a discrete index $p$ rather than a wave number \textit{k}.

\subsection{Input-output formalism} \label{secII:io}
When a cavity is in contact with a photon source such as a MW drive or laser at frequency $\omega_{{\rm D}}$, the coupling term,
\begin{equation}
\hat{H}_{{\rm {\rm D}}}= \sum_p \hbar A_{p}\left( \hat{a}_{p}e^{i\omega_{{\rm D}}t}+\hat{a}_{p}^{\dagger}e^{-i\omega_{{\rm D}}t} \right), \label{eq:H_driving}
\end{equation} should be added to Eq. (\ref{eq:H_free_cavity}) , where the interaction $A_{p}$ with a cavity mode $p$ depends on the driving power $\mathcal{P}$ as $ \vert A_p \vert ^2 \propto \mathcal{P}$, and we use the rotating wave approximation introduced in Sec. \ref{DLC}.  Focusing on a single mode, we can simplify the time dependence by the unitary transformation to the rotating frame $\hat{H} \rightarrow  \hat{U}\hat{H}\hat{U}^{\dagger}-i\hbar\hat{U}\partial\hat{U}^{\dagger}/\partial t$ with $\hat{U}=e^{-i\omega_{{\rm D}}t\hat{a}_{p}^{\dagger}\hat{a}_{p}}$.
The transformed single-mode cavity Hamiltonian including the driving term is
\begin{equation}
\hat{H}_{{\rm c}} + \hat{H}_{\rm D}  \rightarrow
-\hbar\Delta_{p}\hat{a}_{p}^{\dagger}\hat{a}_{p}+\hbar A_{p}(\hat{a}_{p}+\hat{a}_{p}^{\dagger}), \label{eq:RF_trafo}
\end{equation}
where the operators $\hat{a}_{p},\hat{a}_{p}^{\dagger}$ are now in
the rotating frame (denoted by $\hat{\tilde{a}}$ in Sec. \ref{DLC})
When $\Delta_{p}=\omega_{{\rm D}}-\omega_{p}>0$ ($\Delta_{p}<0$) the system is ``blue'' (``red'') detuned.  \cite{Lee2015} discuss the complications occurring when the cavity mode couples to multiple input channels.

The external ports serve to drive and also to probe the cavity, by measuring the transmission or reflection of input photons, while coupling of a closed cavity to the environment induces noise and dissipation. In the following, we introduce the
input-output formalism \cite{Gardiner1985,Walls2008,Clerk2010} that addresses these effects. For technical details we refer to Appendix E of \onlinecite{Clerk2010}.  The total Hamiltonian of the system is given by $\hat{H}_{{\rm tot}}=\hat{H}_{{\rm sys}}+\hat{H}_{{\rm bath}}+\hat{H}_{{\rm int}}$. In this expression, $\hat{H}_{{\rm sys}}$ is the Hamiltonian of the empty cavity $ \hat{H}_{{\rm c}} $ and additionally can contain other terms describing the load such as a magnet (see Sec. \ref{SecIII}). Furthermore, $\hat{H}_{{\rm bath}}$ represents the environment, and $\hat{H}_{{\rm int}}$ its interaction with the system including $\hat{H}_{\rm D}$. Heisenberg equation of motion $\hbar\partial{\hat{a}}/\partial t=i[\hat{H}_{{\rm tot}},\hat{a}]$ then governs the cavity field dynamics.  As discussed in textbooks such as \cite{Meystre2007}, contact with an  environment treated as a large ensemble of harmonic oscillators without memory (Markov approximation) turns  the Heisenberg equation into a stochastic Langevin equation. Focusing on the empty cavity, dropping the mode index $p$, and going to the rotating frame \cite{Aspelmeyer2014}, we write
\begin{equation}
\frac{\partial}{\partial t} \hat{a}(t)=i\Delta\hat{a}(t)-\frac{\kappa}{2}\hat{a}(t)+\sqrt{\kappa_{{\rm ex}}}\hat{a}_{{\rm in}}(t)+\sqrt{\kappa_{{\rm 0}}}\hat{d}_{0}(t). \label{eq:QLE}
\end{equation}
The amplitude decays via the loss term $ \sim -\hat{a}(t)$, while actuation and detection are represented by an input mode that drives or probes the cavity (here $\hat{a}_{{\rm in}}$). The thermal environment introduces noise via $\hat{d}_{{\rm 0}}$. The fluctuation-dissipation theorem governs the statistics of the bosonic operators $\hat{a}_{{\rm in}}$ and $\hat{d}_{{\rm 0}}$
in terms of the extrinsic and intrinsic loss rates $\kappa_{\rm ex}$ and  $\kappa_{\rm 0}$ (see Eq. (\ref{eq:kappa})). In a Fabry-Perot cavity, for example, a semi-transparent mirror can serve as the input and output channel, see Fig. \ref{Fig:RLC_FPC}(b), whereas non-monitored losses through the second mirror would be covered by $\hat{d}_{{\rm 0}}$. Note that $\hat{a}$ and $\hat{a}_{{\rm in}}$, $\hat{d}_{{\rm 0}}$ have different units.  In particular, $\langle\hat{a}_{{\rm in}}^{\dagger}\hat{a}_{{\rm in}}\rangle$ is the \emph{rate} of incoming photons that is proportional to the input power, $\mathcal{P}=\hbar\omega_{{\rm D}}\langle\hat{a}_{{\rm in}}^{\dagger}\hat{a}_{{\rm in}}\rangle$. Furthermore, $\hat{a}_{{\rm in}}$ is a coherent drive with a finite expectation value $\langle\hat{a}_{{\rm in}}\rangle=\alpha_{{\rm in}}$,
whereas $\langle\hat{d}_{0}\rangle=0$ is incoherent. We assume memoryless Markov-like fluctuations for both $\hat{\delta}=\hat{a}_{{\rm in}}-\alpha_{{\rm in}}$ and $\hat{d}_{0}$,
\begin{align}
\langle\hat{\delta}(t')\hat{\delta}^{\dagger}(t'')\rangle & =(\overline{n}_{p}+1)\delta\left(t'-t''\right), \label{eq:correlators}\\
\langle\hat{\delta}^{\dagger}(t')\hat{\delta}(t'')\rangle & =\overline{n}_{p}\delta\left(t'-t''\right), \nonumber 
\end{align}
where $\overline{n}_{p}=n_B( \omega_p)$ is the Planck distribution Eq. (\ref{Bose-Einstein}). The fluctuations of $\hat{d}_0$ obey Eq. (\ref{eq:correlators}) as well. This approximation  holds when the interaction with the bath acts only over a narrow frequency band around $\omega_{p}$. Thermal noise may be disregarded when $\hbar\omega_{p}/k_{{\rm B}}T \gg 1$, which for MWs requires cooling to the temperatures below 1 K. On the other hand, setting $\overline{n}_{p}^{{\rm }}=0$ is allowed for optical cavities even at room temperature. 

The input photons that enter the cavity can be reflected as an output field $\hat{a}_{{\rm out}}$, see Fig. \ref{Fig:RLC_FPC}(b). An equation analogous to Eq.~\eqref{eq:QLE} for $\hat{a}_{{\rm out}}$ is fulfilled by
\begin{equation} \label{losses_cavity_general}
\hat{a}_{{\rm out}}=\hat{a}_{{\rm in}}-\sqrt{\kappa_{{\rm ex}}}\hat{a}\,.
\end{equation}
The expectation value of $\hat{a}_{{\rm out}}$ is the reflection amplitude or  scattering matrix element $S_{11}=\langle\hat{a}_{{\rm out}}\rangle/\langle\hat{a}_{{\rm in}}\rangle$ while the reflected intensity is $\vert S_{11} \vert^2$. In the steady-state defined by $(\partial/\partial t)\langle{\hat{a}}(t)\rangle=0$, Eq. \eqref{eq:QLE} for an empty cavity leads to $\langle\hat{a}\rangle=\sqrt{\kappa_{{\rm ex}}}\alpha_{{\rm in}}/\left(\kappa/2-i\Delta\right)$ and therefore, using Eq. \eqref{losses_cavity_general},
\begin{equation} 
S_{11}(\Delta)=\frac{\langle\hat{a}_{{\rm out}}\rangle}{\langle\hat{a}_{{\rm in}}\rangle}=1+\frac{\kappa_{\rm ex}}{i\Delta-\kappa/2}\,.\label{eq:S11}
\end{equation}
At the resonance ($\Delta=0$) in a high quality cavity with $\kappa_{{\rm ex}}\gg\kappa_{{\rm 0}}$, one has $S_{11}(0)\approx - 1$. When on the other hand $\kappa_{{\rm ex}}=\kappa_{{\rm 0}}$,  $|S_{11}(0)|=0$, i.e. all photons have been absorbed or lost inside the cavity. For a general  $\kappa_{{\rm ex}}$, the reflected intensity $|S_{11}|^2$ has a minimum at the resonance. A two-port cavity has a second input-output field ($\hat{b}_{{\rm in}}, \hat{b}_{{\rm out}}$), leading to the transmission amplitude $S_{21}=\langle\hat{b}_{{\rm out}}\rangle/\langle\hat{a}_{{\rm in}}\rangle$.

\begin{figure}[!t]
\centering
\includegraphics[width=\columnwidth]{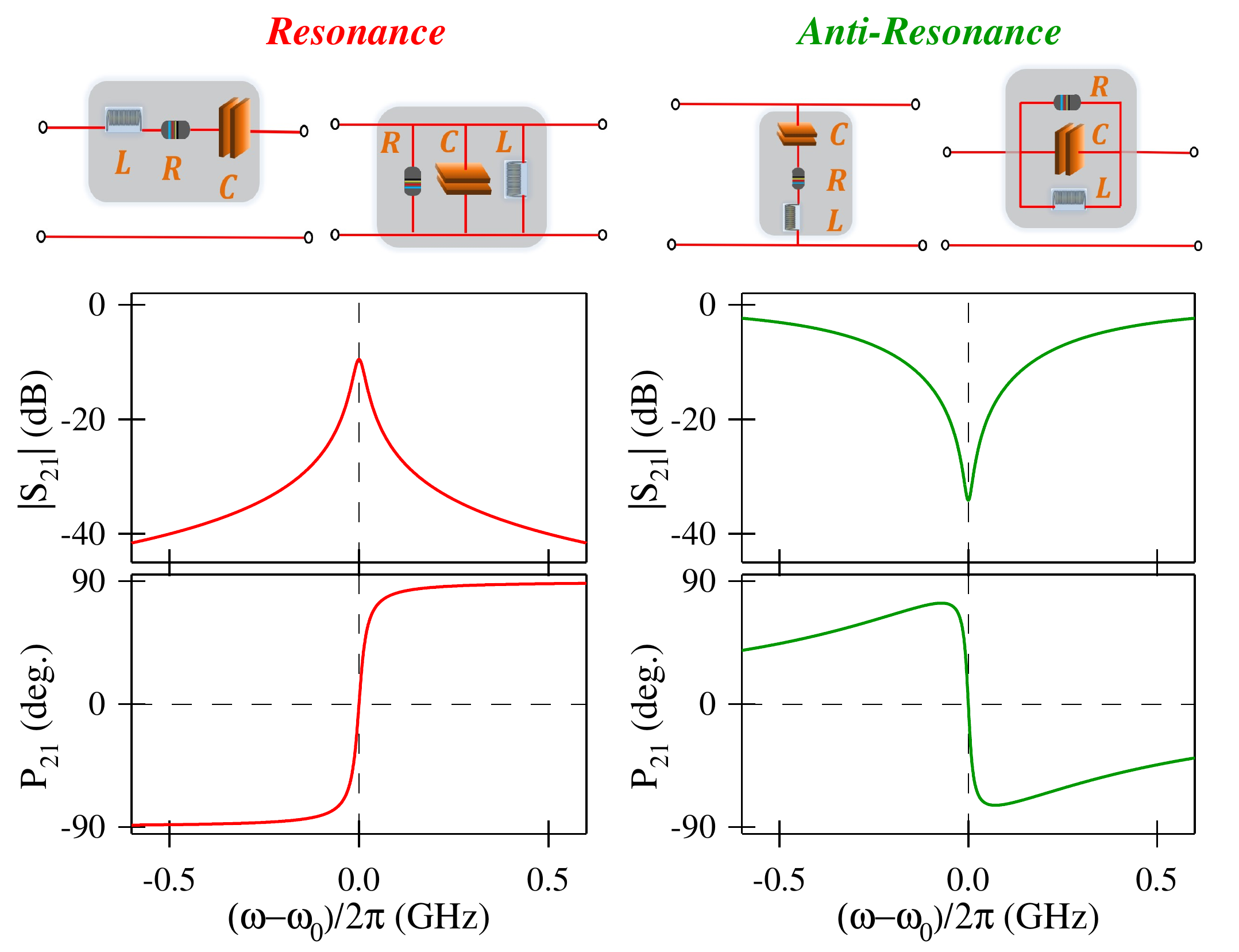}
\caption{\footnotesize{(Color online) Equivalent RCL circuits (top) and their transmission spectra (bottom) that model a cavity resonance and anti-resonance. At resonance, the transmission amplitude exhibits a peak, and the phase jumps by $\pi$. In contrast, at anti-resonance the transmission amplitude dips with phase jumping by $-\pi$ (J. Rao, U. Manitoba, unpublished.}}
\label{Fig:antiR}
\end{figure}

Standing cavity modes are the result of constructive wave interference. In a single-port empty cavity, the resonances always leads to dips in reflection (\ref{eq:S11}) and maxima in the transmission. A two-port (or loaded) cavity can also display {\em anti-resonances} with opposite amplitude and phase characteristics. A resonance (anti-resonance)  is detected as a maximum (minimum) transmission amplitude with a phase jump of $\pi$ ($-\pi$), as shown in Fig. \ref{Fig:antiR}. Both can be modelled by the equivalent RLC circuits in Fig. \ref{Fig:antiR}. In a high-quality closed cavity, input/output ports are weak perturbations, the photons in the cavity have a long dwell time, and constructive interference shows up as resonances. When a cavity is ``lossy", e.g. by invasive input/output ports or internal dissipation, is may become opaque by the destructive interference at anti-resonances.

The discussion above for empty cavities lays the ground for understanding the properties of cavities including magnets or magnetic optical resonators in Sec. \ref{SecIV}.


\section{Magnons} \label{secIIm}

In this review we are interested in describing the interaction of photons with magnons, the elementary excitations of magnetically ordered systems. The simplest example of a magnetically ordered system is a ferromagnet, which can present a finite large magnetization even in the absence of a magnetic field. The magnetization is a result of the presence of permanent magnetic moments in the material, which align to form an ordered state below what is denominated the Curie temperature. These magnetic moments are determined by the spin and orbital angular momentum of the participating atoms. The dynamics of the magnetization is, therefore, that of an angular momentum. In this section we discuss the normal modes of the magnetization dynamics or \textit{spin waves} and their quanta, the \textit{magnons}.

\subsection{Landau-Lifshitz-Gilbert equation of motion}\label{sec:LLG}

 A material with uniform course-grained magnetization $\mathbf{M}$ in the presence of an external magnetic field $\mathbf{H}_0$ gives rise to a Zeeman energy density,
\begin{equation}
h_{\rm Z}=-\mu_{0}\mathbf{H}_0\cdot\mathbf{M},\label{eq:ZeemanSimple}
\end{equation}
and experiences the torque  
\begin{equation}
\dot{\mathbf{M}}=-\gamma\mathbf{M} \times \mu_0 \mathbf{H}_0,\label{eq:LLSimple}
\end{equation}
where $\gamma = g_{\rm Z} \mu_{\rm B}/\hbar$ is the gyromagnetic ratio in which $\mu_{\rm B}$ is the Bohr magneton and $g_{\rm Z}$ is the Land\'e factor. The Landau-Lifshitz (LL) equation can be derived by Poisson bracket algebra in classical mechanics or by quantum mechanical spin commutation rules in the Heisenberg equation of motion. Therefore, the dynamics of classical amplitudes and quantum magnetic operators both obey Eq. \eqref{eq:LLSimple}. Its solution for a homogeneous system describes a precession of the total magnetization vector or ``macrospin'' $\mathbf{S}=-V_{\rm{s}} \mathbf{M}/\gamma$ around the magnetic field, where $V_{\rm{s}}$ is the volume of the magnet. A small angle anti-clockwise precession can be mapped on a harmonic oscillator. Its quantum is the simplest incarnation of the magnon, i.e. the bosonic elementary excitation of the magnetic order.

In real materials dissipation damps the precession. This can be treated  by adding a damping term to Eq. \eqref{eq:LLSimple} that reflects the viscosity by being proportional to \(\dot{\mathbf{M}}\), leads the magnetization back to its equilibrium, and conserves the norm. This is achieved by the Landau-Lifschitz-Gilbert (LLG)
equation \cite{Gilbert2004},
\begin{equation}
\dot{\mathbf{M}}=-\gamma\mu_0\mathbf{M}\times\mathbf{H}_0+\frac{\alpha}{M_{{\rm s}}}\left(\mathbf{M}\times\dot{\mathbf{M}}\right), \label{eq:LLGSimple}
\end{equation}where $\alpha$, the phenomenological  {\it Gilbert damping} constant, approximates possible tensor character, non-locality, and memory effects. The solution of LLG equation without an external drive is an exponentially damped  precession. The linearized LLG equation leads to a resonant response to an ac magnetic field at the ferromagnetic resonance (FMR, see Sec. \ref{subsec:Finite-size-effects}) frequency $\omega_{0}=\gamma\mu_{0}H_{0}$ with a line width $\alpha 
\omega_{0}$. For magnetic metals typically  $\alpha=0.01$ but it can be as small as $10^{-4}$---$10^{-5}$  for YIG thin films and bulk crystals, cf. Sec. \ref{sec:magnon_dissipation}.

The uniform magnetization of Kittel mode is a good description for the FMR at sufficiently high magnetic fields and a homogeneous MW magnetic fields. In general, the magnetization forms equilibrium textures such as domain walls, dynamic textures such as spin waves, and can be driven easily into non-linear regimes. The LLG equation handles these effects by replacing the applied field by an effective $\mathbf{H}_{\rm{eff}}(\mathbf{r})$ that is the functional derivative of the magnetic free energy. Thermal effects can be treated by the stochastic LLG equation in which fluctuating magnetic fields depend on the damping parameter and temperature by the fluctuation dissipation theorem. For more details we refer to Sec. \ref{subsec:Micromagnetics}).

\subsection{Heisenberg Hamiltonian}

According to the Bohr-van Leeuwen theorem, magnetic order does not exist in classical physics. The culprit is the quantum mechanical exchange interaction, a spin-dependent modification of the Coulomb interaction by the effects of Heisenberg uncertainty and Pauli exclusion principle.  A good model for electrically insulating magnets with localized magnetic moments of half-filled 3d or 4f shells is the isotropic Heisenberg
Hamiltonian,
\begin{equation}
\hat{H}_{{\rm H}}=-\frac{1}{2}\sum_{ij}J_{ij} \hat{\mathbf{S}}_{i}\cdot\hat{\mathbf{S}}_{j}, \label{eq:Heisenberg Ham}
\end{equation}
where the operators $\hat{\mathbf{S}}_{i}$ and $\hat{\mathbf{S}}_{j}$ represent spins at lattice sites  $i$ and $j$ that obey angular momentum commutation rules $\left[\hat{S}_{i}^{\alpha},\hat{S}_{j}^{\beta}\right]=i\sum_{\gamma}\epsilon_{\alpha\beta\gamma}\delta_{ij}\hat{S}_{i}^{\gamma}$,
in which $\epsilon_{\alpha\beta\gamma}$ is the Levi-Civita tensor and $\alpha,\,\beta,\,\gamma=\left\{ x,\,y,\,z\right\} $. The exchange parameter $J_{ij}$ is short-ranged and dominated by the nearest neighbor interactions and often approximated by a constant.  When $J\ge 0$, the ground state is then a
ferromagnet (FM). At zero temperature all spins are aligned with total spin $S_{{\rm tot}}=NS$, where $S$ is the spin of a local moment and $N$ is the total number of spins.
The ferromagnetic ground state is an example
of spontaneous symmetry breaking at the critical (Curie) temperature. In the absence of an external magnetic
field it is highly degenerate with $(2S_{{\rm tot}}+1)$ states of equal energy, which corresponds to the classical notion that the energy does not change when rotating the magnetization.

The classical ground state of the antiferromagnetic (AF) Heisenberg model with $J_{ij}=J_{ji}\le0\,\forall_{ i,j}$,
the \emph{classical} ground state on a square bipartite sublattice is the N\'{e}el state --- a state with staggered magnetization with opposite spin directions of the
two sublattices. However, this is not the ground state of the quantum model \cite{Nolting2009}. It rather is a non-degenerate spin singlet, $\langle\hat{\mathbf{S}}_{{\rm tot}}\rangle=\langle\sum_{i}\hat{\mathbf{S}}_{i}\rangle=0$. This statement is known as Marshall's theorem \cite{Auerbach1994}. The actual form of this singlet depends on the, for example, the lattice structure and the interaction range. The {\em quantum} magnetic ground state of a general Heisenberg model in three dimensions is simply not known \cite{Auerbach1994}. 

The local moments in the following chapters are ``large'',  Fe\(^{3+}\) ions which have a half-filled 3d-shell with ordered spins that add up to \(S=5/2\). For our purposes it is then an excellent approximation to interpret the local moments as classical vectors with fixed modulus $|\mathbf{S}_{i}|=S$ that obey coupled LL equations in the external magnetic and local exchange fields. This model is analogous to that for lattices of classical ions, in which quantum effects appear only in the collective dynamics.

The Heisenberg Hamiltonian is usually augmented by symmetry breaking terms, such as the Zeeman interaction with an effective magnetic field \(\mathbf{B}_\mathrm{eff}\),
\begin{equation}
\hat{H}_{{\rm Z}}=-g_{\rm Z} \mu_{{\rm B}}\mathbf{B}_\mathrm{eff}\cdot\sum_{i}\hat{\mathbf{S}}_{i},\label{eq:Zeeman}
\end{equation}
where $\mathbf{B}_\mathrm{eff}$ represents applied and dipolar fields, Dzyaloshinskii--Moriya spin-orbit interactions with neighboring moments, magnetoelastic  
interactions, and the magnetocrystalline  anisotropies. The competition between different interactions depends on materials, geometry, temperature, etc., and can favor magnetic textures such as domain walls or skyrmions. A sufficiently strong uniform external magnetic field always recovers a homogeneous ferromagnetic ground state. The LL equation can be recovered in the continuum limit of the classical Heisenberg model.  

\subsection{Micromagnetic theory\label{subsec:Micromagnetics}}

The field of micromagnetics addresses the ground state and time-dependence of magnetic textures by the solving the LLG equation. When the relevant length scale of the magnetic texture is much larger than atomic distances, the discrete local magnetic moments become a smooth magnetization field $\mathbf{M}(\mathbf{r})$. Since the exchange energy cost of changes of its modulus is very high, it may taken to be constant $|\mathbf{M}_{{\rm s}}(\mathbf{r})|=M_{\rm s}$ \cite{Braun2012}. 

The equilibrium configuration of the magnetization minimizes the free energy functional (disregarding magnetoelastic, antisymmetric exchange, and other contributions),
\begin{eqnarray}\label{EnMag}
E  =  \int_{V}{\rm d}^{3}r \left[ \frac{A}{M_{{\rm s}}^2}
\sum_{i=x,y,z}|\nabla M_{i}|^{2} + U_{{\rm an}}[\mathbf{M}] \right. \nonumber  \\
  \qquad - \left. \mu_{0}\mathbf{M}\cdot\mathbf{H}_{0}-\frac{\mu_{0}}{2}\mathbf{M}\cdot\mathbf{H}_{{\rm d}}[\mathbf{M}]\right].
\end{eqnarray}The first term in the integral is the exchange energy density, since it follows from a gradient expansion of the Heisenberg Hamiltonian and $A\propto J$ \cite{Stancil2009}. $U_{{\rm an}}[\mathbf{M}]$ is the anisotropy energy density. An ``easy axis'' anisotropy along $z$, for instance, takes the form $-K_{{\rm e}}M_{z}^{2}$ with a positive constant $K_{\rm e}$. Furthermore, $-\mu_{0}\mathbf{M}\cdot\mathbf{H}_{0}$ is the Zeeman interaction
induced by an applied magnetic field $\mathbf{H}_{0}$. The dipolar or demagnetization self energy by the stray field $\mathbf{H}_{{\rm d}}[\mathbf{M}]$ is a functional of the entire magnetization, and the factor $1/2$ corrects for double counting. The scalar potential $\phi$ defined as
\begin{equation}\label{Hdem}
\mathbf{H}_{{\rm d}}=-\nabla\phi\,
\end{equation}
obeys the Poisson equation
\begin{equation}\label{Poisson}
\nabla^{2}\phi=\nabla\cdot\mathbf{M},
\end{equation}
 where the right-hand side is the magnetic charge density. The integral representation 
\begin{equation}
\phi=-\frac{1}{4\pi}\int_{V}{\rm d}^{3}r'\frac{\nabla\cdot\mathbf{M}}{|\mathbf{r}-\mathbf{r'}|}+\frac{1}{4\pi}\int_{\partial V}{\rm d}^{2}r'\frac{\mathbf{\hat{n}}(\mathbf{r}')\cdot\mathbf{M}}{|\mathbf{r}-\mathbf{r'}|},\label{eq:phipot}
\end{equation}
has contributions from the volume $\nabla\cdot\mathbf{M}$  and surface $\mathbf{\hat{n}}\cdot\mathbf{M}$
charges at the sample boundaries $\partial V$. The dipolar energy depends strongly on the sample geometry and thereby causes  ``shape anisotropies". According to
\begin{equation}\label{Hd}
E_{{\rm d}}=-\frac{\mu_{0}}{2}\int_{V}{\rm d}^{3}r\mathbf{M}\cdot\mathbf{H}_{{\rm d}}=\frac{\mu_{0}}{2}\int_{{\rm all\,space}}{\rm d}^{3}r|\mathbf{H}_{{\rm d}}|^{2},
\end{equation}
the dipolar energy can be minimized by suppressing the stray field outside the sample by magnetic configuration without surface charges, i.e. when $\mathbf{M}$ is parallel to the surface.  Flux-closure configurations often come at the expense of the exchange energy cost of introducing domain walls. The crossover scale is the \emph{exchange length}  $l_{{\rm ex}}=\sqrt{2A/\mu_0M_s^2}$, obtained by comparing the exchange energy cost of a domain wall of width $l_{{\rm ex}}$, $\epsilon_{{{\rm ex}}}\sim A/(l_{{\rm ex}}^2)$, with the dipolar energy cost of its absence, $\epsilon_{{{\rm d}}}\sim \mu_0M_s^2/2$. Samples smaller than the exchange length of typically a few tens of nm are usually uniformly magnetized.

At equilibrium $\delta E/\delta\mathbf{M}(\mathbf{r})=0,\,$ where
\begin{align}
\delta E= & -\mu_{0}\int_{V}{\rm d}^{3}r\mathbf{H}_{{\rm eff}}\cdot\delta\mathbf{M}  - l_{{\rm ex}}^2  \oint_{\partial V}{\rm d}^{2}r \frac{\partial\mathbf{M}}{\partial\mathbf{n}}\cdot\delta\mathbf{M},   \label{eq:delta E}
\end{align}
and \(\mathbf{H}_{{\rm eff}}=\mathbf{H}_{0}+\mathbf{H}_{{\rm d}}+\mathbf{H}_{{\rm an}}+\mathbf{H}_{{\rm ex}}\) with 
\begin{equation} \label{anisotropy_field_texture}
\mathbf{H}_{{\rm an}} =-\frac{1}{\mu_{0}}\frac{\partial U_{{\rm an}}}{\partial\mathbf{M}}\,, \mathbf{H}_{{\rm ex}}=\frac{2A}{\mu_{0}M_{{\rm s}}^{2}}\nabla^{2}\mathbf{M}.
\end{equation}
Since $|\mathbf{M}(\mathbf{r})|=M_{{\rm s}}$, the variation $\delta\mathbf{M}(\mathbf{r})$ must be transverse,
\begin{equation}
\delta\mathbf{M}=\mathbf{M}\times\delta\mathbf{v},\label{eq:constraint}
\end{equation}
where $\delta\mathbf{v}(\mathbf{r})$ is a small arbitrary vector.
Substituting Eq.~(\ref{eq:constraint}) into Eq. (\ref{eq:delta E})
and using $\mathbf{v}\cdot(\mathbf{w}\times\mathbf{u})=\mathbf{u}\cdot(\mathbf{v}\times\mathbf{w})$, we get
\begin{align}
\mathbf{M}\times\mathbf{H}_{{\rm eff}}[\mathbf{M}] & =0, & \left.\frac{\partial\mathbf{M}}{\partial\mathbf{n}}\right|_{\partial V}=0. \label{eq:EqM}
\end{align}
In second equation we assumed absence of a surface
anisotropy. The nonlinear equations (\ref{eq:EqM}) paint a complex energy landscape with possibly multiple local minima that correspond to (meta) stable magnetic textures such as magnetic vortices in thin-film magnetic disks \cite{Guslienko2008}.

The generalization of the LLG equation \eqref{eq:LLGSimple},
\begin{equation}
\dot{\mathbf{M}}=-\gamma\mu_0\mathbf{M}\times\mathbf{H}_{{\rm eff}}[\textbf{M}]+\frac{\alpha}{M_{{\rm s}}} \mathbf{M}\times\dot{\mathbf{M}}, \label{eq:LLG}
\end{equation}
is the self-consistent and nonlinear problem for the local magnetization dynamics $\mathbf{M}(\mathbf{r})$ that evolves under the effective magnetic field $\mathbf{H}_{{\rm eff}}[\textbf{M}]$ governed by an integral over the entire magnetization. Analytic solutions can be obtained only in limiting cases. In general, the LLG equation must be solved numerically by \textit{micromagnetic simulations}. Thermal noise perturbs the magnetization by a stochastic magnetic field $\textbf{h}(t)$ whose correlation function is linked to the Gilbert damping and temperature by the fluctuation-dissipation theorem  \cite{Brown1963}. The noise power is colored by the Planck distribution function \cite{Barker2020}, but becomes white in the high-temperature limit. In particular, for the homogeneous case,  $\mathbf{M}(\mathbf{r},t) \rightarrow \mathbf{M}(t)$, we have
\begin{equation}
\langle h_p (t)h_q(0) \rangle _\omega =  \frac{2 \alpha \delta_{pq} }{\gamma M_sV}  \frac{\hbar \omega} {e^{\frac{\hbar \omega} {k_BT}}-1}  \xrightarrow[k_B T \gg \hbar \omega] \,  \frac{2 \alpha k_B T}{\gamma M_sV} \delta_{pq}. 
\end{equation}

\subsection{Magnons }

A {\em magnon} is the quantum of a spin wave excitation in a magnetically ordered ground state, i.e.  a coherent precession of the spins around their equilibrium direction. In extended systems, the precession phase is a plane wave with wave vector \textbf{k}. The Kittel mode described in Sec. \ref{sec:LLG} corresponds to $\mathbf{k}=0$.
The magnon frequency dispersion \(\omega _\textbf{k}\) is affected by all interactions that govern the ground state. The exchange energy cost to twist the magnetizations leads to a quadratic dispersion  \(\propto J k^2\) that dominates at large and is negligible at small wave numbers \(k\) compared to other contributions. When the size and shape of the sample are of the order of $k^{-1}$, the dipolar energy is important. 

Bloch \cite{Bloch1930} introduced the first microscopic model for spin waves in a ferromagnet.
Holstein and Primakoff \cite{Holstein1940} included the effects of an external magnetic field and dipolar interactions. They introduced the nonlinear transformation of a spin Hamiltonian to bosonic magnons that carries their name. In terms of the raising and lowering spin operators  $\hat{S}_{i}^{\pm}=\hat{S}_{i}^{x}\pm i\hat{S}_{i}^{y}$  relative to a quantization axis 
along $z$, the isotropic Heisenberg Hamiltonian with Zeeman term and nearest neighbour exchange $J_{ij}=J$ for nearest neighbors (nn) and zero otherwise reads
\begin{equation}
\hat{H}=-\frac{J}{2}\sum_{ij={\rm nn}}\left[\hat{S}_{i}^{+}\hat{S}_{j}^{-}+\hat{S}_{i}^{z}\hat{S}_{j}^{z}\right]+g_{\rm Z}\mu_{{\rm B}} B_0\sum_{i}\hat{S}_{i}^{z}. \label{eq:Heis ladder}
\end{equation}
The Holstein-Primakoff (HP) transformation for a local spin then reads
\begin{align}
\hat{S}_{i}^{+} & =\sqrt{2S}\sqrt{1-\frac{\hat{m}_{i}^{\dagger}\hat{m}_{i}}{2S}}\hat{m}_{i}, & \hat{S}_{i}^{z}=\left(S-\hat{m}_{i}^{\dagger}\hat{m}_{i}\right),\nonumber \\
\hat{S}_{i}^{-} & =\sqrt{2S}\hat{m}_{i}^{\dagger}\sqrt{1-\frac{\hat{m}_{i}^{\dagger}\hat{m}_{i}}{2S}},\label{eq:Holstein Primakoff}
\end{align}
where $\hat{m}_{i}$ and $\hat{m}_{i}^{\dagger}$ are bosonic creation and annihilation operators that
act on the ground state  $|0_{i}\rangle$ with spin aligned along \(z\) as
\begin{eqnarray}
&& \hat{m}_{i}|0_{i}\rangle =0, \qquad \hat{m}_{i}^{\dagger}|n_{i}\rangle  =\sqrt{n_{i}+1}|n_{i}+1\rangle, \label{eq:a ops}\\
&& \hat{n}_i|n_{i}\rangle=\hat{m}_{i}^{\dagger}\hat{m}_{i}|n_{i}\rangle  =n_{i}|n_{i}\rangle, \qquad  \hat{m}_{i}|n_{i}\rangle  =\sqrt{n_{i}}|n_{i}-1\rangle, \nonumber 
\end{eqnarray}
where $n_{i}=S-S_{i}^{z}$ and $\vert n_i \rangle$ is the Fock state with spin projection $S_{i}^{z}$, i.e. 
$\hat{S}_{i}^{z}\vert n_i \rangle=S_{i}^{z}\vert n_i \rangle$. Hence, $n_{i}$ counts the quanta of the spin projection relative to its maximum value \(S\).
The creation operator $\hat{m}_{i}^{\dagger}$ decreases the spin projection, while the annihilation operator $\hat{m}_{i}$ increases it. 
The square root can be expanded into a power series $\sqrt{1-\hat{m}_{i}^{\dagger}\hat{m}_{i}/2S}=1-\hat{m}_{i}^{\dagger}\hat{m}_{i}/4S+...$ in the number operator $\hat{n}_i$. A weakly excited state contains only few magnons, i.e. \(n_i \ll 1\) so that $\sqrt{1-\hat{m}_{i}^{\dagger}\hat{m}_{i}/2S}\approx1$
and $\hat{S}_{i}^{z}\approx S$. In this limit, the Heisenberg Hamiltonian reduces to that of a harmonic oscillator.

To leading order in the HP expansion, the Hamiltonian for a ferromagnetic crystal with \(N\) spins can be diagonalized by the plane-wave ansatz
\begin{align}
\hat{m}_{\mathbf{k}} & =\frac{1}{\sqrt{N}}\sum_{\mathbf{R}_{i}}e^{-i\mathbf{k}\cdot\mathbf{R}_{i}}\hat{m}_{i}\,, & \hat{m}_{\mathbf{k}}^{\dagger}=\frac{1}{\sqrt{N}}\sum_{\mathbf{R}_{i}}e^{i\mathbf{k}\cdot\mathbf{R}_{i}}\hat{m}_{i}^{\dagger}, \label{eq:mk}
\end{align}
where $\mathbf{R}_{i}$ denotes the position of lattice site $i$. The spin wave Hamiltonian then reduces to
\begin{equation}
\hat{H}_{{\rm sw}}=E_{0}(B_{0})+\sum_{\mathbf{k}}\hbar\omega(\mathbf{k})\hat{m}_{\mathbf{k}}^{\dagger}\hat{m}_{\mathbf{k}}, \label{eq:SW Ham}
\end{equation}
where $E_{0}\left(B_{0}\right)=-(S^{2}/2)J\sum_{ij=\rm{nn}}\,-g_{\rm Z} \mu_{{\rm B}}B_{0}NS$
is the energy of the fully polarized ground state. The operator $\hat{m}_{\mathbf{k}}^{\dagger}$ ($\hat{m}_{\mathbf{k}}$) creates (annihilates) a magnon with momentum
$\mathbf{k}$ and energy
\begin{equation} \label{magnon_intermediate}
\hbar\omega(\mathbf{k})=g_{\rm Z} \mu_{{\rm B}}B_{0}-S\left[J(\mathbf{k})-J(\mathbf{k}=0)\right],
\end{equation}
where $J(\mathbf{k})=J\sum_{j} e^{i\mathbf{k}\cdot(\mathbf{R}_{i}-\mathbf{R}_{j})}\,$, and the sum runs over $j$ such that $ij = \rm{nn}$. Note that this expression does not depend on $i$.
For a cubic lattice with constant \(a\), $J(\mathbf{k})=2J\left(\cos(k_{x}a)+\cos(k_{y}a)+\cos(k_{z}a)\right)$. When $ka\ll1$, one obtains parabolic dispersion,
\begin{equation}
\hbar\omega(\mathbf{k})\approx g_{\rm Z} \mu_{{\rm B}}B_{0} + JSa^{2}k^{2}. \label{eq:dispersionFM}
\end{equation}
A magnon is a collective excitation that spreads the flip of a single electron with angular momentum change \(\hbar\) over the entire lattice. The non-interacting {\em spin wave approximation} holds when the magnon numbers  \(\left<\hat{n}_{\mathbf{k}} \right> \ll N\) for all \(\textbf{k}\), where $\hat{n}_{\mathbf{k}}=\hat{m}_{\mathbf{k}}^{\dagger}\hat{m}_{\mathbf{k}}$. Higher order terms in the expansion of the HP transformation in the magnon density operators, or non-linearities, generate interactions between the magnons (see below).

The HP transformation for a single local moment described by Eq. \eqref{eq:LLGSimple} can be employed in principle to handle arbitrary magnetic configurations. Disregarding subtleties associated with the exact quantum ground state, the nearest-neighbor Heisenberg model with $J<0$ describes an antiferromagnet (AFM) with staggered ground state magnetization, i.e. a sublattice $A$ with spin ``up'' and another one 
($B$) with spins pointing ``down''. The sublattice creation, $\hat{m}_{A\mathbf{k}}^{\dagger}$ and $\hat{m}_{B\mathbf{k}}^{\dagger}$, and annihilation, $\hat{m}_{A\mathbf{k}}$ and $\hat{m}_{B\mathbf{k}}$, operators
\begin{align}
\hat{S}^{+}_{i\in A} & =\sqrt{\frac{2S}{N}}\sum_{\mathbf{k}}e^{-i\mathbf{k}\cdot\mathbf{R}_{i}}\hat{m}_{A\mathbf{k}},\nonumber\\
\hat{S}^{+}_{j\in B} & =\sqrt{\frac{2S}{N}}\sum_{\mathbf{k}}e^{-i\mathbf{k}\cdot\mathbf{R}_{i}}\hat{m}_{B\mathbf{k}}^{\dagger},\nonumber \\
S^{z}_{i\in A} & =S-\frac{1}{N}\sum_{\mathbf{k}\mathbf{k}'}e^{i(\mathbf{k}-\mathbf{k}')\cdot\mathbf{R}_{i}}\hat{m}_{A\mathbf{k}}^{\dagger}\hat{m}_{B\mathbf{k}'},\nonumber \\
S^{z}_{j\in B} & =-S+\frac{1}{N}\sum_{\mathbf{k}\mathbf{k}'}e^{-i(\mathbf{k}-\mathbf{k}')\cdot\mathbf{R}_{i}}\hat{m}_{B\mathbf{k}}^{\dagger}\hat{m}_{B\mathbf{k}'},\label{eq:HP_Trannsf_AFM} 
\end{align}
where the total number of spin is \(2N\). When substituted into Eq. (\ref{eq:Heis ladder}) terms
such as $\hat{m}_{A\mathbf{k}}\hat{m}_{B\mathbf{k}}$ and $\hat{m}_{A\mathbf{k}}^{\dagger}\hat{m}_{B\mathbf{k}}^{\dagger}$
remain, that can be eliminated by a {\em Bogoliubov transformation}
$\hat{\alpha}_{\mathbf{k}}=u_{\mathbf{k}}\hat{m}_{A\mathbf{k}}-v_{\mathbf{k}}\hat{m}_{B\mathbf{k}}^{\dagger}$,
$\hat{\beta}_{\mathbf{k}}=u_{\mathbf{k}}\hat{m}_{B\mathbf{k}}-v_{\mathbf{k}}\hat{m}_{A\mathbf{k}}^{\dagger}$
with real $u_{\mathbf{k}}$ and $v_{\mathbf{k}}$  and $u_{\mathbf{k}}^{2}-v_{\mathbf{k}}^{2}=1$. 
The transformed Hamiltonian is diagonal,
\begin{align}\label{eq:AFM_0}
\hat{H}_{\rm AF} & =E_{0}^{{\rm AF}}\\
&+\hbar\sum_{k}\left[\omega_{\mathbf{k}}^{+}\left(\alpha_{\mathbf{k}}^{\dagger}\alpha_{\mathbf{k}}+\frac{1}{2}\right)+\omega_{\mathbf{k}}^{-}\left(\beta_{\mathbf{k}}^{\dagger}\beta_{\mathbf{k}}+\frac{1}{2}\right)\right],\nonumber
\end{align}
 with $\hbar\omega_{\mathbf{k}}^{\pm}=\sqrt{(JZS)^{2}\left(1-\gamma_{\mathbf{k}}^{2}\right)}\pm g_{\rm Z}\mu_{{\rm B}}B_{0}$, $Z$ is the lattice coordination number, and $\gamma_{\mathbf{k}}=Z^{-1}\sum_{j} e^{i\mathbf{k} \cdot (\mathbf{\mathbf{R}_{i} - \mathbf{R}_j)}}$, where again $ij = \rm{nn}$. Here, $E_{0}^{{\rm AF}}=NJZS(S+1)+\hbar/2\sum_{k}\left(\omega_{\mathbf{k}}^{+}+\omega_{\mathbf{k}}^{-}\right)$ is the zero-point energy.
In the limit $ka\ll1$, the dispersion is linear with
\begin{equation}
\hbar\omega_{\mathbf{k}}^{\pm}=\pm g_{\rm Z}\mu_{{\rm B}}B_{0} + 2J\sqrt{3}Sak\,,\label{eq:dispersionAFM}
\end{equation}
where the factor $\sqrt{3}$ is a geometrical factor for the simple cubic lattice.

The magnons in ferrimagnets with sublattice magnetizations that do not cancel  can be treated analogously.

\subsection{Finite size effects\label{subsec:Finite-size-effects}}

The broken translational symmetry normal to magnetic films and in small magnetic particles leads to standing spin wave modes with a discrete spectrum. As described in Sec. \ref{subsec:Micromagnetics}, the dipolar interaction in ferro/ferrimagnets then generates effective demagnetizing fields that depend on the shape and magnetization direction. The demagnetization field of homogeneously magnetized ellipsoids (including needles and pancakes) \cite{Osborn1945} reads
\begin{equation}
\mathbf{H}_{{\rm d},0} =-\left(N_{x}M_{{\rm s}}^{x},N_{y}M_{{\rm s}}^{y},N_{z}M_{{\rm s}}^{z}\right),\label{eq:Hdem}
\end{equation}
where $N_{x,y,z}$ are the so-called demagnetization factors along the principal
axes, and $N_{x}+N_{y}+N_{z}=1$. Limiting cases are a film (in $xz$-plane: $N_{x}=N_{z}=0\,,\,N_{y}=1$), a cylindrical wire (along $z$: $N_{x}=N_{y}=1/2\,,\,N_{z}=0$), and a sphere ($N_{x}=N_{z}=N_{y}=1/3$). The effective field in the LL equation  ${\mathbf H}_{{\rm eff}}={\mathbf H}_{0}+{\mathbf H}_{{\rm d},0}$, where ${\mathbf H}_{0}=H_{0}\hat{\textbf{z}}+H_{x} (t)\hat{\textbf{x}}$,  leads to $H_{x}^{{\rm eff}}=H_{x}-N_{x}M_{x}$, $H_{y}^{{\rm eff}}=-N_{y}M_{y}$,
and $H_{z}^{{\rm eff}}=H_{0}-N_{z}M_{z}$. This is the typical setup in a ferromagnetic resonance (FMR) experiment.  The LL equation Eq. \eqref{eq:LLSimple} with this effective field modifies the frequency $\omega_{0}$ to leading order in \(M_{x/y}\) to the Kittel formula \cite{Kittel1948}
\begin{equation}
\omega_{0}=\gamma\mu_0\sqrt{\left[H_{z}+(N_{y}-N_{z})M_{z}\right]\left[H_{z}+(N_{x}-N_{z})M_{z}\right]}. \label{eq:WKittel}
\end{equation}
Magnetic anisotropies lead therefore to a spin-wave gap, i.e. a finite resonance frequency for zero applied field. 

\begin{figure}[tb!]
\includegraphics[width=8.5cm]{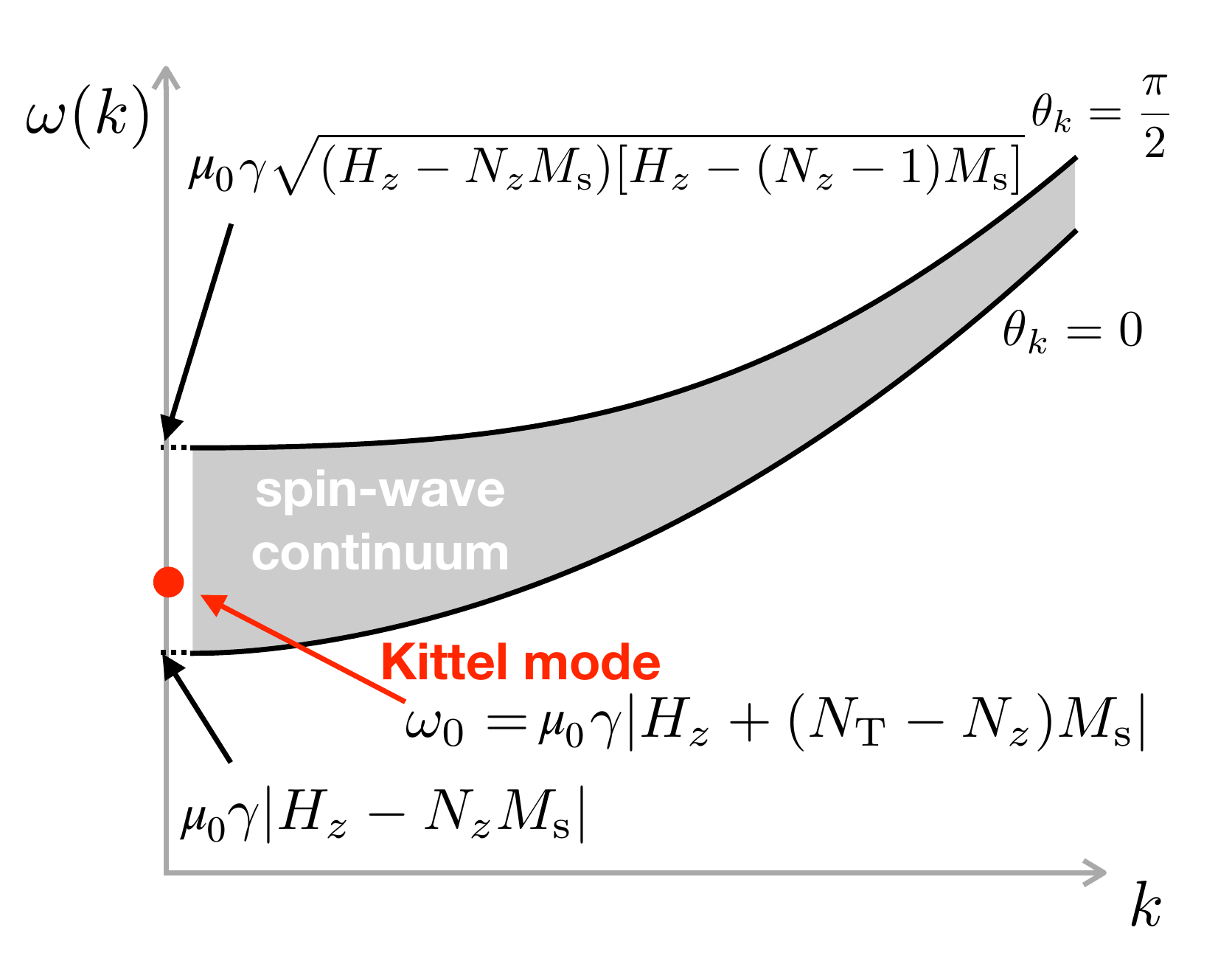}
\caption{(Color online) Schematic magnon spectra $\omega (k)$ for a magnetic prolate ellipsoid   with $N_{{\rm T}}\equiv N_{x}=N_{y}$ as a function of momentum $k$ and propagation direction in terms of the angle \(\theta_k\), see Eqs.~\eqref{eq:WKittel} and (\ref{eq:w_finite_k})..  The shaded area indicates the presence of spin waves. Between the dotted lines at small \(k\) the wave length is of the order of the size of the magnet and the spectrum is discrete \cite{Walker1958}.  } \label{fig:sec2}
\end{figure}

The magneto-dipolar interaction affects not only the Kittel mode, but the entire magnon dispersion \(\omega_{\textbf{k}}\) at small wave vectors. The micromagnetic Landau-Lifshitz equation can be derived from a microscopic 
Heisenberg Hamiltonian with a dipolar interaction  $D_{jk}$ \cite{Clogston1956}, 
\begin{eqnarray}
\hat{H}_{\rm LL}& =& \hat{H}_{{\rm H}}+\hat{H}_{{\rm Z}} \label{eq:MicroHamiltonian} \\
  & + & \frac{1}{2}\sum_{jk}D_{jk}\left[\mathbf{S}_{j}\cdot\mathbf{S}_{k}-\frac{3}{R_{jk}^{2}}\left(\mathbf{R}_{jk}\cdot\mathbf{S}_{j}\right)\left(\mathbf{R}_{jk}\cdot\mathbf{S}_{k}\right)\right]\,, \nonumber
\end{eqnarray}
where  \(\mathbf{R}_{jk} = \mathbf{R}_k - \mathbf{R}_j\) is the vector between spins, and in the absence of spin-orbit interactions one has
\begin{equation}
D_{jk} =\frac{\mu_0 \gamma^2}{4 \pi |R_{jk}|^{3}}\,.
\end{equation}
Let us consider the limit of sufficiently large ellipsoidal magnet in which the eigenstates may be labeled by a continuous wave vector.  When axially symmetric with $N_{{\rm T}}\equiv N_{x}=N_{y}$ and $\textbf{M}_{{\rm s}}$ and $\mathbf{H} \parallel \hat{\textbf{z}}$, the dispersion relation as a function of  $\theta_{k}$ is the angle between the wave vector $\mathbf{k}$ and the quantization axis $z$ reads
\begin{equation}
\omega(\mathbf{k}\neq0)=\gamma\mu_0\sqrt{\omega_{\rm d}(k)\left(\omega_{\rm d}(k)+M_{{\rm s}}\sin^2\theta_{k}\right)}\,,\label{eq:w_finite_k}
\end{equation}
where $\omega_{\rm d}(k)=H_{z}-N_{z}M_{{\rm s}}+ (2/3) JzSa^{2}k^{2}$.
We can recast the Kittel mode frequency Eq. \eqref{eq:WKittel} as  $\omega_{0}=\gamma\mu_0|H_{z}+(N_{{\rm T}}-N_{z})M_{{\rm s}}|$.  In the absence of dipolar forces ($N_{z}=N_{{\rm T}}=0$), $\omega_{0}$ lies at the bottom of the band. However, in general the Kittel mode $\omega_{0}$ can be degenerate with spin waves at finite $k$, as sketched in Fig. \ref{fig:sec2}. 

The degeneracy of the Kittel mode with a manifold of spin waves at finite wavelength creates extra dissipation channels through the magnon interactions in higher order terms of the Holstein-Primakoff expansion. These nonlinearities are captured by the LLG equation of motion, but lost in its linearized version. When allowed, the decay of a small \(k\) spin waves into two large \(k\) ones with half its frequency is very efficient even in a nominally linear regime  \cite{Kurebayashi2011}. Other non-linearities become increasingly important with the number of excited magnons. At a critical value of the pumping power or cone angle of the Kittel-mode precession so-called \emph{Suhl instabilities} occur \cite{Suhl1957}. These dissipation channels also relevant in the quantum regime, see Sec. \ref{SecVI}.

The dispersion given by Eq.~(\ref{eq:w_finite_k}) holds for $2\pi/L \ll k \ll 2\pi/a$, where $L$ is a characteristic diameter of the ellipsoid. When this condition is not fulfilled, the magnons become standing waves with a discrete spectrum. The exchange interaction may be disregarded for particles with $L\gg l_{{\rm ex}}$ and/or wave numbers \(k \ll 1/l_{ex}\).  The solutions in that regime are the \emph{Walker modes} \cite{Walker1957, Walker1958}, i.e. the solutions of the LLG equation with magnetic field $\mathbf{H}^{{\rm eff}}=\left(H_{0}-N_{z}M_{{\rm s}}\right)\hat{\mathbf{e}}_{z}$
where $\mathbf{H}_{0} = \mathbf{B}_0/\mu_0$ is a static applied field. Applying a MW field with frequency \(\omega\) and amplitude $\mathbf{\delta H}$, we write
\begin{eqnarray}
\mathbf{H} & = & \mathbf{H}^{{\rm eff}}+\mathbf{\delta H}e^{-i\omega t}, \label{full_demagnetization_fields} \\
\mathbf{M} & = & \mathbf{M}_{{\rm s}}+\mathbf{\delta M}e^{-i\omega t}, \nonumber
\end{eqnarray}
where $\mathbf{\delta M}\times\mathbf{M}=0$. To leading order in the small $\mathbf{\delta M}$ we obtain
\begin{equation}
-i\omega\mathbf{\delta M}=\gamma\mu_0\left[\hat{\mathbf{e}}_{z}\times\left(M_{{\rm s}}\mathbf{\delta H}-H^{{\rm eff}}\mathbf{\delta M}\right)\right].\label{eq:EOMdeltaM}
\end{equation}
This is basically a Maxwell equation that can be solved using the magnetostatic potential $\psi$, $\delta\mathbf{H}=-\nabla\psi$, and invoking
Poisson's equation $\nabla^{2}\psi=\nabla\cdot\delta\mathbf{M}$. Inside the magnet,
\begin{equation} \label{magnetic_potential_equation}
\left(1+\frac{\Omega_{{\rm H}}}{\Omega_{{\rm H}}^{2}-\Omega^{2}}\right)\left(\frac{\partial^{2}}{\partial x^{2}}+\frac{\partial^{2}}{\partial y^{2}}\right)\psi_{{\rm }}+\frac{\partial^{2}}{\partial z^{2}}\psi=0,
\end{equation}
with $\Omega_{{\rm H}}=H_{{\rm in}}/M_{{\rm s}}$ and $\Omega_{{\rm }}=\omega/\gamma M_{{\rm s}}$, while $\nabla^{2}\psi=0$ otherwise. Imposing
the boundary conditions of (i) continuity of $\psi$ and the normal component
of $\mathbf{\delta H}+\mathbf{\delta M}$ at the surface 
and (ii) $\psi\rightarrow0$ at infinity, leads to characteristic
equations for the magnetostatic resonance frequencies and modes. \onlinecite{Walker1957, Walker1958} showed that
these discrete-, long-wavelength modes also become degenerate with the 
the Kittel mode.

Magnetic thin films are limiting case of the ellipsoid with a continuous but also strongly anisotropic magnon dispersion for small in-plane wave vectors \cite{Kalinikos1986}. Spin waves with $\mathbf{k}\parallel \mathbf{M} $ in in-plane magnetized films are called {\em Backward Moving Volume Waves}, because their negative group velocity for small \(k\) and an suppressed surface amplitude. The exchange interaction bends these modes upward at some finite wave number forming two degenerate low frequency ``valleys''. In the presence of magnon-conserving energy relaxation that is much faster than their decay, magnons may accumulate in these valleys and eventually form a condensate \cite{Demokritov2006}. 

For $\mathbf{k}\perp \mathbf{M}$  the dispersion increases monotonically with $k$.  When  \( \mathbf{M}\) is normal to the plane the spin waves have an isotropic dispersion that starts from a Kittel mode that is pushed to lower frequency by the static demagnetizing field. These are the {\em Forward moving volume waves} because of their positive group velocity and amplitude in the bulk of the film. 

Spin waves with $\mathbf{k}\perp \mathbf{M}$ in the film  plane are exponentially localized to the surface. 
These \emph{Damon-Eshbach} modes propagate with wave vector $\mathbf{k}/k=\mathbf{M}_{{\rm s}}/M_{{\rm s}}\times\mathbf{n}$,
where $\mathbf{n}$ is the outer normal to the magnetic surface~\cite{Damon1961,Gurevich1996}. These waves are therefore ``unidirectional'', i.e. propagate only in one direction that is opposite on the upper and lower surfaces~\cite{Stancil2009}. When the skin depth of the Damon-Eshbach mode is much larger than the film thickness the Damon-Eshbach modes merge into two degenerate counter-propagating modes with equal amplitude. 

\subsection{Normalization of the magnon modes\label{subsec:Magnon-normalization}}

In the absence of dissipation, the magnon eigenmodes $\mathbf{w}_{\eta}(\mathbf{r})$ with frequencies \(\omega_\eta\) solve the  LL equation in the limit of small \(\delta\mathbf{M}\)  in the expansion $\mathbf{M}(\mathbf{r},t)=\mathbf{M_{{\rm s}}}(\mathbf{r})+\delta\mathbf{M}(\mathbf{r},t)$,
around the equilibrium texture $\mathbf{M_{{\rm s}}}(\mathbf{r})$ that is governed by Eqs. (\ref{eq:EqM}). The magnetization is written as
\begin{equation} 
\delta \hat{\mathbf{M}}(\mathbf{r},t)\rightarrow\frac{M_{{\rm s}}}{2}\sum_{\eta}\left[\mathbf{w}_{\eta}(\mathbf{r})\hat{m}_{\eta}+\mathbf{w}_{\eta}^{*}(\mathbf{r})\hat{m}_{\eta}^{\dagger}\right],\label{eq:delta_m}
\end{equation}
where $\hat{m}_{\eta}$ ($\hat{m}_{\eta}^{\dagger}$) is the annihilation (creation) operator of the magnon mode $\eta$ and $\mathbf{w}_{\eta}(\mathbf{r})$
is the corresponding (dimensionless) mode amplitude. It is convenient to normalize the modes to the energy of a single magnon obtained by substituting the amplitude into Eq. \eqref{EnMag} to compute the excess energy relative to $E(\mathbf{M_{{\rm s}}})$ in the limit  $|\delta\mathbf{M}|\ll|\mathbf{M_{{\rm s}}}|$ \cite{Graf2018}, and equating the results to $\hbar \omega_\eta$. This leads to a free magnon Hamiltonian,
\begin{equation}
\hat{H}_{\rm{m}}=\hbar\sum_\eta\omega_\eta \hat{m}_\eta^\dagger\hat{m}_\eta. \label{eq:freemagnonham}
\end{equation}

This normalization can be expressed as  \cite{Sharma2019,Graf2021}
\begin{equation}\label{eq:norm1}
\int d\mathbf{r}\left[{w}_{x}(\mathbf{r}){w}_{y}^{\ast}(\mathbf{r})-{w}_{x}^{\ast
}(\mathbf{r}){w}_{y}(\mathbf{r})\right]  =\frac{4\gamma\hbar}{M_s}.
\end{equation}
In the alternative normalization,
\begin{equation}
\int{\rm d}\mathbf{r}\,|\mathbf{w}_{\eta}(\mathbf{r})|^{2}=\frac{4\gamma\hbar}{M_{\rm s}}\,,\label{eq:m_norm}
\end{equation}
each mode carries one Bohr magneton \cite{Graf2018}. It is equivalent to the energy normalization only for circularly polarized magnon modes, because anisotropies reduce the angular momentum of a magnon \cite{Kamra2016}. It is physically appealing to adopt an effective mode volume as in optics, 
\begin{equation}
V_\textrm{m}^\eta=\frac{\int{ |\mathbf{w}_{\eta}(\mathbf{r})|^2 {\rm{d}}^3 r}}{\max{|\mathbf{w}_{\eta}(\mathbf{r})|^2}}, \label{eq:V_mag_eff}
\end{equation}
which is a measure of the spatial extent of the magnon mode in the whole sample, see Section \ref{SecV}.

\subsection{Magnon dissipation} \label{sec:magnon_dissipation}

We now discuss dissipation mechanisms that cause magnons to decay at a rate $\kappa_{\mathrm{m}}$. 

The material of choice to study the interaction of magnons with cavity photons is YIG, a ferrimagnetic insulator with high critical temperature and record magnetic and acoustic quality \cite{Wu2013}. YIG has \(N=40\)  magnetic moments (with \(S=5/2\)) in a unit cell with volume  $V=(1.24\, \mathrm{nm})^3$, with density $N/V=2\times10^{22}$~cm$^{-3}$  \cite{Gilleo1958}. This is smaller than that of most metallic ferromagnets, but much larger than that of paramagnetic spin ensembles, with $N/V \sim 10^{15}\sim10^{18}$~cm$^{-3}$ \cite{Schuster2010, Kubo2010, Abe2011}. The reported values of Gilbert damping in YIG are in the range between $3 \times 10^{-5}$--$10^{-4}$ with lower values for single crystals and thick films \cite{Klingler2017,Schmidt2020}.

We focus here on the dissipation of the Kittel mode in a millimeter-sized spherical YIG crystal. For more details we refer to \cite{Sparks1964,Gurevich1996}. 

The first of three mechanisms identified in the 1960s~\cite{Sparks1964} is the elastic scattering of a Kittel magnon into degenerate modes with finite wave numbers through the demagnetization field caused by surface roughness (pits). This two-magnon scattering does not depend on temperature and limits the FMR line width at low temperatures. In the so-called slow-relaxation mechanism, the stray field associated with the precessing magnetization modulates the energies of magnetic impurities, which leads to a non-monotonous temperature dependence that peaks between 15~K and 100~K.  It is most efficient when $\omega_{0}T_{\mathrm{hop}} \sim 1$, where  $1/T_{\mathrm{hop}}\propto \exp(-E_{b}/(k_{\mathrm{B}}T))$ is the temperature-dependent hopping rate between energy minima separated by a energy barrier $E_{b}$. The third so-called Kasuya-LeCraw mechanism is intrinsic, viz. a Kittel mode magnon inelastically scatters at thermally-excited phonons or magnons even in otherwise perfect samples. These three-boson processes increase linearly with temperature, i.e. with the number of phonons or magnons that contribute to the scattering.

\onlinecite{Tabuchi2014} observed that the line width of the Kittel mode in YIG spheres below 1~K decreases with decreasing temperature down to \(\sim\)1 K, but increases again at even lower temperatures. The non-monotonous dependence can be a signature of the transverse relaxation by rare earth impurities that can be modeled as two-level systems (TLSs)~\cite{van_vleck1964}: The Kittel magnon decays by exciting an ensemble of near-resonant TLSs with a temperature dependent magnetization that follows the Brillouin function. The decay rate of the Kittel mode can then be estimated to be $\kappa_{\mathrm{TLS}}(T) = \kappa_{\mathrm{TLS}}(0) \tanh \left( \hbar \omega_{0} / 2 k_{\mathrm{B}}T \right)$, where $\kappa_{\mathrm{TLS}}(0)$ is a constant. At low temperatures $\kappa_{\rm m}=\kappa_{\mathrm{mm}}+\kappa_{\mathrm{TLS}}(T)$, where $\kappa_{\mathrm{mm}}$ is the surface roughness contribution, with saturated $\kappa_{\mathrm{TLS}}/2\pi \approx 0.63$ MHz and $\kappa_{\mathrm{mm}}/2\pi\approx$ 0.39 MHz, respectively~\cite{Tabuchi2014}. The TLS contribution dominates the Gilbert damping of thin YIG films at temperatures below 1 K as well  \cite{Kosen2019}, 

A Gilbert damping of $5 \times 10^{-5}$ reported for example by \onlinecite{Kajiwara2010} in YIG films corresponds to a lifetime of 300 ns for a 10 GHz mode. A \(\sim\)100-ns lifetime is short compared to other systems considered for quantum technological applications. It is three orders of magnitude shorter than the lifetimes of state-of-the-art superconducting qubit ~\cite{IBM2017}, and even six orders of magnitude shorter than coherence times of state-of-the-art paramagnetic impurities~\cite{Morton2012}. While dramatic improvements in the magnetic quality are not very likely in the new future, the advantages of magnets such as strong coupling to MWs and easy accessibility of non-linear dynamics more than outweigh this drawback.  

\subsection{Squeezing and non-linearities} \label{subsec:Beyond-Heisenberg}

Magneto-crystalline or dipolar anisotropies cause spin non-conserving terms
such as $\hat{m}_{\mathbf{k}}\hat{m}_{\mathbf{-k}}$ in the Hamiltonian for ferromagnets -- similar to those discussed for AFM around Eq. \eqref{eq:HP_Trannsf_AFM} and compromise the rotating wave approximation. This effect of a shape anisotropy on the Kittel mode macrospin approximation is obvious in  the expression for the demagnetizing field  $\mathbf{H}_{{\rm d},0}$ \eqref{eq:Hdem}: Using Eq. \eqref{EnMag}, the anisotropy contribution to the magnetic energy reads
\begin{equation}
H_{{\rm sa}}=\frac{\mu_{{0}}}{2}\int {\rm{d}}\mathbf{r}\left( N_x M_x^2+ N_y M_y^2 +N_z M_z^2\right),\
\end{equation}
that after a HP transformation generates terms that are quadratic in the magnon operators.  Diagonalizing the Hamiltonian with a Bogoliubov transformation generates a spectrum with a magnon energy gap and eigenstates that carry non-integer spin
\cite{Kamra2016, Kamra2016a, Kamra2017a,Kamra2020}. 
In the quantum limit, the magnons are ``squeezed'', with anisotropic quantum mechanical uncertainties in their amplitudes, i.e. reduced quantum mechanical fluctuations in one magnetization direction at the expense of the other \cite{Walls2008}.  In the classical limit of many magnons, the precession is elliptic, which is a linear combination of counter-precessing states. 

We can illustrate these notions for a prolate magnetic ellipsoid (cigar) subject to a perpendicular magnetic field $H_0 $. The Hamiltonian comes down to of a harmonic oscillator plus squeezing terms  \cite{Sharma2021}
\begin{equation}
	\hat{H}_{\rm {sq}} = \hbar \omega_0\hat{m}^{\dagger}\hat{m} + \frac{\hbar\omega_{\rm m}}{2}\left(\hat{m}^2 + \hat{m}^{\dagger2}\right),\label{eq:Hmag_ellipsoid} 
\end{equation} 
where $\omega_{\rm m} = \left(3N_T - 1\right)\gamma\mu_0M_s/2$ and $\omega_0 = \gamma\mu_0H_0 - \omega_{\rm m}$.
The ground state of the system is the vacuum that can be squeezed by the external field $H_0 $.
The coupling to MW cavity photons can generate macroscopic (involving a large number of spins $> 10^{18}$) ``cat states'', i.e.  quantum mechanical superpositions of two semiclassical magnetizations that point in different directions. At cryogenic temperatures a difference of up  $5\hbar$ should be observable. 

Retaining higher order terms in the Holstein-Primakoff expansion generates magnon interactions and thereby a many-body problem. The expansion of the crystal or dipolar  anisotropy energy parameterized with constant $K$  causes a nonlinearity of the Kerr type, which is quadratic in the magnon numbers. The complex dynamics of the Kittel mode with Kerr nonlinearity 
\begin{equation}
	\hat{H}_{\rm {Kerr}} =  \frac{1}{2} K\left(\hat{m}^{\dagger}\hat{m} \right)^2.\label{eq:HKerr} 
\end{equation} 
can be mapped on that of a Duffing oscillator as was done by \onlinecite{Elyasi2020} in the classical and quantum regimes.  The anisotropy field $\mathbf{H}_{{\rm an}}$ (see Sec. \ref{subsec:Micromagnetics}) depends on the direction of the external magnetic field  $\mathbf{H}_{0}$ with respect to the crystallographic axes of the material \cite{Macdonald1951}. According to \onlinecite{Zhang2019},  a change in the sign of $K$  can be achieved by rotating the applied field.

Atomistic models of coupled LLG equations for individual local moments do not rely on the HP expansion and therefore include the non-linearities and magnon-magnon interaction to all orders. The calculated broadening of the lines in power spectra with temperature can be interpreted in terms of magnon-scattering induced decrease of the magnon lifetimes  \cite{Barker2017}. 

In AFMs, the exchange interaction alone squeezes magnons 
through the terms $\hat{m}_{A\mathbf{k}}\hat{m}_{B\mathbf{k}}$ and $\hat{m}_{A\mathbf{k}}^{\dagger}\hat{m}_{B\mathbf{k}}^{\dagger}$, where A and B refer to different sublattices. The vacuum state of Eq. \eqref{eq:AFM_0} can be obtained from the N\'{e}el
state by a squeezing transformation \cite{Kamra2019}. 

The fabrication of high-quality cavities for THz radiation is still a formidable technological challenge. The cavity magnonics of antiferromagnets is therefore still in its infancy and not a central theme of this review. Some issues of AFM dynamics can be observed also in the GHz regime.  \onlinecite{Johansen2018} predict that AFM magnon can couple non-locally to those in a FM via MW cavity photons. A strong coupling of antiferromagnetic fluctuations to cavity states in MW resonator has been recently demonstrated by  \onlinecite{Mergenthaler2017}.   \onlinecite{Parvini2019} studied the coupling of AFM magnons to  photons in an optical interferometer.

After having discussed  cavity photons, Sec. \ref{secII}, and magnons,  Sec. \ref{secIIm}, separately, we turn in the following Section to their coupling.


\section{Light-matter interaction}\label{SecIII}

Quantum electrodynamics (QED) is the field that covers the quantum aspects of light-matter interactions. We addressed quantization of the cavity field and of the magnetic excitations in Sections \ref{secII} and \ref{secIIm}. The usually weak coupling between light and matter can be treated by perturbation theory or Fermi's golden rule. In EM cavities, such a treatment fails at resonance frequencies with strongly enhanced photon density of states.  In this Section we discuss the physics of both perturbative and non-perturbative magnon-photon interactions.

\subsection{Models for cavity-matter coupling}

We first introduce basic models for the light-matter interaction with emphasis on magnon-photon interactions in hybrid cavity-magnet systems.

\subsubsection{Coupled harmonic oscillators}

\begin{figure}[!t]
\centering
\includegraphics[width=\columnwidth]{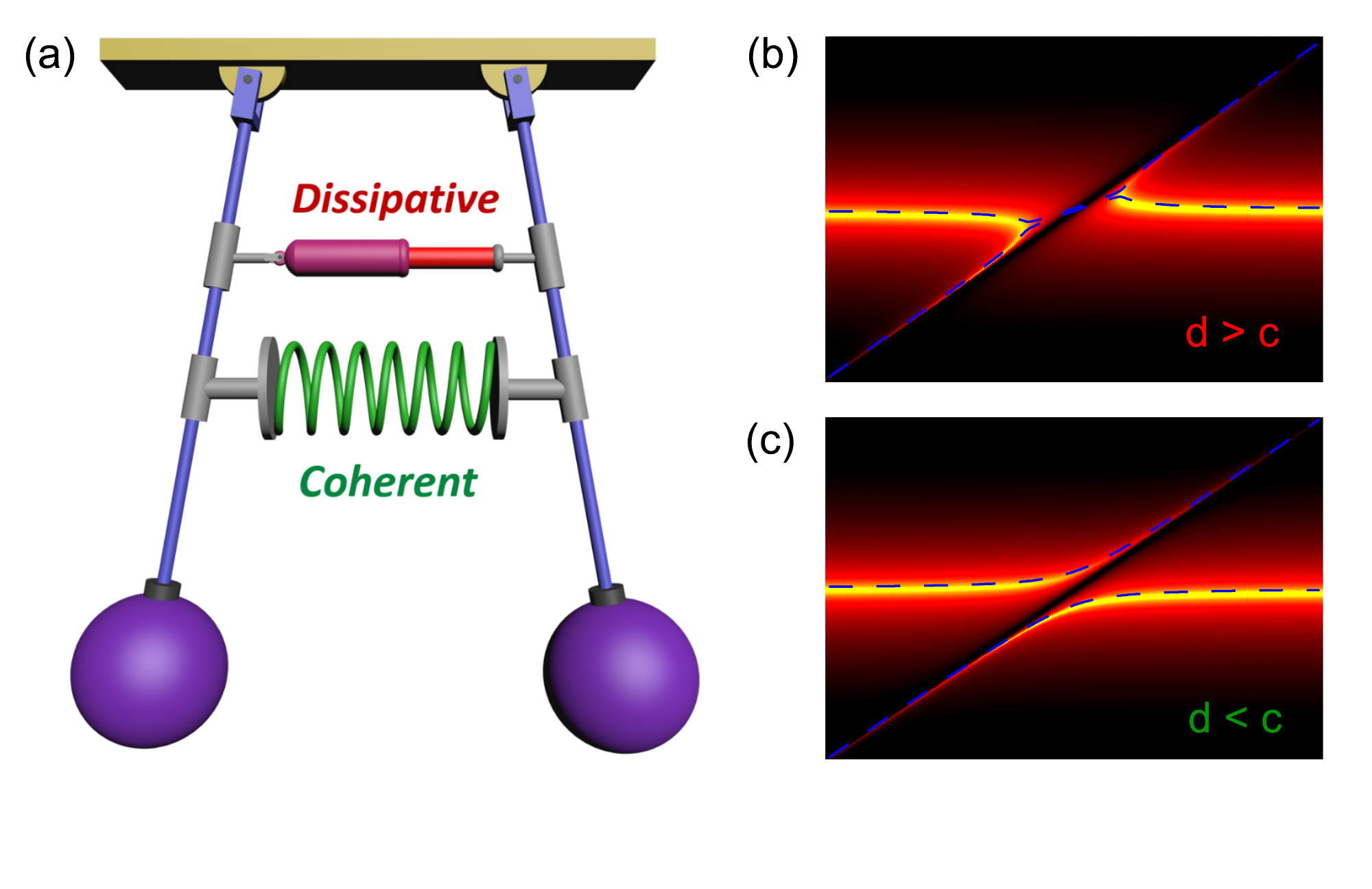}
\caption{(Color online) (a) Two harmonic oscillators connected by a spring with force constant \(c\) and a dashpot with friction constant \(d\). (b,c) show the amplitude of oscillator 1 of the coupled system as a function of  the resonance frequency difference $\omega_2 - \omega_1$ (abscissa) and the drive frequency $\omega_{\rm D} - \omega_1$ (ordinate) for (b) dominantly dissipative and (c) coherent couplings. In units of the resonant frequency $\omega_1 \approx \omega_2$ the parameters are for (b) $d = 0.02\;\omega_1$, $c = 1.6\times 10^{-3}\; \omega_1$ and for (c) $c = 0.02\;\omega_1$, $d = 1.6\times 10^{-3}\; \omega_1$.  with damping rates $\kappa_{1} = 1.6\times 10^{-3}\;\omega_1$, $\kappa_{2} = 2.0\times10^{-4}\;\omega_1$. Results by J.W. Rao (unpublished).
\label{FigIV2}}
\end{figure}

According to Sections \ref{secII} and \ref{secIIm}, both cavity and magnet can be approximated as damped harmonic oscillators. The coherent light-matter interaction introduces a coupling between them that can be pictured by a spring that connects two mechanical pendula as in Fig. \ref{FigIV2}a. A dissipative coupling can  modelled by a ``dashpot", i.e. a damper that resists motion by viscous friction.  When we drive only the first oscillator, the linearized equations of motion for the deflection angles \(\theta_{1,2}\) read
\begin{eqnarray}
\ddot{\theta_1}&+&\omega_1^2\theta_1+\kappa_1\dot{\theta_1}-2c\omega_1\theta_2-2d\dot{\theta_2}=fe^{-i\omega_{\rm D} t}, \nonumber \\
\ddot{\theta_2}&+&\omega_2^2\theta_2+\kappa_2\dot{\theta_2}-2c\omega_2\theta_1-2d\dot{\theta_1}=0,
\label{Eq19}
\end{eqnarray}
where $\kappa_1$ and $\kappa_2$ are the respective damping rates. Here, $c$/\textit{d} are the coupling force/friction constants, respectively, and $fe^{-i\omega_{\rm D} t}$ is a time-periodic force. The conservative elastic force is proportional to the phase difference  $\theta_1 - \theta_2$ between the pendula, while the dashpot acts on the velocity difference $\dot{\theta}_1 - \dot{\theta}_2$. The dynamics depends sensitively on the ratio $c/d$. In frequency space and near the resonance with $\omega_{1,2} \approx \omega_{\rm D}$, Eqs.~(\ref{Eq19}) becomes 
\begin{equation}
 \left[
  \begin{array}{cc}
  \omega_{\rm D}-\omega_1+i \frac{\kappa_1}{2} & c-id\\
  c-id & \omega_{\rm D} -\omega_2+i \frac{\kappa_2}{2}
  \end{array}
  \right]\left[
  \begin{array}{cc}
  \theta_1\\
  \theta_2
  \end{array}
  \right]=-\frac{1}{2\omega_{\rm D}}\left[
  \begin{array}{cc}
  f\\
  0
  \end{array}
  \right].
 \label{Eq:IV3}
\end{equation}

Fig.~\ref{FigIV2} shows the response of the coupled system, $R_1(\omega_{\rm D}) \propto \theta_1/f$, as a function of \(\omega_{\rm D}\) and \(\omega_2\). The anticrossing observed for $d<c$  can be interpreted as a ``level repulsion", while dominantly dissipative coupling,  $d>c$ causes a ``level'' attraction. 

Before addressing the microscopic interaction between photons and magnons, we survey a few frequently used models.

\subsubsection{Resonant coupling}

According to Sec. \ref{secII}, a cavity efficiently modulates the EM density of states at wave lengths of the order of its spatial dimension by maximizing the spacing between the cavity modes.  When the characteristic frequency of the load is close to resonance with a certain cavity mode, and the mode splittings exceed all other relevant energy scales, the system dynamics reduces to that of three levels.  

The Rabi Hamiltonian for a two level system interacting with a single photon mode reads~\cite{Rabi1937,ScullyZubairy}
\begin{align}
\hat{H}_{\rm Rabi}=&\hbar\omega_{\rm a}\hat{a}^{\dagger}\hat{a} + \frac{1}{2} \hbar\omega_{\rm sys}\hat{\sigma}_{z}+\hbar g\left( \hat{a} \hat{\sigma}_{+} + \hat{a}^{\dagger} \hat{\sigma}_{-}\right)\nonumber\\
&+\hbar g\left( \hat{a} \hat{\sigma}_{-} + \hat{a}^{\dagger} \hat{\sigma}_{+}\right), \label{qRabiHam}
\end{align}
where $\omega_{\rm a}$ and $\omega_{\rm sys}$ denote the cavity mode frequency and the level splitting, respectively, $\hat{a}$ ($\hat{a}^{\dagger}$) is the annihilation (creation) operator of the cavity mode, and $g$ the coupling strength. The two-level system can be a real spin $1/2$ in which case $\hat{\sigma}_{-}=\hat{\sigma}_x - i \hat{\sigma}_y$ ($\hat{\sigma}_{+}=\hat{\sigma}^{\dagger}_{-})$ is the lowering (raising) operator and  $\mathbf{\hat{\sigma}}=(\hat{\sigma}_{x},\hat{\sigma}_{y},\hat{\sigma}_{z} )$ is the vector of Pauli matrices. Pseudo spin 1/2 are other two-level systems that obey the spin-1/2 algebra. The last term on the right-hand side of Eq. (\ref{qRabiHam}) is ``counter-rotating". It changes the number of excitations $\hat{n}_{\rm exc}=\hat{a}^{\dagger}\hat{a}+\hat{\sigma}_{+}\hat{\sigma}_{-}$ by two, but conserves parity $\mathcal{\hat{P}}=\exp(i\pi \hat{n}_{\rm exc})$.  Since the Hamiltonian is Hermitian, energy is conserved, but it does not commute with $\hat{n}_{\rm exc}$. \onlinecite{Braak2011} reported an analytic solution of the Rabi model.  

When \(g \ll \omega_{\mathrm{a,sys}}\), the counter-rotating terms oscillate rapidly compared to other length scales and average out efficiently. Disregarding them is the rotating wave approximation (RWA, see Sec.~\ref{secII}), which reduces the Rabi model to the Jaynes-Cummings model \cite{Jaynes1963} with much simpler solutions and a conserved $\hat{n}_{\rm exc}$~\cite{Shore1993}. The Jaynes-Cummings model captures the Rabi oscillations between the two-level system and the cavity mode, but it breaks down when the coupling becomes ultra-strong, see below.

The Dicke model extends the quantum Rabi model to multiple two-level systems coupled to and by a cavity field~\cite{Dicke1954}. \onlinecite{Braak2013} and \onlinecite{Peng2013} solved the $N=3$ and $N=2$ Dicke model exactly, respectively. The Dicke model with RWA or Tavis-Cummings model ~\cite{Tavis1968} describes the cavity QED of multiple two-level systems~\cite{Retzker2007}  including quantum dots~\cite{Fink2009} that resonate with a cavity mode.

The Hopfield model~\cite{Hopfield1958} addresses the interaction between two bosonic modes and can be understood as variation of the Rabi model. In our case 
\begin{align}
\hat{H}_{\rm Hopf}=&\hbar\omega_{a}\hat{a}^{\dagger} \hat{a} + \hbar\omega_{\rm sys} \hat{b}^{\dagger} \hat{b} + \hbar g\left(\hat{a} \hat{b}^{\dagger} + \hat{a}^{\dagger} \hat{b} \right)\nonumber\\
&+\hbar g\left(\hat{a}^{\dagger}\hat{b}^{\dagger} + \hat{a} \hat{b}\right) +H_{\rm dia},\label{HopfieldHam}
\end{align}
where $\hat{b}$ ($\hat{b}^{\dagger}$) is a boson annihilation (creation) operator of an oscillator with frequency $\omega_{\rm sys}$. $H_{\rm dia}\propto A^{2}\propto(\hat{a}^{\dagger}+ \hat{a})^{2}$ is a photon scattering term that is important only in the ultra- and deep strong coupling regimes~\cite{Liberato2014}.

The Hamiltonians above are Hermitian and thereby conserve energy, as appropriate for (nearly) closed systems. The environment is taken into account by  the theory of open quantum systems, in which the Heisenberg equations of motion are replaced by master kinetic equations, such as the Lindblad equation~\cite{Breuer2007}. Their solution can be computationally demanding when the Hilbert space of the combined system is large. The bosonic input-output theory described in Sec. \ref{secII}  integrates the internal dynamics out  to obtain the scattering matrix between the coherent modes in the leads to source and detector. The calculation of the time-dependent operator $\hat{a}(t)$ in the input-output relation $\hat{a}_{\rm out}(t)=\hat{a}_{\rm in}(t)-\sqrt{\kappa_{\rm ex}}\hat{a}(t)$, Eq. (\ref{losses_cavity_general}),  becomes very cumbersome for all but quadratic Hamiltonians~\cite{Gardiner2004}, however. In a third approach the interaction with the environment is included by dissipative terms into the Hamiltonian, that thereby becomes non-Hermitian.

\subsubsection{Off-resonant coupling}

When interacting with infrared photons, the frequency of the cavity mode is very different from the magnetic ones, $\vert\Delta\vert\equiv\vert\omega_{\rm sys}-\omega_{a}\vert\gg g$, which is the  {\em dispersive regime}. The Hopfield Hamiltonian (\ref{HopfieldHam}) originally holds only when the frequencies of the two bosonic modes are nearly equal, but can be adopted to include large detuning, see Sec. \ref{SecVI}. The large detuning limit has been extensively studied in optomechanics \cite{Aspelmeyer2014}, in which the frequency $\omega_{\rm a}$ of  both MW and optical cavity modes is much higher than the vibration frequency $\omega_{\rm sys}$ of a macroscopic mechanical membrane or cantilever. The coupled system is then well represented by the Hamiltonian
\begin{align}
\hat{H}_{\rm OM}=&\hbar\omega_{\rm a}\hat{a}^{\dagger} \hat{a} + \hbar\omega_{\rm sys} \hat{b}^{\dagger} \hat{b} + \hbar g \hat{a}^{\dagger} \hat{a} \left( \hat{b}^{\dagger} + \hat{b}\right),
\label{optomechHam}
\end{align}
where $\hat{b}^{\dagger}$ creates a phonon and $ \hat{b}^{\dagger} + \hat{b}$ is proportional to the displacement operator. The coupling term is the radiation pressure proportional to the number of photons $\hat{a}^{\dagger} \hat{a}$. The constant $g$ is the {\em single-photon coupling rate}.

We are not aware of analytic solutions of the nonlinear Eq. (\ref{optomechHam}). The problem is simplified when the number of photons \textit{N} in the cavity mode is large and the EM field can be treated classically. In this strongly driven limit  $\langle \hat{a} \rangle \approx \langle \hat{a}^{\dagger}\rangle \approx \sqrt{N}$ (see Sec. \ref{secII}). Introducing the fluctuation operator $\hat{a} = \sqrt{N} + \delta \hat{a}$, disregarding higher powers of $\delta \hat{a}$  in Eq. (\ref{optomechHam}), and applying the RWA, the problem reduces to that of two coupled harmonic oscillators
\begin{align}
\hat{H}_{\rm OM} \rightarrow &\hbar\omega_{\rm a}\hat{a}^{\dagger} \hat{a} + \hbar\omega_{\rm sys} \hat{b}^{\dagger} \hat{b} + \hbar g_{N} \left( \delta \hat{a}^{\dagger} \hat{b} + \delta \hat{a} \hat{b}^{\dagger} \right) ,
\label{optomechHam_lin}
\end{align}
where $g_{N} = g \sqrt{N}\gg g$ is the enhanced {\em multi-photon coupling rate}. 

\subsection{Interaction parameters and regimes}

The physics of light-matter systems is governed by the transition and dissipation rates. Different regimes dictate the approximations and techniques of theoretical treatments. Cavities can tailor the coupling strength via the density of states of the EM environment \cite{Purcell1946}, which forms the basis of cavity QED \cite{Mabuchi2002}. The light-matter coupling strength is also proportional to the number of electric and magnetic dipole moments that interact with the cavity photons, which in collective modes scales with the sample size. The dissipation depends on the nature and quality of the cavity and the lifetime of the quasiparticles that depend, e.g., on temperature and disorder in the sample.  The mapping of this entire parameter space is an important task~\cite{Haroche2013}, also for cavity magnonics.



In the \textit{weak coupling} limit the ratio between the coupling strength and the total decoherence rate is small and hybridization cannot take off. The light-matter interaction may treated by perturbation theory or Fermi's Golden Rule. Conventional FMR with Lorenztian MW absorption spectra falls into this class. 

An ensemble of Rydberg atoms in a resonant MW cavity is an early system that can be tuned into the {\em strong coupling} regime~\cite{Kaluzny1983} in which the coupling is larger than the level broadening. Light and matter modes hybridize at the resonance to form polaritons. An injected photon is not an eigenstate and its amplitude oscillates between matter and wave modes within its lifetime. This process is referred to as vacuum Rabi oscillation with a frequency governed by the coupling constant, even though is often a classical wave phenomenon. The minimum mode splitting at the anticrossing is twice the vacuum Rabi frequency. When the coupling is non-resonant, full hybridization cannot be achieved, and strong coupling corresponds to a ``lamp'' shift that is larger than the decay rates. 

Strong coupling has been observed subsequently in single atoms interacting with MW \cite{Meschede1985} and optical \cite{Thompson1992} cavities, atomic Bose-Einstein condensates in optical cavities \cite{Colombe2007}, and excitons in quantum dots (QDs) coupled to photonic resonators~\cite{Hennessy2007}. The quantum manipulation of single atoms in photonic resonators may lead to applications~\cite{Georgescu2012} such as quantum information processing~\cite{Wendin2017,Blais2020} and sensing~\cite{Degen2017}. Circuit QED seeks to couple artificial atoms, such as superconducting qubits~\cite{Blais2020} and semiconductor QDs~\cite{Lodahl2015,Burkard2020} to MW resonators, that in contrast to natural atoms can be tuned into different coupling regimes~\cite{Clarke2008,Devoret2013}.

The \textit{ultra-strong coupling} (USC) regime corresponds to a coupling parameter that approaches the mode frequencies $g/\omega_{\rm c} \lesssim 1$, irrespective of the loss rates, so it does not require the strong coupling condition defined above~\cite{Liberato2017}. The break-down of the RWA in the USC leads, e.g., to light-matter hybridization in the ground state. In the \textit{deep-strong coupling} regime  $g/\omega_{\rm c} \gtrsim 1$ \cite{Casanova2010,Bayer2017,Yoshihara2016} the addition of just one photon may affect the system properties.

\onlinecite{Ciuti2005} suggested to study the USC regime by cyclotron resonance of intersubband transitions in semiconductor quantum wells in a cavity. \onlinecite{Anappara2009}  reported corresponding experiments with $g/\omega_{\rm c} \gtrsim 0.1$. The USC for microwaves was also observed in quantum well inter-Landau level resonances \cite{Muravev2011, Scalari2012}, in two-level systems formed in superconducting circuits \cite{Niemczyk2010, Forn-Diaz2010}, and in optomechanical systems \cite{Benz2016,Pirkkalainen2015}. Optical photons can ultra-strongly couple to molecular excitons at room temperature~\cite{Schwartz2011, Kena-Cohen2013}.

\subsection{Cavity-magnet coupling}

Here we consider specifically magnetizations that interact resonantly with MW and non-resonantly with infrared photons.

\subsubsection{Resonant coupling with microwave cavity modes}

We consider magnons in a closed MW cavity, i.e. an anti-clockwise precession of the magnetization vector around its equilibrium direction defined by a static magnetic field. The  ac MW magnetic field in the cavity is governed by the Maxwell equations of Eqs.~(\ref{eq:Maxwell_free}). The instantaneous energy of the cavity-magnet system is $\hat{H}=\int d{\bf r}\hat{\mathcal{H}} ({\bf r}, t)$ with energy density $\hat{\mathcal{H}}=(\varepsilon_{0}{\bf E}^{2}+\mu_{0}{\bf H}^{2})/2$, where ${\bf H}={\bf B}/\mu_{0}-{\bf M}$, or
\begin{eqnarray}
\hat{\mathcal{H}}({\bf r}, t)=\frac{\varepsilon_{0}}{2}{\bf E}^{2}+\frac{1}{2\mu_{0}}{\bf B}^{2}-{\bf M}\cdot{\bf B}+ \hat{\mathcal{H}}_{\rm m}. \label{TotEDensity}
\end{eqnarray}
The first two terms account for the empty cavity-field Hamiltonian \(\hat{H}_{\rm c}\), the third term is the Zeeman interaction and \( \hat{\mathcal{H}}\) is the magnetic energy density. Using Eqs. (\ref{eq:E op}) and (\ref{eq:M op})
\begin{eqnarray}
\hat{H}_{\rm c}=\sum_{p}\hbar\omega_{p}\left(\hat{a}_{p}^{\dagger}\hat{a}_{p}+\frac{1}{2}\right),
\end{eqnarray}
where $\omega_{p}$ and $\hat{a}_{p}$ are the frequency and annihilation operator of a photon in mode $p$, respectively.

\begin{figure}[!t]
\centering
\includegraphics[width=\columnwidth]{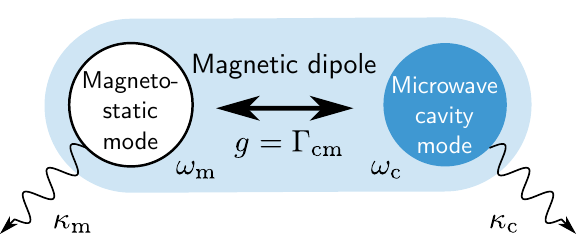}
\caption{\footnotesize{(Color online) The parameters in the interaction Eq. (\ref{ResonantCouplingHam}) between a single cavity and magnon mode \cite{Lachance-Quirion2019}.}}
\label{FigIIIa}
\end{figure}

Substituting the lowest order HP expansion of the magnetization field, Eq.~(\ref{eq:delta_m}), the Zeeman interaction $-{\bf M}\cdot{\bf B}$ in the RWA reads
\begin{eqnarray} \label{Cavity_magnon_coupling_RWA}
\hat{H}_{\rm cm}=\hbar\sum_{p\eta}\left(\Gamma_{p\eta} \hat{a}_{p} \hat{m}^{\dagger}_{\eta} + \Gamma_{p\eta}^{*} \hat{a}^{\dagger}_{p} \hat{m}_{\eta}\right),
\end{eqnarray} 
where $p$ and $\eta$ label cavity and magnon modes, respectively, and the coupling rate $\Gamma_{p\eta}$ is \cite{Soykal2010, Soykal2010a, Flower2019, Bourhill2019}
\begin{eqnarray} \label{resonant_coupling_arb_mode}
\hbar\Gamma_{p\eta}=-\frac{M_s}{2}\sqrt{\frac{\hbar}{2V\varepsilon_{0}\varepsilon\omega_{p}}} \int d{\bf r}\left[\nabla\times{\bf u}_{p}({\bf r})\right]\cdot{\bf w}_{\eta}^{*}({\bf r}).\qquad
\end{eqnarray}
Here, ${\bf u}_{p}({\bf r})$ is a cavity mode amplitude that obeys the wave equation (\ref{eq:Helmholtz})
subject to the boundary conditions of the cavity with volume \(V\), while  ${\bf w}_{\eta}({\bf r})$ is the magnon mode amplitude (\ref{eq:delta_m}). In the total Hamiltonian,
\begin{eqnarray}
\hat{H}&=&\sum_{p}\hbar\omega_{p} \hat{a}_{p}^{\dagger} \hat{a}_{p}+\sum_{\alpha}\hbar\omega_{\eta}\hat{m}_{\eta}^{\dagger}\hat{m}_{\eta}\nonumber\\
&&+\hbar\sum_{p\eta}\left(\Gamma_{p\eta}\hat{a}_{p} \hat{m}^{\dagger}_{\eta}+\Gamma_{p\eta}^{*} \hat{a}^{\dagger}_{p} \hat{m}_{\eta}\right), \label{ResonantCouplingHam}
\end{eqnarray}
we disregard the zero-point energies of field and magnet. Figure \ref{FigIIIa} illustrates the different terms for single cavity and magnon modes. In the following we focus on the situation in which the Kittel mode couples to one or two cavity modes, using $g$ for the magnon-photon coupling constants, $\omega_{\rm c}$ and $\kappa_{\rm c}$ for the frequency and damping rate of the cavity mode and $\omega_{\rm m}$ and $\kappa_{\rm m}$ for the magnon mode, see Fig. \ref{FigIIIa}. Inhomogeneous magnetic fields introduce coupling to more than one magnon mode \cite{Weichselbaumer2019} in terms of form factors \cite{Flower2019,Bourhill2019}.

The coupling of photons to magnetic dipoles is smaller than that to electric dipoles by the fine structure constant 1/137. However, coherent spin ensembles enjoy a collective enhancement of the coupling that scales with the square root of the total number of spins $\sqrt{N}$  \cite{Dicke1954}. If everything else is kept constant, the strong coupling regime can be reached simply by increasing the effective coupling $g_{N}=g\sqrt{N}$ by the sample size. Non-interacting spins form paramagnetic ensembles,  such as nitrogen-vacancy (NV) centers in diamond~\cite{Zhu2011, Kubo2011}, cold atomic clouds~\cite{Verdu2009}, molecules~\cite{Eddins2014}, and dilute magnetic ion-doped oxides~\cite{Schuster2010, Longdell2005, Tkalcec2014, Probst2013}. In MW cavities these systems can reach the strong coupling regime. 

\begin{figure}[tb!]
\includegraphics[width=8.6cm]{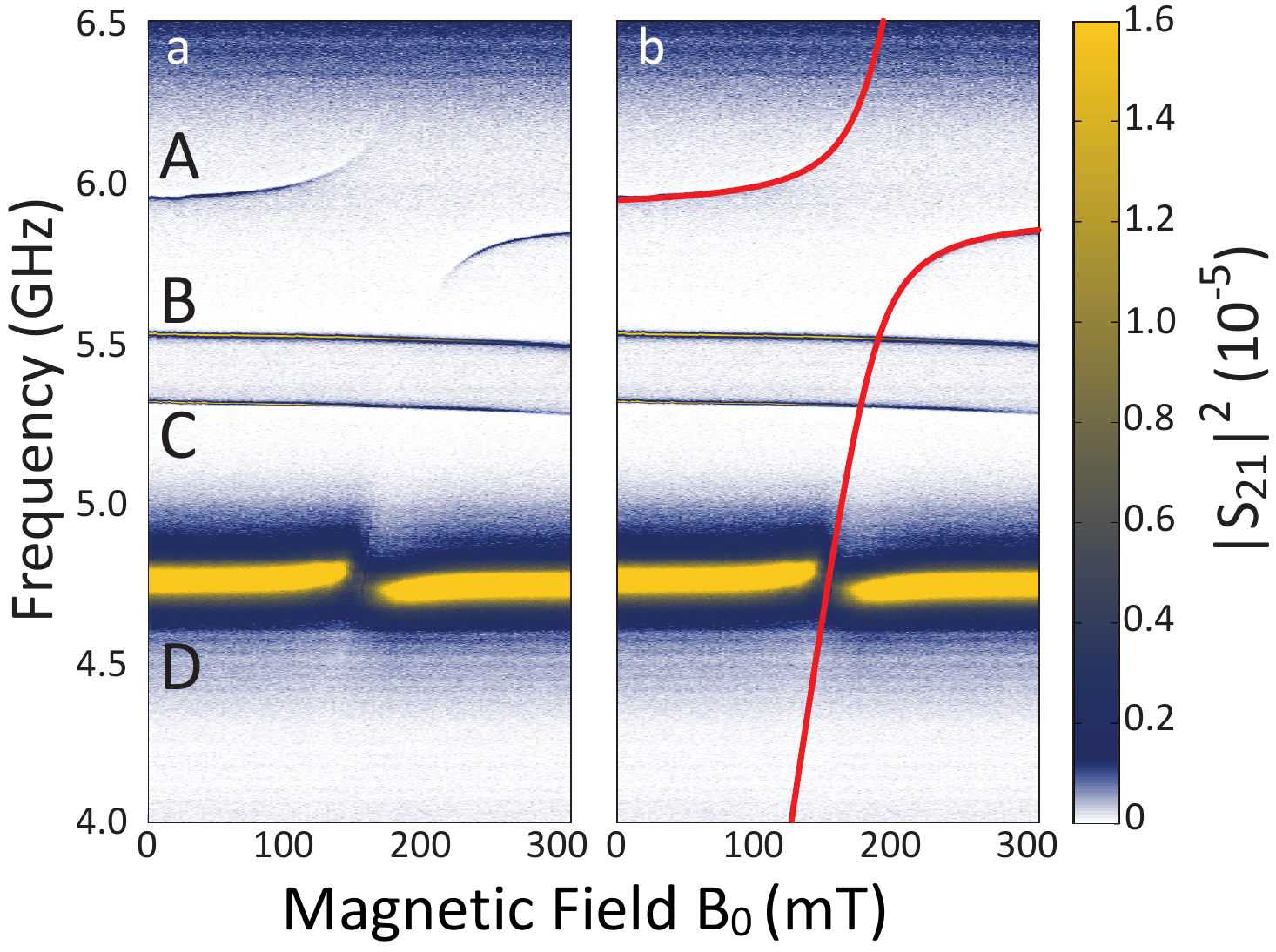}
\caption{(Color online) MW transmission probability $\left|S_{21}\right|^2$ as a function of external magnetic field $B_z^{ext}$ and probe frequency for a YIG film coupled to a superconducting MW coplanar resonator. A denotes the magnon polariton, i.e., the anticrossing between Kittel and cavity mode frequencies. B and C are uncoupled resonators, and D is a parasitic mode. The red curve in (b) is a fit the coupled harmonic oscillator model. Adapted from~\onlinecite{Huebl2013}.}\label{figIII_1}
\end{figure}

The emphasis of this review article is on magnetic materials below the critical temperature at which the magnetic moments are spontaneously ordered at much higher spin densities than those of paramagnetic ensembles. Large spin-photon couplings can be achieved in magnets at weaker applied magnetic fields, higher temperatures, and smaller samples, but at the cost of larger intrinsic damping. Ferromagnetic resonance (FMR), see Section \ref{secIIm}, is a traditional technique to characterize magnetic materials, but without cavity enhancement the coupling is weak. MW photons couple predominantly to the collective Kittel mode, but standing spin waves may also resonate \cite{Hillebrands2001} when magnetic fields are not homogeneous and/or magnetization is pinned at the sample boundaries.

Another strategy to reach the strong coupling regime is the improvement of the cavities and using magnetic materials with low damping rates. YIG has been the material of choice, but with the right cavity design metallic ferromagnets, with higher spin densities  but also higher Gilbert damping, can also be pushed into the strong coupling regime \cite{Li2019,Hou2019}. 

\begin{figure}[tb!]
\includegraphics[width=8.6cm]{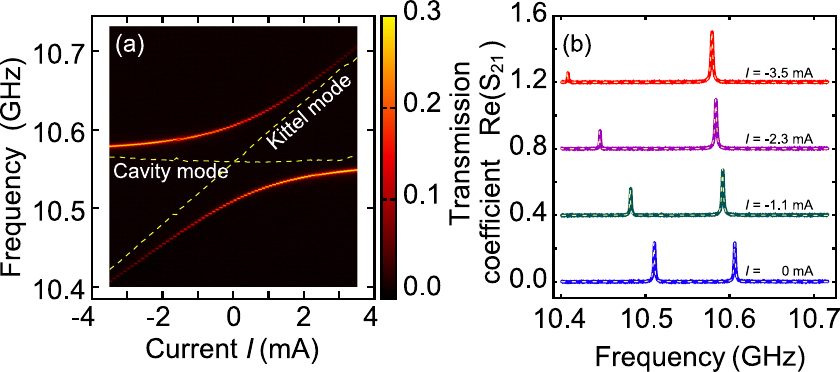}
\caption{(Color online) (a)  Real part of the  transmission coefficient $S_{21}$ of  cavity machined from high-purity Cu as a function of microwave frequency and a current $I$ through a coil close to a magnetic YIG sphere that modulates the magnetic field at the sample. The anticrossing between the cavity and Kittel modes signals the strong coupling regime. (b) Cross-sections of panel (a) at different coil currents. The splitting at $I=0$ corresponds to twice the magnon-photon coupling rate. Adapted from~\onlinecite{Tabuchi2014}.}\label{figIII_2}
\end{figure}

\onlinecite{Soykal2010, Soykal2010a} computed the coupled quantum dynamics of MW photons in a cavity mode and the Kittel mode of magnetic spheres over the full Bloch sphere. \onlinecite{Huebl2013} reported the anticrossing signature of strong coupling for YIG films in coplanar waveguides, see Fig.~\ref{figIII_1}, followed by its observation in split-ring resonators~\cite{Stenning2013, Bhoi2014}.  \onlinecite{Tabuchi2014, Zhang2014, Goryachev2014} found strong coupling for YIG spheres in 3D MW cavities, see Fig.~\ref{figIII_2}. These experiments focused on the microwave transmission coefficient $S_{21}$ as a function of frequency and applied magnetic field that does not vanish when the hybrid polariton has a significant photon contribution.

\onlinecite{Cao2015, Zare2015, Maksymov2015} modelled the magnon-cavity photon system semiclassically taking multiple magnon and cavity modes into account in different geometries. \onlinecite{Zhang2014} investigated different parameter regimes and reported a Purcell effect when $\kappa_{\rm c}<g<\kappa_{\rm m}$, and magnetically induced transparency for $\kappa_{\rm m}<g<\kappa_{\rm c}$. 

By increasing the ratio between magnet and cavity sizes the ultra-strong coupling regime can be entered \cite{Zhang2014}, but we are not aware of experimental reports that address the exotic consequences predicted by theory. 

\subsubsection{Off-resonant coupling of magnons to light}\label{OffresonantMO}

Magnetooptics studies the interaction of magnets with infrared and visible light~\cite{Pershan1967, Shen1966, Gall1971, Cottam1986}  at frequencies $\omega_{\rm c}/2\pi \sim100-800$ THz by means of several established phenomena.  The Faraday (Kerr) effect is the rotation of the polarization plane of linearly polarized light upon transmission through (reflection by) a material with magnetization component parallel to the beam.  The torque on the light field exerted by a magnet implies that there is a reaction, viz. the inverse Faraday effect or light-induced torque on the magnetization~\cite{Kirilyuk2010}. The Cotton-Mouton effect is the birefringence caused by a magnetization normal to the wave vector of the incoming light. The inelastic scattering of light with emission or absorption of magnons is known as Brillouin light scattering (BLS). 

The Zeeman interaction between the ac magnetic field component and magnetization governs the resonant interaction in the GHz regime. A large mismatch between the frequencies suppresses the Zeeman coupling and the second order interaction of magnetization with the ac electric field takes over. Without spin orbit coupling the electron spin is not affected by electric fields, so we may expect larger magnetooptical couplings for heavier elements. The magnetic permeability approaches that of vacuum $\mu_{0}$. We note that the optomechanical coupling in Eq. (\ref{optomechHam}) is even stronger detuned but does not invoke the spin-orbit interaction.

The interaction between the magnetization and electric field is a relativistic effect \cite{Elliott1963}, but for most purposes the coupling can be parameterized in terms of a small number of symmetry-related empirical constants \cite{Borovik-Romanov1982}. The starting point of most theories is the macroscopic dielectric tensor as a function of the magnetization ${\bf M}$ in the displacement field \cite{Fleury1968},
\begin{equation} \label{displacement+field}
{\bf D}=\overleftrightarrow{\varepsilon}({\bf M})\cdot{\bf E}.
\end{equation}
The magneto-optical effects are captured by the leading-order expansion of \(\overleftrightarrow{\varepsilon}\) in ${\bf M}$ in the low-frequency limit \cite{Gall1971a, Wettling1975}. For a cubic crystal such as YIG $\overleftrightarrow{\varepsilon}({\bf M})=\overleftrightarrow{\varepsilon_0}+\overleftrightarrow{\varepsilon_1}({\bf M}),$ where  \(\overleftrightarrow{\varepsilon_0}= \varepsilon_{\rm s}\overleftrightarrow{1}\), $\varepsilon_{\rm s}$ is the scalar static dielectric constant and \(\overleftrightarrow{1}\) the unit matrix. The leading perturbation $\overleftrightarrow{\varepsilon_1}({\bf M})$ describes the response to the optical electric field as a function of the magnetization direction as measured by, for instance, the Faraday effect. For cubic crystal and an equilibrium ${{\bf M}_{\rm s} \parallel  \bf z}$, one has
\begin{eqnarray}
\overleftrightarrow{\varepsilon_1}=\begin{pmatrix}
0 & -ifM_{\rm s}  & ifM_{y} \\
ifM_{\rm s} & 0 & -ifM_{x} \\
-ifM_{y} & ifM_{x} & 0
\end{pmatrix},\label{permitivitty}
\end{eqnarray}where $M_{x,y}$ are the small dynamical components. In the presence of a slowly varying  magnetic texture, these components refer to a local coordinate system with \(z'\)-axis along the magnetization. The parameter $f$ can be fitted to experiments. The second order term in the expansion reads $\overleftrightarrow{\varepsilon_2}({\bf M})=\varepsilon_0 v_{ijkl} M_k M_l ,$ where $v_{ijkl}$ parameterizes the Cotton-Mouton effect \cite{Wettling1976}. The second order term competes with Eq. (\ref{permitivitty}) when one of the $M_{i}$'s in $\overleftrightarrow{\varepsilon_2}$ is $M_{\rm s}$, which introduce two more parameters on top of \(f\). \onlinecite{Liu2016} calculated the contribution of the Cotton-Mouton effect to the optomagnonic coupling for a YIG waveguide with constant equilibrium magnetization, while \onlinecite{Graf2021} addressed arbitrary magnetic textures. In the following, we disregard the Cotton-Mouton effect since it is not essential for the configurations discussed later.

We obtain the Hamiltonian for magneto-optical effects replacing $\varepsilon_{0}{\bf E}^{2} /2$ by ${\bf E}\cdot{\bf D}/2$ in the vacuum energy density Eq.~(\ref{TotEDensity}). The magnetization-dependent contribution is $H_{\rm OM}=\frac{1}{8}\int d{\bf r}{\bf E}^{*}\cdot\overleftrightarrow{\varepsilon_1}({\bf M})\cdot{\bf E}+\frac{1}{8}\int d{\bf r}{\bf E}\cdot\overleftrightarrow{\varepsilon_1}^{*}({\bf M})\cdot{\bf E}^{*}$, where ${\bf E}$ and ${\bf E}^{*}$ are the real and imaginary parts of the complex electric field amplitude. Inserting the permittivity tensor Eq. (\ref{permitivitty}) and using the second quantized form of the electric field of Eq. (\ref{eq:E op}) (note that e.g. ${\bf E}^{*}/2\rightarrow {\bf E}^{+}$) and the magnetization of Eq. (\ref{eq:delta_m}) gives the optomagnonic interaction Hamiltonian. It contains two terms. 
\begin{eqnarray} \label{Faraday}
\hat H_{\rm{Faraday}}=\hbar\sum_{pq}\Omega_{pq}\hat{a}_{p}^{\dagger}\hat{a}_{q}+{\rm h.c.},
\end{eqnarray} describing the static Faraday effect by an equilibrium magnetization \(\textbf{M} = M_{\rm s} \hat{\textbf{z}}\), where $\Omega_{pq}=\mathcal{G}M_{\rm s}\int d{\bf r}\left[{\bf u}^{*}_{p}(\bf r)\times{\bf u}_{q}(\bf r)\right]_{z}$ and  $\mathcal{G}=-if\sqrt{\omega_{p}\omega_{q}}/(4\sqrt{V_{p}V_{q}}\varepsilon_{0}\varepsilon)$, $V_{p(q)}$ being the optical mode volume as defined in Section \ref{secII} [see discussion after Eq.~\eqref{eq:M op}]. In the RWA and discarding photon non-conserving terms such $\hat{a}_{p}^{\dagger}\hat{a}_{q}^{\dagger}\hat{m}_{\eta}$ that contribute only to higher order, the second term becomes
\begin{eqnarray} \label{optomagnonicHam_dispersive}
\hat{H}_{\rm BLS}=\hbar\sum_{pq\eta}\hat{a}_{p}^{\dagger}\hat{a}_{q}\left(G_{pq\eta}^{+}\hat{m}_{\eta}+G_{pq\eta}^{-}\hat{m}_{\eta}^{\dagger}\right) +{\rm{h.c.}}\qquad
\end{eqnarray}
The matrix elements $G_{pq\eta}^{\pm}$ are the anti-Stokes (Stokes) scattering amplitude of a photon from mode $q$ to $p$ by the annihilation (creation) of a magnon in mode $\eta$,
\begin{eqnarray}
  G_{pq\eta}^{+}&=&\frac{\mathcal{G}M_{\rm s}}{2}\int {\bf w}_{\eta}({\bf r})\cdot\left[{\bf u}_{p}^{*}({\bf r})\times{\bf u}_{q}({\bf r})\right]d{\bf r}, \label{G_pqalpha} \\
  G_{pq\eta}^{-}&=&\frac{\mathcal{G}M_{\rm s}}{2}\int {\bf w}_{\eta}^{*}({\bf r})\cdot\left[{\bf u}_{p}^{*}({\bf r})\times{\bf u}_{q}({\bf r})\right] d{\bf r}. \label{G+pqalpha}
\end{eqnarray}
These expressions may be used for arbitrary magnetic textures $\textbf{M}(\textbf{r}) = M_{\rm s} \hat{\textbf{z}}(\textbf{r})$ by re-defining the static Faraday contribution $\Omega_{pq}$. 

The static Faraday effect does not interfere with Brillouin light scattering. The latter is governed by the Hamiltonian $\hat{H}=\hat{H}_{\rm m}+\hat{H}_{\rm opt}+\hat{H}_{\rm BLS}$, where \(\hat{H}_{\rm opt}=\sum_{p}\hbar\omega_{p}\hat{a}^{\dagger}_{p}\hat{a}_{p}\) is the photon energy in a magnetic medium with static magnetization, while $\hat{H}_{\rm m}=\sum_{\eta}\hbar\omega_{\eta}\hat{m}^{\dagger}_{\eta}\hat{m}_{\eta}$, as in Eq. (\ref{eq:freemagnonham}). 

For a homogeneous ground state magnetization, the normal modes ${\bf w}_{\eta}({\bf r})$ and ${\bf u}_{p,q}({\bf r})$ can be expanded into ${\bf e}_{\pm}=({\bf e}_{x}\pm i{\bf e}_{y})/\sqrt{2}$ , i.e. left- and right-hand circular polarized waves with spin \(\pm \hbar \). The degeneracy of the \(\pm\)-polarization is broken by the boundaries of waveguides and cavities in favor of linearly polarized TE and TM modes, as discussed in Section \ref{secII}. A single inelastic scattering process conserves the sum of energy and angular momenta of the quasiparticles in the initial and final states, which implies that the dielectric tensor $\overleftrightarrow{\varepsilon_1}$  does not contain diagonal elements and that an incident TM (TE) mode must be scattered into a TE (TM) mode. These simple rules do not hold for magnetic textures such as vortices, however \cite{Graf2018, Graf2021}.  Since the coupling constants $\mathcal{G}$ are small, the BLS experiments reviewed in Section \ref{SecV} are well described to leading second order.

The optomagnonical, Eq. (\ref{optomagnonicHam_dispersive}), and  optomechanical, Eq. (\ref{optomechHam}), Hamiltonians are very similar, so these systems share a number of effects. An example is the {\em electromagnetically induced transparency}  \cite{Liu2016} in a cavity driven at the red sideband with $\omega_{{\rm D}} = \omega_{\rm c} - \omega_{\rm m}$ when the magnetic dissipation rate $\kappa_{\rm m}$ is the smallest energy scale: A probe beam at the cavity resonance burns a spectral hole in the form of a deep and sharp dip with a linewidth of the order of $\kappa_{\rm m}$ by a destructive interference of probe and pump photons. 

At a very strong drive of the blue sideband, a large number of magnons can be injected that cannot be assumed non-interacting anymore.  In the macrospin approximation, \onlinecite{Kusminskiy2016} predicted a rich dynamics in that regime, including optically-induced magnetization switching and self-oscillations.
 
\subsubsection{Dissipative coupling}

In open cavities that support both standing and traveling waves, the radiation loss into the environment may lead to dissipative coupling between cavity photon and magnon modes as observed in a MW Fabry-P\'{e}rot cavity~\cite{Harder2018} and split-ring resonator~\cite{Bhoi2019}. The dissipative coupling can be controlled by the matrix elements between the magnon and the traveling waves~\cite{Yao2019}. In contrast to the avoided crossing between coherently coupled modes, dissipative coupling causes a level attraction, i.e., the coalescence of cavity-magnet normal modes. In general, both coherent and dissipative coupling may both contribute to the mode spectrum.

In the presence of both coherent and dissipative coupling, Eq.~(\ref{ResonantCouplingHam}) becomes
\begin{eqnarray}
\hat{H}&=&\sum_{p}\hbar\omega_p \hat{a}_{p}^{\dagger}\hat{a}_{p}+\sum_{\eta}\hbar\omega_{\eta}\hat{m}_{\eta}^{\dagger}\hat{m}_{\eta}\nonumber\\
&&+\hbar\sum_{p\eta}\left(\Gamma_{p\eta}e^{i\Phi}\hat{a}_{p}\hat{m}^{\dagger}_{\eta}+\Gamma_{p\eta}^{*}\hat{a}^{\dagger}_{p}\hat{m}_{\eta}\right), \label{DissipativeCouplingHam}
\end{eqnarray}
where $\Phi$ is a tunable phase factor that describes the competition between resonant and dissipative couplings. $\Phi=0$ corresponds to a purely coherent coupling and the formation of magnon polaritons. When $\Phi=\pi$ the coupling is imaginary and thereby purely dissipative. \onlinecite{Yao2019} control the phase $\Phi$ by the direction of the external magnetic field. A drive (anti-damping) can be included on the same footing ~\cite{Boventer2019}.
The cavity-magnet Hamiltonian with both coherent and dissipative couplings is not Hermitian with complex eigenvalues. The corresponding Heisenberg equation of motion should lead to the same result as the input-output model in which dissipation and drive are added {\em a posteriori}.

\subsubsection{Input-output relations} \label{secIII:io}

Here we generalize the input-output relations for an empty cavity as introduced in Section \ref{secII:io} to a cavity loaded with a magnet.

Hybrid systems of spin ensembles coherently coupled to a MW cavity mode are usually described by Eq. (\ref{ResonantCouplingHam}), with a collective coupling that is proportional to the square root of the number of identical spins $\sqrt{N}$. By using this model Hamiltonian, the quantum Langevin equations of motion, in the frame rotating with the drive frequency $\omega_{\rm D}$, read
\begin{align}
&\hspace*{-7pt}\frac{d\hat{a}_{p}}{dt}=i\Delta_{p}\hat{a}_{p}-i\sum_{\eta}\Gamma_{p\eta}^{*}\hat{m}_{\eta}-\frac{\kappa_{\rm c}}{2}\hat{a}_{p}-\sqrt{\kappa_{0}}\hat{d}_{0}+\sqrt{\kappa_{\rm ex}}\hat{a}_{\rm in},\\
&\hspace*{-7pt}\frac{d\hat{m}_{\eta}}{dt}=i\Delta_{\rm m}\hat{m}_{\eta}-i\sum_{p}\Gamma_{p\eta}\hat{a}_{p}-\frac{\kappa_{\rm m}}{2}\hat{m}_{\eta}-\sqrt{\kappa_{m}}\hat{d}_{\rm m},
\end{align}
where $\kappa_{\rm c}=\kappa_{0}+\kappa_{\rm ex}$ is the total linewidth of the cavity in terms of the intrinsic (extrinsic) loss rate, $\Delta_{p}=\omega_{\rm D}-\omega_{p}$, $\Delta_{\rm m}=\omega_{\rm D}-\omega_{\eta}$, and $\kappa_{\rm m}$ is the magnetic relaxation rate. Furthermore, $\hat{d}_{\rm m}$ is a stochastic magnetic field satisfying $\langle\hat{d}_{\rm m}\rangle=0$, $\langle\hat{d}_{\rm m}^{\dagger}(t)\hat{d}_{\rm m}(t^{\prime})\rangle=n_{\rm m}\delta(t-t^{\prime})$, and $\langle\hat{d}_{\rm m}(t)\hat{d}_{\rm m}^{\dagger}(t^{\prime})\rangle=(n_{\rm m}+1)\delta(t-t^{\prime})$. The steady state solutions are 
\begin{align}
&\left\langle \hat{a}_{p}\right\rangle=i\sum_{\eta}\frac{\Gamma_{p\eta}^{*}}{i\Delta_{p}-\kappa_{\rm c}/2}\left\langle \hat{m}_{\eta}\right\rangle-\frac{\sqrt{\kappa_{\rm ex}}}{i\Delta_{p}-\kappa_{\rm c}/2}\left\langle \hat{a}_{\rm in}\right\rangle, \\
&\left\langle \hat{m}_{\eta}\right\rangle=i\sum_{p}\frac{\Gamma_{p\eta}}{i\Delta_{\rm m}-\kappa_{\rm m}/2}\left\langle \hat{a}_{p}\right\rangle.
\end{align}
The input-output theory yields a MW transmission amplitude $S_{21}=\langle\hat{a}_{\rm out}\rangle/\langle \hat{a}_{\rm in}\rangle$ between ports 1 and 2,
\begin{align}
S_{21} & =\frac{2\kappa_{\rm ex}}{\kappa_{\rm c}} \frac{1}{\frac{2i\Delta_{p}}{\kappa_{\rm c}}-1+i\sum_{p\eta}\frac{C_{p\eta}}{1-2\Delta_{\rm m}/\kappa_{\rm m}}},
\end{align}
The  {\it cooperativity} $C_{p\eta}=4\vert\Gamma_{p\eta}\vert^{2}/(\kappa_{\rm c}\kappa_{\rm m})$ is a ratio between coupling strength and dissipation. The magnon-photon coupling appears in the form of a self-energy. Its real part shifts the photon frequency and the imaginary part represents magnetic damping. The anticrossing between the bare magnon and photon modes in the transmission amplitude of the loaded cavity in Figs. \ref{figIII_1} and \ref{figIII_2} is resolved when  $\Gamma_{p\eta}\gg\kappa_{\rm c}, \kappa_{\rm m}$.

\subsubsection{Classical wave theory}

The quantum language used above is convenient and essential in quest of quantum mechanics. However, many phenomena are purely classical, or, when magnons and photons are simple harmonic oscillators, cannot be distinguished between classical and quantum, analogous to the classical vs. quantum description for LC circuits introduced in Sec. \ref{secII}. Here we look at interaction of magnons and cavity MW radiation from the viewpoint of a classical field theory, i.e. using the coupled LLG and Maxwell equations. We then do not have to invoke the RWA or magnetostatic approximation, in principle without restrictions for the photon and magnon amplitudes. Linearized solutions account for multiple cavity and magnon modes and their interactions~\cite{Cao2015, Zare2015}. The same results can be obtained in a dynamical phase correlation approach in terms of finite-element circuit \cite{Harder2016} in which the the LLG equation generates a MW dynamics by the Faraday law. 

A small ac field \(\bf{h}(\bf{r},t)\) drives a small magnetization amplitude \(\delta \bf{M} (\bf{r},t)\),
\begin{subequations}
\begin{align}
{\bf M}({\bf r},t)  &  ={\bf M}_{s}+ {\bf \delta M}({\bf r},t),\\
{\bf H}({\bf r},t)  &  ={\bf H}_{0}+{\bf h}({\bf r},t).
\end{align}
\end{subequations}
To leading order the LLG equation reads
\begin{eqnarray}
\delta\dot{{\bf M}} & = & -\gamma\mu_0 \left({\bf M}_{s}\times{\bf H}_{\rm eff}^{(1)}+\delta{\bf M}\times{\bf H}_{\rm eff}^{(0)}\right) \nonumber \\
  & + & \frac{\alpha}{M_{s}}{\bf M}_{s}\times\delta\dot{{\bf M}}, \label{LinearziedLLG}
\end{eqnarray}
where ${\bf H}_{\rm eff}^{(0)}={\bf H}_{\rm ext}$ and ${\bf H}_{\rm eff}^{(1)}={\bf H}_{\rm ex}+{\bf h}$. In frequency and momentum space, Eq. (\ref{LinearziedLLG}) can be recast into ${\bf \delta M}=\overleftrightarrow{\chi}\cdot{\bf h}$, where $\overleftrightarrow{\mu}=\mu_{0}(\overleftrightarrow{\rm I}+\overleftrightarrow{\chi})$ is the magnetic permeability tensor. Substituting $\overleftrightarrow{\mu}$ into Eq.~(\ref{WaveEquation}) and boundary conditions of the EM field across interfaces, Eq.~(\ref{BCs}) leads to the mode amplitudes in the cavity and ultimately the scattering matrix. The cavity modes are also affected by magnetic loads~\cite{Macedo2020}. The modulation of the electric field by the large dielectric constant can cause significant distortions of the mode spectrum when the YIG sample size approaches that of the photon wave length~\cite{Cao2015, Zare2015}. 

\begin{figure}[tb!]
\includegraphics[width=8.5cm]{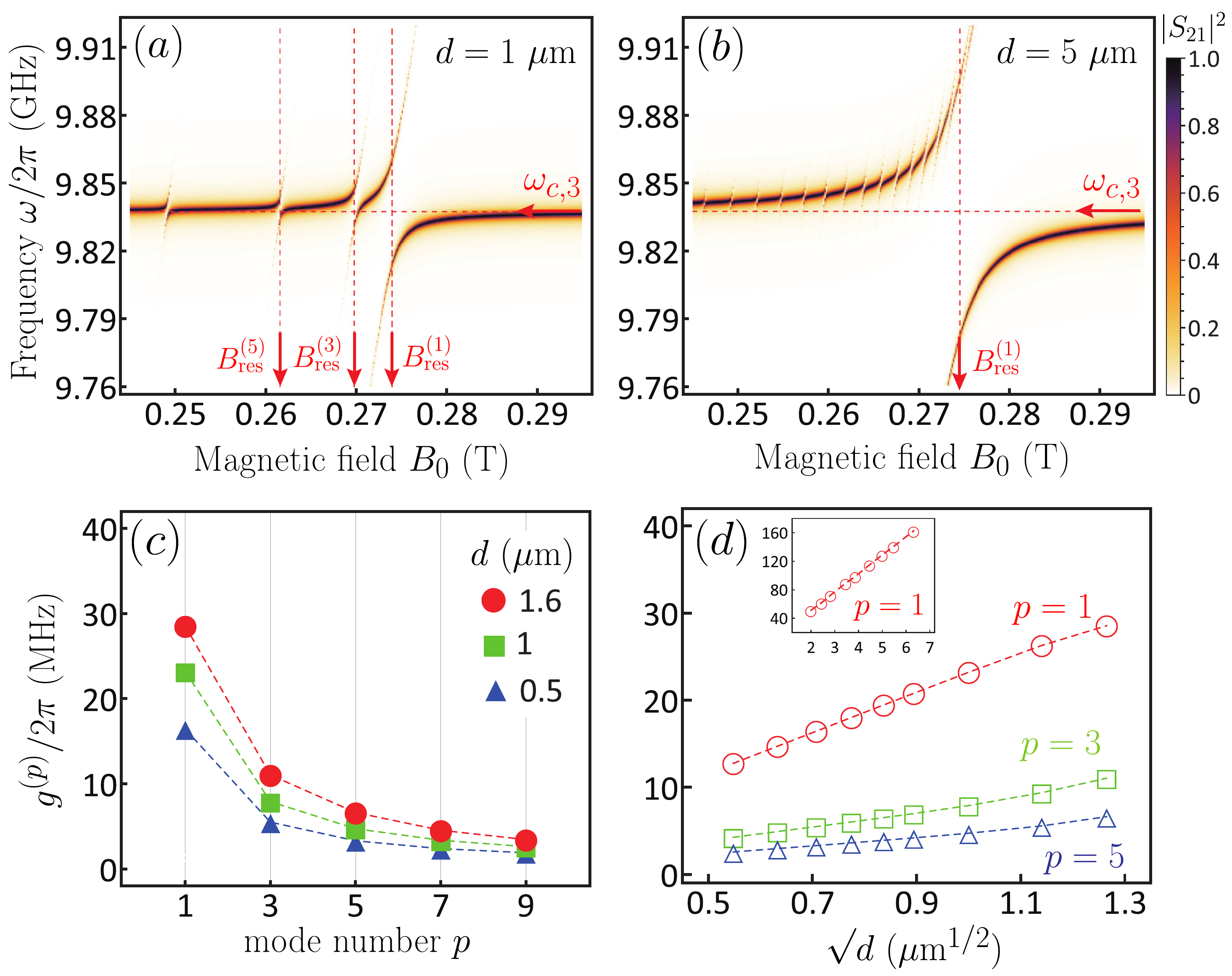}
\caption{(Color online) Theoretical results for (a,b) MW transmission spectra as function of magnetic field and frequency for a YIG film placed in a 1D cavity; (c,d) mode-dependent coupling rates. The YIG parameters used for the calculation are as follows: Gilbert damping $\alpha=10^{-5}$~\cite{Kajiwara2010}, ferromagnetic exchange constant $J=3\times10^{-16}\;{\rm m^{2}}$~\cite{Serga2010}, relative dielectric constant $\varepsilon/\varepsilon_{0}=15$~\cite{Sadhana2009}, gyromagnetic ratio  $\gamma/2\pi=28\,$GHz/T, and saturation magnetization $\mu_{0}M_{s}=175\,$mT~\cite{Manuilov2009}. Here, ${B}_{\rm res}^{(p)}$ denotes the resonance field for mode $p$. Adapted from~\onlinecite{Cao2015}.}
\label{figIII_3}
\end{figure}

\onlinecite{Cao2015} considered a YIG film in a planar MW cavity by a classical version of  scattering theory, reporting strong coupling for the Kittel mode and even for spin waves, see Fig.~\ref{figIII_3}, which has been experimentally confirmed~\cite{Bai2015, Maier-Flaig2016}. Parameters can easily be tuned, thereby capturing magnetically induced transparency, Purcell effect. Since the RWA approximation is not implied, the USC can be handled on equal footing. The MW-driven magnetization can be also detected by spin pumping from the ferromagnet into an adjacent metallic contact with large spin Hall angle, which serves as an interface between electronics and cavity magnonics \cite{Maier-Flaig2016}.

\begin{figure}[tb!]
\includegraphics[width=8.5cm]{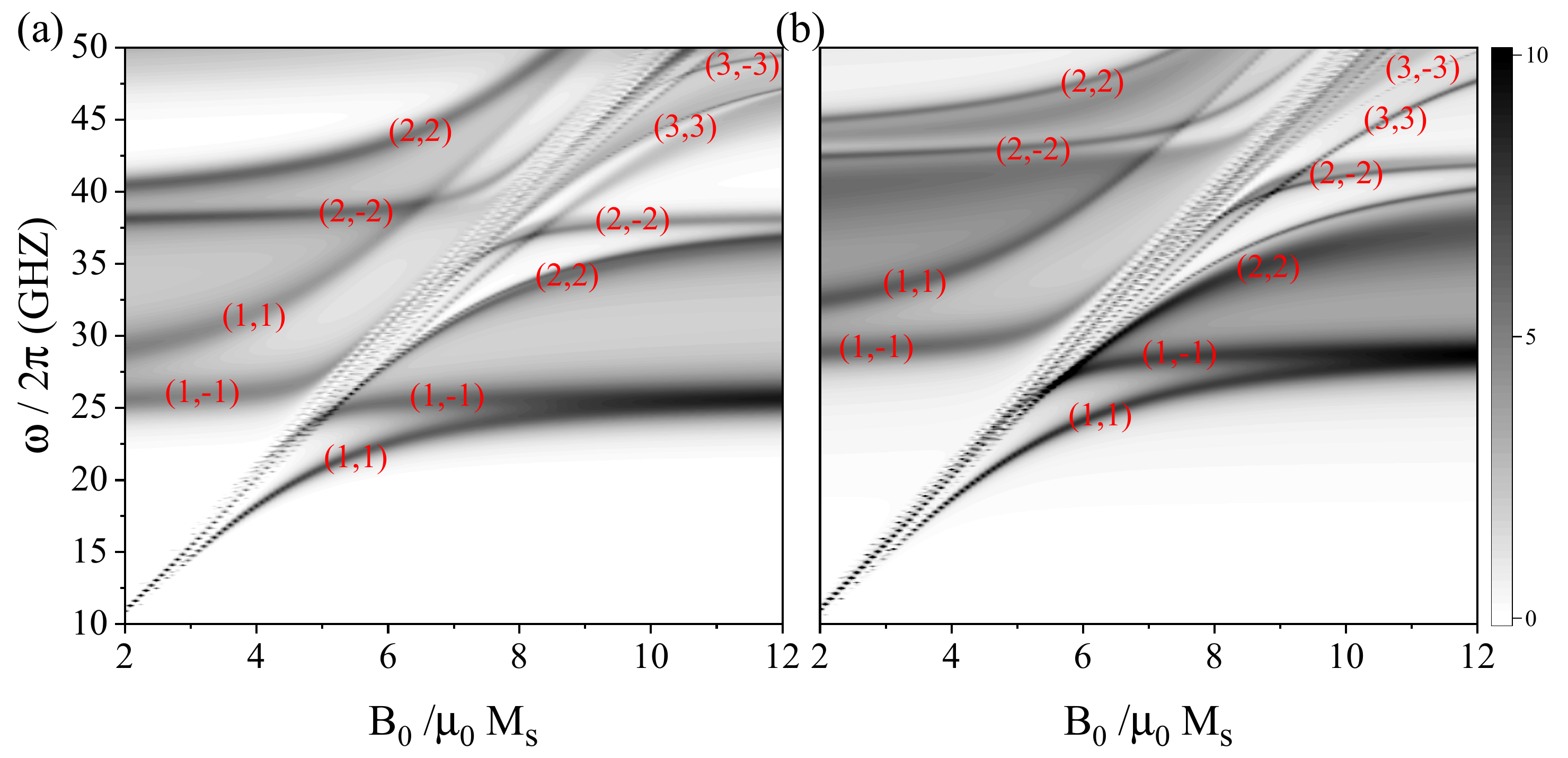}
\caption{(Color online) Calculated scattering efficiency of a YIG sphere for the same parameters as Fig. \ref{figIII_3}, as a function of magnetic field $B_0$ and frequency, (a) with and (b) without the cavity. The MW modes are labeled by the spherical harmonic indices $(n,m)$. Adapted from~\onlinecite{Zare2015}.}\label{figIII_4}
\end{figure}

Practically all experiments use either spheres or films. High quality samples of the former are commercially available at radii down to half a micron and positioned freely inside 3D MW cavities, which allows realization of the strong coupling with relative ease. The spherical symmetry allows an expansion into spherical harmonics. MW input-output relations can be  mapped on Mie scattering theory, which leads to semi-analytic results for the properties of dielectric/magnetic spheres in MW cavities~\cite{Zare2015} beyond the weak coupling regime~\cite{Arias2005} and the magnetostatic approximation, and can be extended to treat the collective dynamics of multiple spheres~\cite{Zare2018}. By acting as an antenna for EM fields, large YIG spheres trap the MWs by their large dielectric constant, even without external cavities~\cite{Zare2015,Neuman2020} . This prediction illustrated by Fig. \ref{figIII_4} was confirmed experimentally~\cite{Bourhill2016}.

\section{Magnons in microwave cavities}\label{SecIV}
In Section \ref{SecIII} we explained the basics of interactions of magnons with EM radiation and illustrated them illustrated by few seminal experiments. Here, we turn to the design of MW cavities and their magnetic loading, focusing on the strong coupling regime. We introduce different MW cavities and mode families and turn then to experimental results on coherent and dissipative magnon-photon couplings, nonlinearities, and cavities filled with more than a single magnet.

\subsection{Microwave cavities}\label{SecIV-A1}
MW cavities are usually passive, i.e. the observable is the scattering matrix between incoming and outgoing MWs. Active cavities, on the other hand, contain a feedback loop that controls the MW response. For the basic properties of cavity resonances and anti-resonances and their phase characteristics we refer to Sec. \ref{secII:io}.

Since the first experiments \cite{Huebl2013}, the magnon-photon interaction has been studied by several cavity designs, such as metallic 3D MW cavities, 2D planar cavities, and 1D Fabry-Pérot type of cavity (Table \ref{tbl:cavity}), while magnetic samples have been either spheres or films. In passive cavities, the photon dissipation rate $\kappa_{\rm c}$ determines the cavity quality factor $Q = \omega_{\rm c}/\kappa_{\rm c}$ at the mode resonance frequency $\omega_{\rm c}$, see Sec. \ref{secII}, while the magnon-photon coupling $g_{N} = g \sqrt{N}$ is enhanced by the number of spins \(N\), see Sec. \ref{SecIII}.
The Gilbert damping constant  $\alpha$ is in YIG of the order of $10^{-4}$, which corresponds to a line broadening $\kappa_{\rm m}/2\pi = \alpha \omega_{\rm m}/2\pi \approx$ 1 MHz at $\omega_{\rm m}/2\pi$ of 10 GHz, as we discussed in Sec. \ref{secIIm}. The coupling (\ref{resonant_coupling_arb_mode}) of the Kittel mode of a magnet on an anti-node of a cavity mode becomes $g \sim \eta\gamma\sqrt{\mu_0\hbar\omega_{\rm c} /V_{\rm c}}$, where $\gamma/2\pi \approx 28$ GHz/T is the electron gyromagnetic ratio and $\eta<1$ is a ``magnon filling" factor that takes the spatial overlap and polarization matching into account \cite{Zhang2014}. The cavity mode volume is $V_{\rm c} = \int \vert \mathbf{H} \vert^2 dV / \vert \mathbf{H} \vert_{\max}^2$, cf. Eq. (\ref{eq:modevolume}),  where $\vert \mathbf{H} \vert_{\max}^2$ is the maximum value the MW intensity~\cite{Zhang2014}. The largest reported coupling is $g_{N}/2\pi$ = 3.06 GHz for a 15.5-GHz MW cavity mode, corresponding to a cooperativity of $C$ = 1.5 $\times$ 10$^7$. It was achieved by nearly filling a 3D cavity with a 5-mm-diameter YIG sphere~\cite{Bourhill2016} and approaches the ultra-strong coupling regime, see Sec. \ref{SecIII}.

Closed 3D cavities can be chosen to have different shapes and input-output ports. High quality 3D cavities are usually machined from high-purity copper that reflects MWs with very small loss. The one shown in Table \ref{tbl:cavity} has inner dimensions of 43.0 $\times$ 21.0 $\times$ 9.6 mm$^3$~ \cite{Zhang2014}. Its lowest TE$_{101}$ mode is at $\omega_{\rm c}/2\pi$ = 7.875 GHz with a linewidth of a few MHz, which corresponds to a quality factor of about 1000 at room temperature. Commonly, a small YIG sphere is placed at a local maximum of the intensity of chosen mode of a much larger cavity such that the field distribution over the sphere is nearly constant. Re-entrant cavities with multiple posts~\cite{Creedon2015} focus the cavity magnetic field on to the YIG crystal, thereby enhancing the effective filling factor far beyond the geometrical  one~\cite{Goryachev2014}. Standing waves in 3D cavities are simply defined by the reflecting boundaries. 2D planar cavities are defined by superconducting resonators~\cite{Huebl2013, Morris2017}  or  microstrips of normal metals~\cite{Bhoi2014, Kaur2016} with simplified fabrication, design, sample placement, and tunability that offset the weaker confinement and often reduced quality factor. 

Superconducting resonators play an important role in circuit QED by facilitating strong coupling, for example, to a superconducting qubit~\cite{Wallraff2004}.
\onlinecite{Huebl2013}  investigated a slab of YIG by placing it on top of a superconducting Nb resonator.
The sensitivity of a resonator can be improved in the form of a lumped element consisting of a small inductor shunted by a large capacitance that reduces the impedance, thereby detecting the much smaller number of spins in micro- and nanoscale cavities~\cite{MckenzieSell2019}. Similar superconducting resonators with low mode volumes in all-on-chip devices strongly couple MW photons and magnons in nanometer thick permalloy  structures~\cite{Li2019, Hou2019}. These studies demonstrate that scaling down the cavity dimensions allows MW control of metallic magnets and, in the future, spintronic devices with higher damping. For instance, a noncollinear magnetic insulator, Cu$_2$OSeO$_3$, was investigated in the MW cavity~\cite{Abdurakhimov2019}.

Planar cavities fabricated with normal metal microstrips in the shape of a split-ring~\cite{Bhoi2014, Kaur2016}, a T (notch filter)~\cite{Castel2017}, or a cross~\cite{Yang2019} operate at room temperature. They can be interpreted as magnetically tunable metamaterials~\cite{Bhoi2014} and enable magnon controlled logic devices~\cite{Rao2019} and non-reciprocal MW isolators~\cite{Wang2019}.

3D cavities with large aspect ratios and standing waves with relatively small frequency splittings in one direction come down to the (quasi) 1D Fabry-Pérot type introduced in Sec. \ref{SecII-MWOcavities}. In a waveguide  the end-points are open and do not reflect the waves, which leads to a continuous spectrum and very different physics that can be captured by an external coupling rate $\kappa_{\rm ex}$  much larger than the intrinsic dissipation rate $\kappa_{0}$. The quasi-1D cavity in Table \ref{tbl:cavity} illustrates a partially closed design with both standing and traveling waves, consisting of a waveguide with a circular cross-section connected to the MW source and detector by two non-circular transition regions that are rotated by an angle $\theta$. This device resembles musical instruments such as a flute --- consisting of a resonating body with tunable  cavity modes that are coupled to propagating, audible sound waves~\cite{Yao2015}. Open cavities support cavity anti-resonances~\cite{Rao2019} and have modified magnon-photon couplings~\cite{Harder2018, Bhoi2019, Yang2019, Yao2019, Wang2019, Yang2020}. The classical wave physics of open cavities can be described by the formalism of open quantum systems as discussed in the next subsection.

The feedback between MW output and input in active cavities can reveal the cooperative dynamics of a polariton ensemble~\cite{Yao2017}. The example in Table \ref{tbl:cavity} consists of a planar passive (straight) cavity in proximity with a magnet and an active (curved) cavity that contains a MW amplifier with voltage-controlled gain. Both cavities are high-quality half-wavelength strip line resonators that form a 2$\pi$ phase loop. The active cavity acts as a feedback loop that compensates the loss of the passive cavity, with a gain of up to 360,000; the effective cavity quality factor can reach  \(Q=81,500\) at room temperature, which is about 3 orders of magnitude higher than that of conventional planar cavities. The feedback photons thereby enhance the magnon-photon coupling analogous to the superradiance of the Dicke model~\cite{Dicke1954}.

\newpage

\begin{widetext}
\begin{table}[ht!]
\parbox{\textwidth}{\caption{Typical MW cavities used to study magnon-photon couplings. The parameters are for room temperature, with the exception of the 2D planar cavity where parameters are for cryogenic temperatures.} \label{tbl:cavity}}
\centering
\begin{tabular}{ccc}
\toprule
Cavity Structure & Key Features & Reference \\
\toprule
  3D cavity & & \\
\hline
      \begin{minipage}[m]{.45\textwidth}\centering\vspace*{5pt}
      \includegraphics[width=4.5cm]{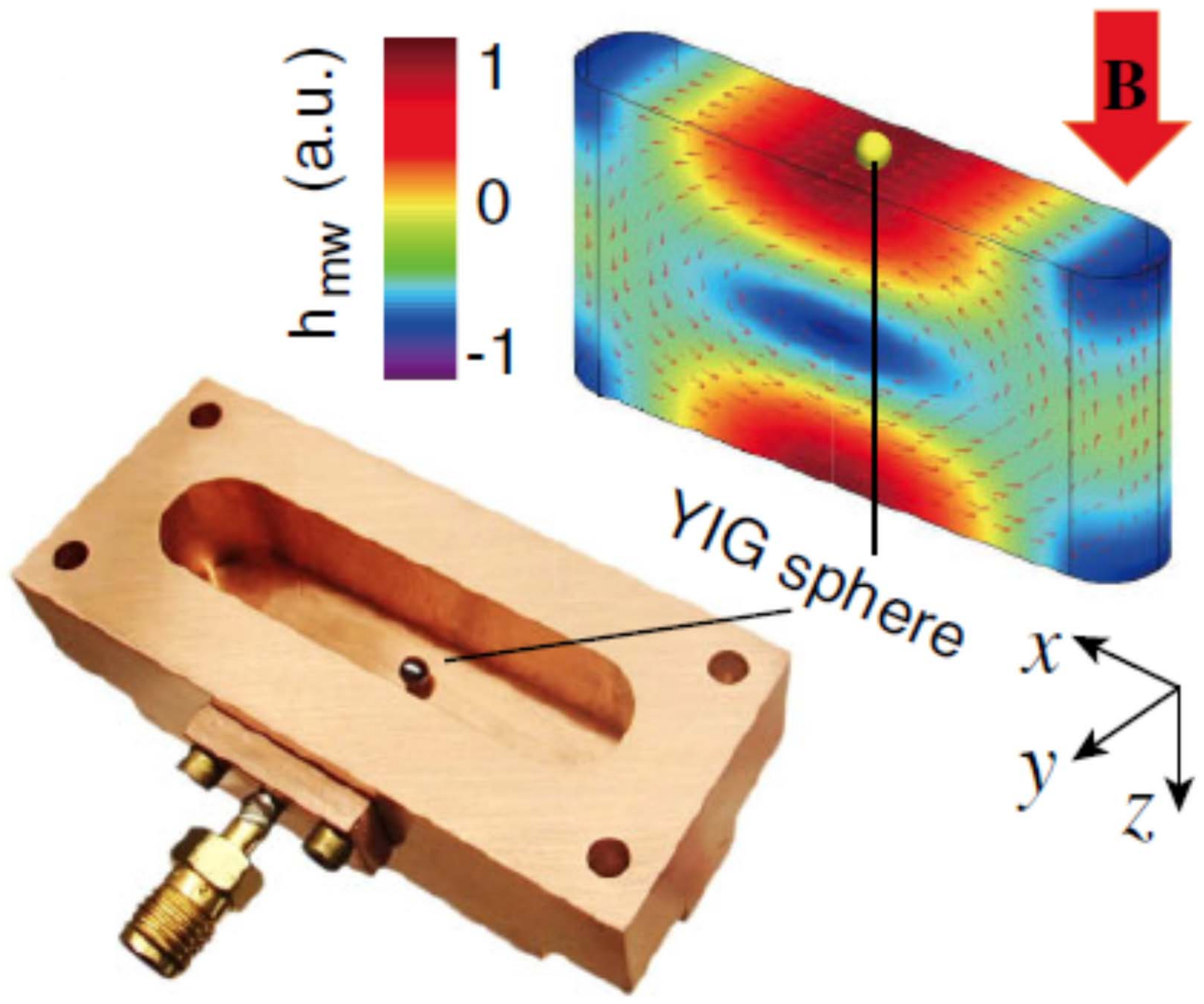}\vspace*{5pt}
      \end{minipage} &
      \begin{minipage}[m]{6cm}
      \begin{itemize}
      \item standing cavity modes
      \item intrinsic damping: $\kappa_{\rm in} /2\pi\sim$1--10 MHz
      \item magnon filling factor: $\ll$1\%
      \end{itemize}
      \end{minipage} &
\cite{Zhang2014}
\\
\toprule
  3D lumped-element cavity & & \\
\hline
      \begin{minipage}{.45\textwidth}\centering\vspace*{5pt}
      \includegraphics[width=4.5cm]{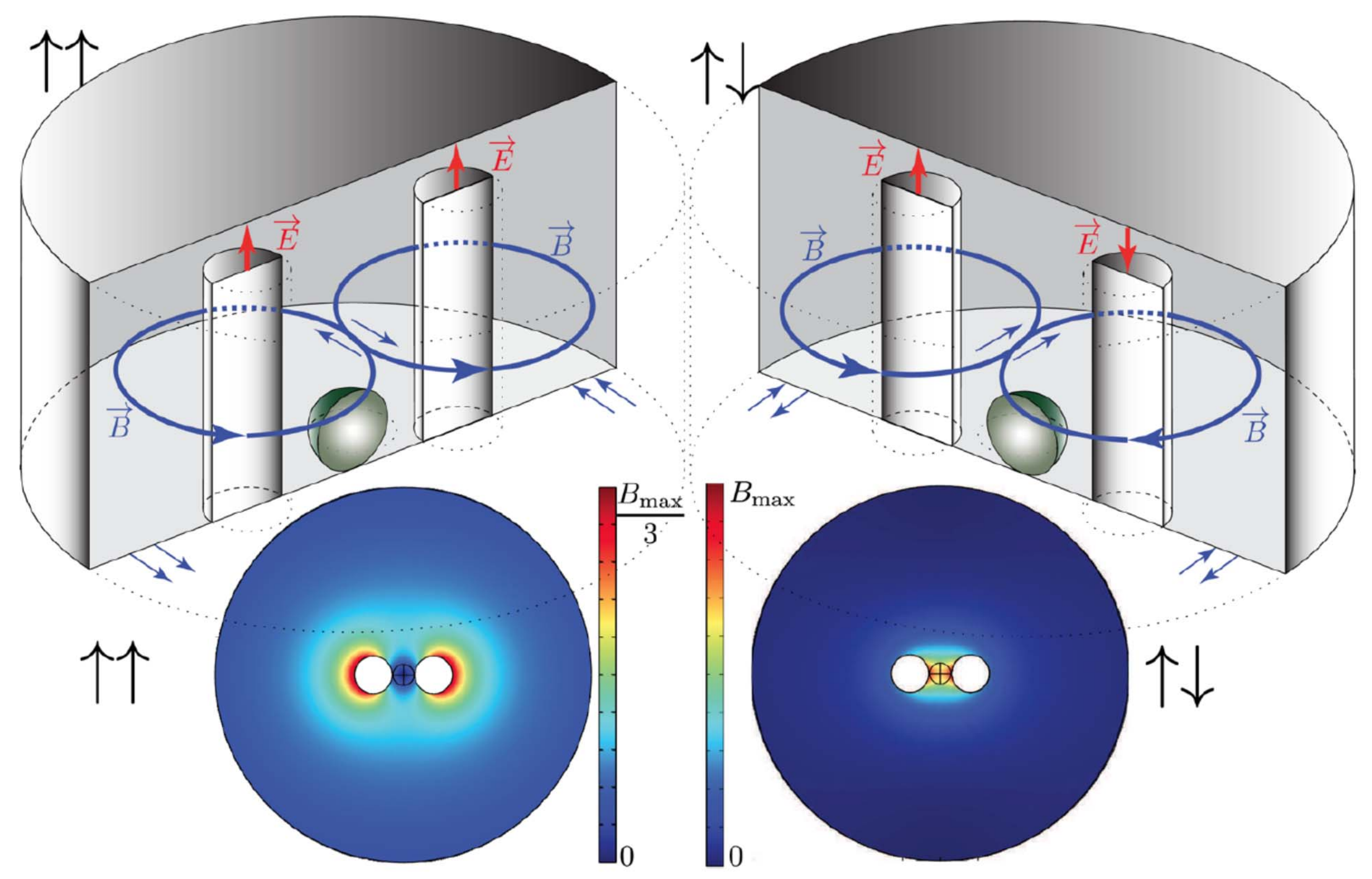}\vspace*{5pt}
      \end{minipage} &
      \begin{minipage}[m]{6cm}
      \begin{itemize}
      \item standing cavity modes
      \item intrinsic damping: $\kappa_{\rm in} /2\pi \sim$10 MHz
      \item magnon filling factor: $\sim$1\%
      \item localized magnetic field enhancement
      \end{itemize}
      \end{minipage} &
      \cite{Goryachev2014}
\\
\toprule
  2D planar cavity & & \\
\hline
      \begin{minipage}{.45\textwidth}\centering\vspace*{5pt}
      \includegraphics[width=4.5cm]{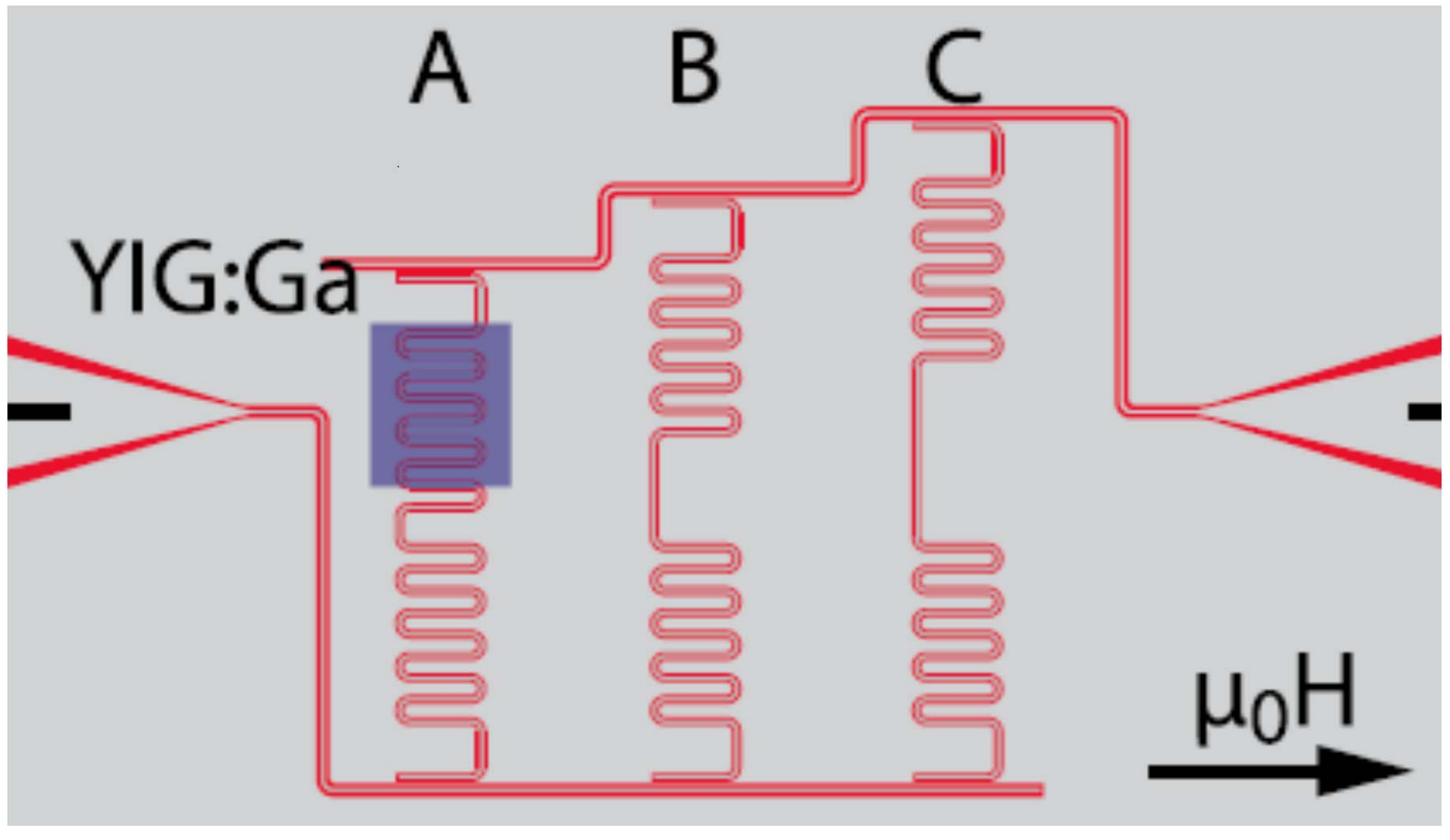}\vspace*{5pt}
      \end{minipage} &
      \begin{minipage}[m]{6cm}
      \begin{itemize}
      \item standing cavity modes
      \item intrinsic damping: $\kappa_{\rm in} /2\pi\sim$1 MHz
      \item magnon filling factor: $\sim$1\%
      \end{itemize}
      \end{minipage} &
      \cite{Huebl2013}
\\
\toprule
  Quasi-1D cavity & & \\
\hline
      \begin{minipage}{.45\textwidth}\centering\vspace*{5pt}
      \includegraphics[width=5.5cm]{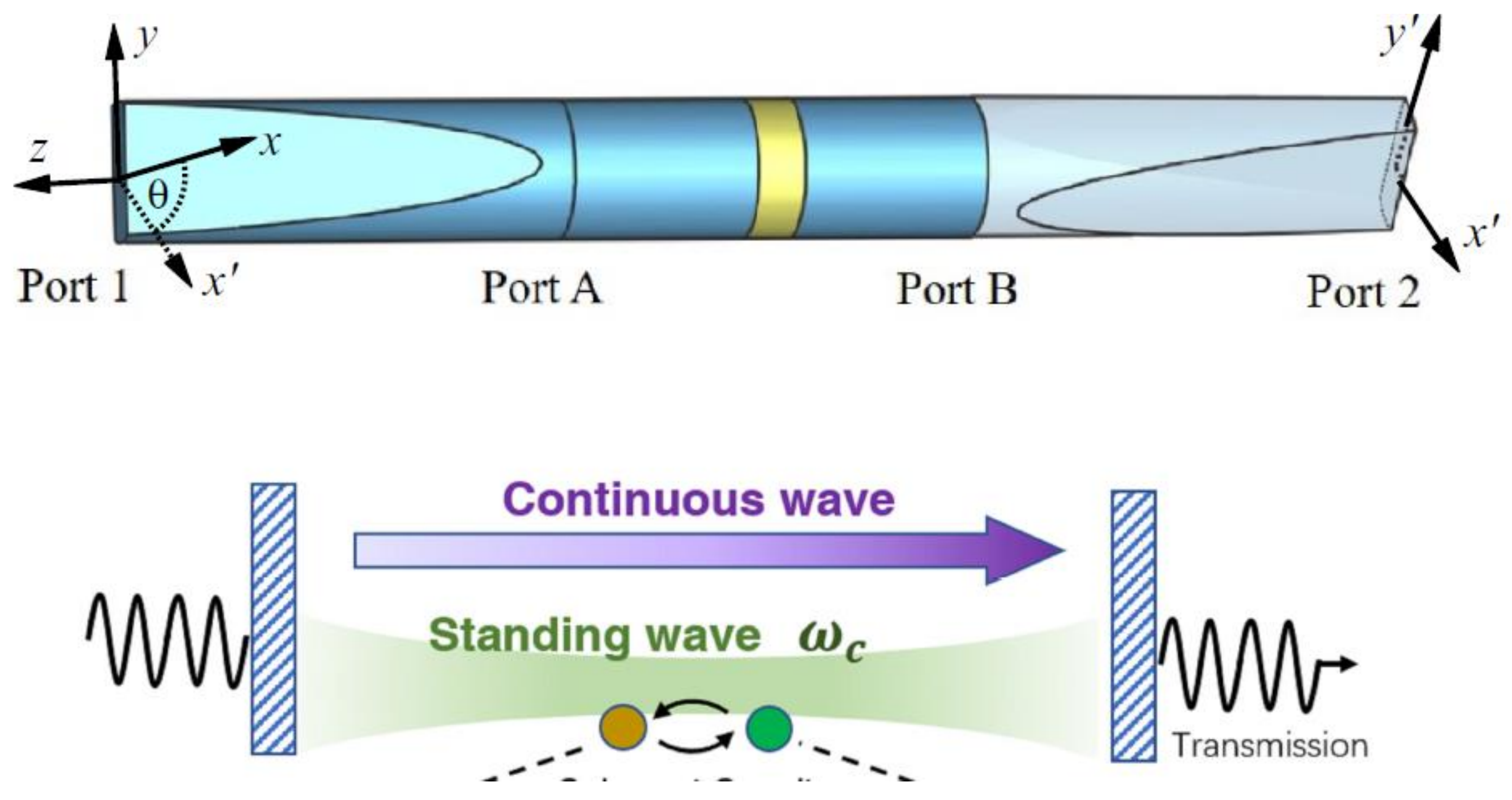}\vspace*{5pt}
      \end{minipage} &
      \begin{minipage}[m]{6cm}
      \begin{itemize}
      \item standing cavity and travelling waveguide modes
      \item intrinsic damping: $\kappa_{\rm in} /2\pi\sim$10 MHz
      \item extrinsic damping: $\kappa_{\rm ex} /2\pi\sim$100 MHz
      \end{itemize}
      \end{minipage} &
      \cite{Harder2018}
\\
\toprule
  Active cavity & & \\
\hline
      \begin{minipage}{.45\textwidth}\centering\vspace*{5pt}
      \includegraphics[width=5.5cm]{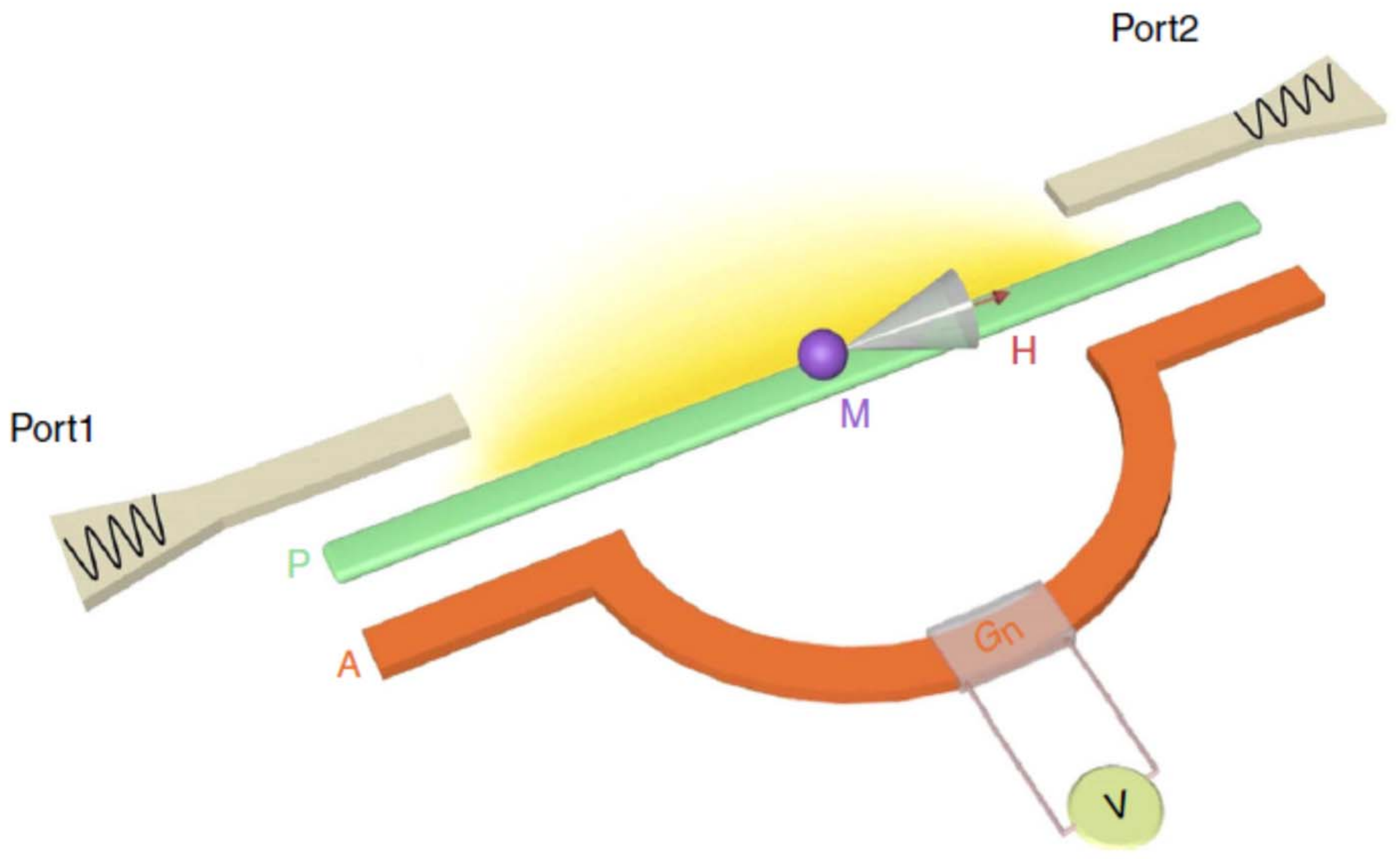}\vspace*{5pt}
      \end{minipage} &
      \begin{minipage}[m]{6cm}
      \begin{itemize}
      \item standing cavity modes
      \item intrinsic damping: $\kappa_{\rm in} /2\pi\sim$0.01 MHz
      \item feedback gain
      \end{itemize}
      \end{minipage} &
      \cite{Yao2017}
\\
\hline
\end{tabular}
\end{table}
\newpage
\end{widetext}

%

\subsection{Coherent and dissipative coupling}\label{SecIV-B1}

\subsubsection{Coherent coupling and level repulsion}

Coherent magnon-photon coupling in MW cavities has been observed by MW transmission (or reflection) spectroscopy~\cite{Huebl2013, Tabuchi2014, Goryachev2014, Bhoi2014, Haigh2015, Yao2015, Bourhill2016, Li2019, Hou2019} time-domain measurement~\cite{Zhang2014, Loo2018, Match2019}, electrical detection~\cite{Bai2015, Maier-Flaig2016, Bai2017}, and Brillouin light scattering experiments~\cite{Klingler2016,Hisatomi2016}.

\begin{figure}[!t]
\centering
\includegraphics[width=0.7\columnwidth]{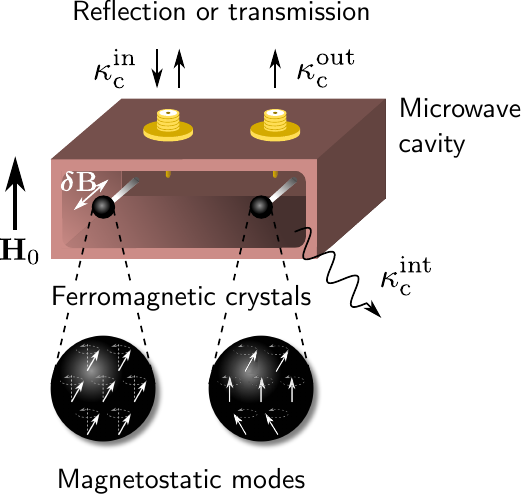}
\caption{\footnotesize{(Color online) Schematic diagram of a general hybrid system for creating cavity-magnon polaritons. The vacuum fluctuations $\mathbf{\delta B}$ of the MW magnetic field of a cavity mode overlaps with one or more ferromagnetic crystals. An external magnetic field  $\mathbf{H_0} = \mathbf{B_0}/\mu_0$ is applied either uniformly or locally to each magnet. Depending on the uniformity of the MW magnetic field of the cavity mode, different magnetostatic modes can be coupled. The MW cavity can be probed either by transmission or reflection through coupling rates $\kappa_{\rm c}^{\rm in}$ and $\kappa_{\rm c}^{\rm out}$ to input and output ports. The internal loss of the MW cavity mode is given by $\kappa_{\rm c}^{\rm int}$. Adapted from \cite{Lachance-Quirion2019}}.}
\label{FigIV3}
\end{figure}

Figures \ref{FigIIIa}, \ref{figIII_2}, and \ref{FigIV3} illustrate the concept, the typical signatures of strong magnon-photon coupling, and the setup to measure them in MW transmission (or reflection) spectra. A schematic experimental setup~\cite{Lachance-Quirion2019} is shown in Fig. \ref{FigIV3}. The MW magnetic field  $\mathbf{\delta B}$ of a cavity mode interacts with one or more ferromagnets or other loads. The external magnetic field $\mathbf{H_0} = \mathbf{B_0}/\mu_0$ can be applied either uniformly or locally to each sample. When the MW magnetic field or the ground state magnetization are not uniform, magnons other than the Kittel mode can interact with the photons. The MW cavity can be probed either by the reflection amplitudes at the input or transmission to the output ports that are characterized by coupling rates $\kappa_{\rm c}^{\rm in}$ and $\kappa_{\rm c}^{\rm out}$. Figure \ref{figIII_2} shows the real part of a typical transmission coefficient $S_{21}$ as a function of the probe frequency $\omega_{\rm D}$ and the current $I$ in a coil that controls the amplitude $H_0 = \vert \mathbf{H_0} \vert$ of the static magnetic field. The anticrossing gap is much larger than the linewidths,  proof of the strong and coherent coupling between the Kittel mode of a YIG sphere and a standing MW cavity mode. As discussed in Sec. \ref{SecIII}, the coupling mechanism is the Zeeman interaction between the macroscopic magnetic dipole and the MW magnetic field. The minimum splitting of the two modes (right panel) gives a coupling strength \(g/2\pi=22.9\) MHz. Horizontal and diagonal dashed lines indicate the frequencies of the uncoupled cavity and Kittel modes. The coupling between MW photons and magnon modes other than the Kittel mode depends on the overlap between the magnon and cavity modes and can be strong in spheres \cite{Zhang2014} as well as films \cite{Maier-Flaig2016}.

\begin{figure}[!t]
\centering
\includegraphics[width=\columnwidth]{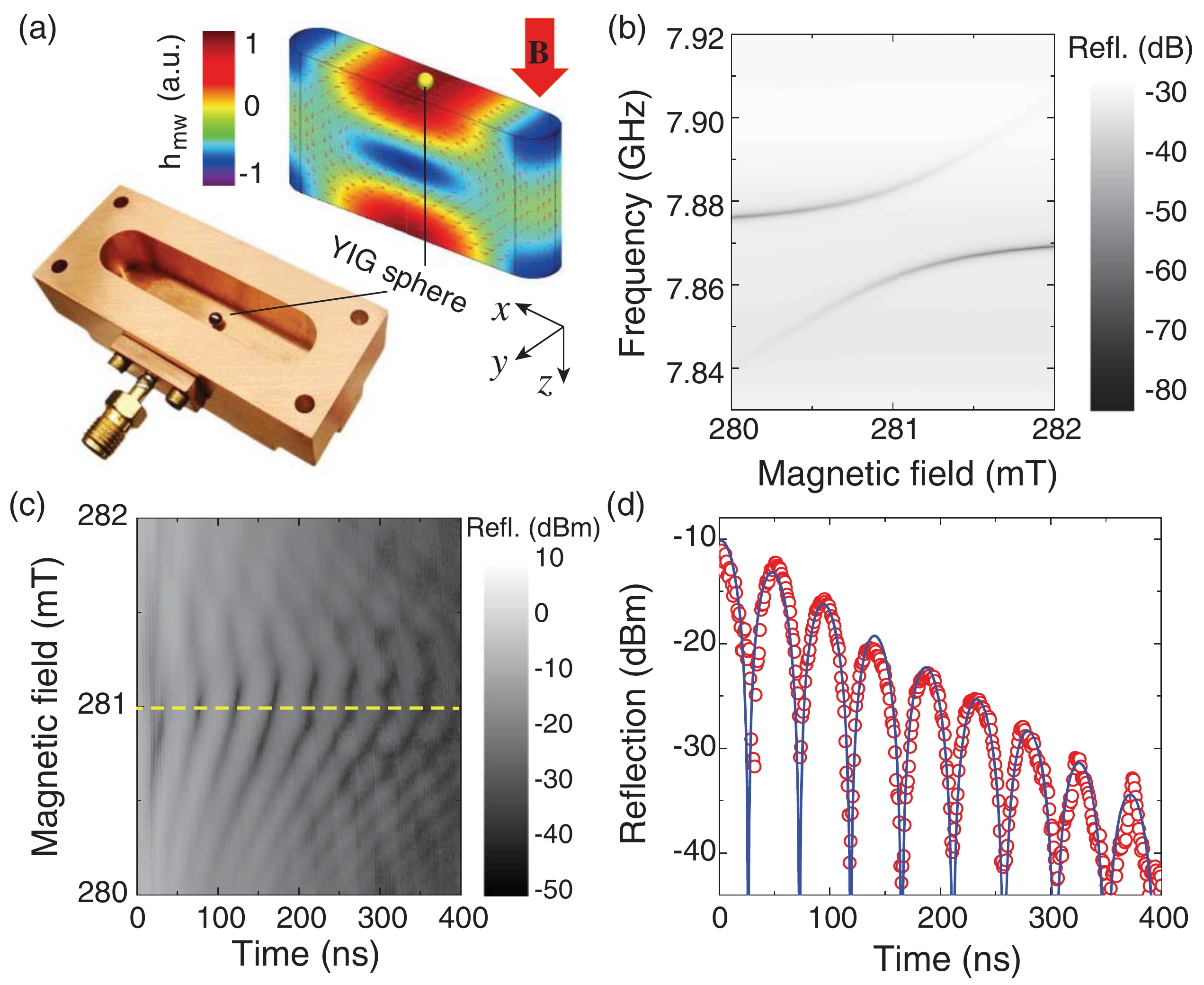}
\caption{\footnotesize{(Color online) Strong magnon-photon interaction and Rabi oscillations observed in the reflection amplitude of a YIG sphere in a 3D MW cavity (explanations in the text). Adapted from \cite{Zhang2014}.}}
\label{FigIV4}
\end{figure}

The transient response of the cavity after pulsed excitation also gives direct information on CMP~\cite{Zhang2014}. Figure~\ref{FigIV4} shows experimental results for a YIG sphere (0.36 mm in diameter) on the magnetic field antinode of the TE$_{101}$ cavity mode. The MW reflection spectra in Fig. \ref{FigIV4}b as a function of magnetic field demonstrate an anticrossing corresponding to the level repulsion. Figures \ref{FigIV4}c and \ref{FigIV4}d monitor the time evolution of the reflection amplitude after populating the cavity mode by a short MW pulse. The observed time traces show the Rabi-like oscillations between magnon-like and photon-like excitations. At resonance (yellow dashed line) the signal contrast is highest, indicating a nearly complete periodic energy exchange between the two systems. The maximum oscillation period in Fig.~\ref{FigIV4}c corresponds to the smallest gap in the avoided crossing of the reflection spectrum in Fig.~\ref{FigIV4}d.  The oscillations at resonance as Fig. \ref{FigIV4}d, plotted on a logarithmic scale, show a slow exponential decay that is governed by dissipation. The contrast between maxima and minima is \(>20\) dB, while the period of 46 ns agrees well with the coupling strength $\pi/g$ = 46.3 ns. The signal according to the two-level model (solid line) agrees very well with the measured time trace (circles).

\begin{figure}[!t]
\centering
\includegraphics[width=\columnwidth]{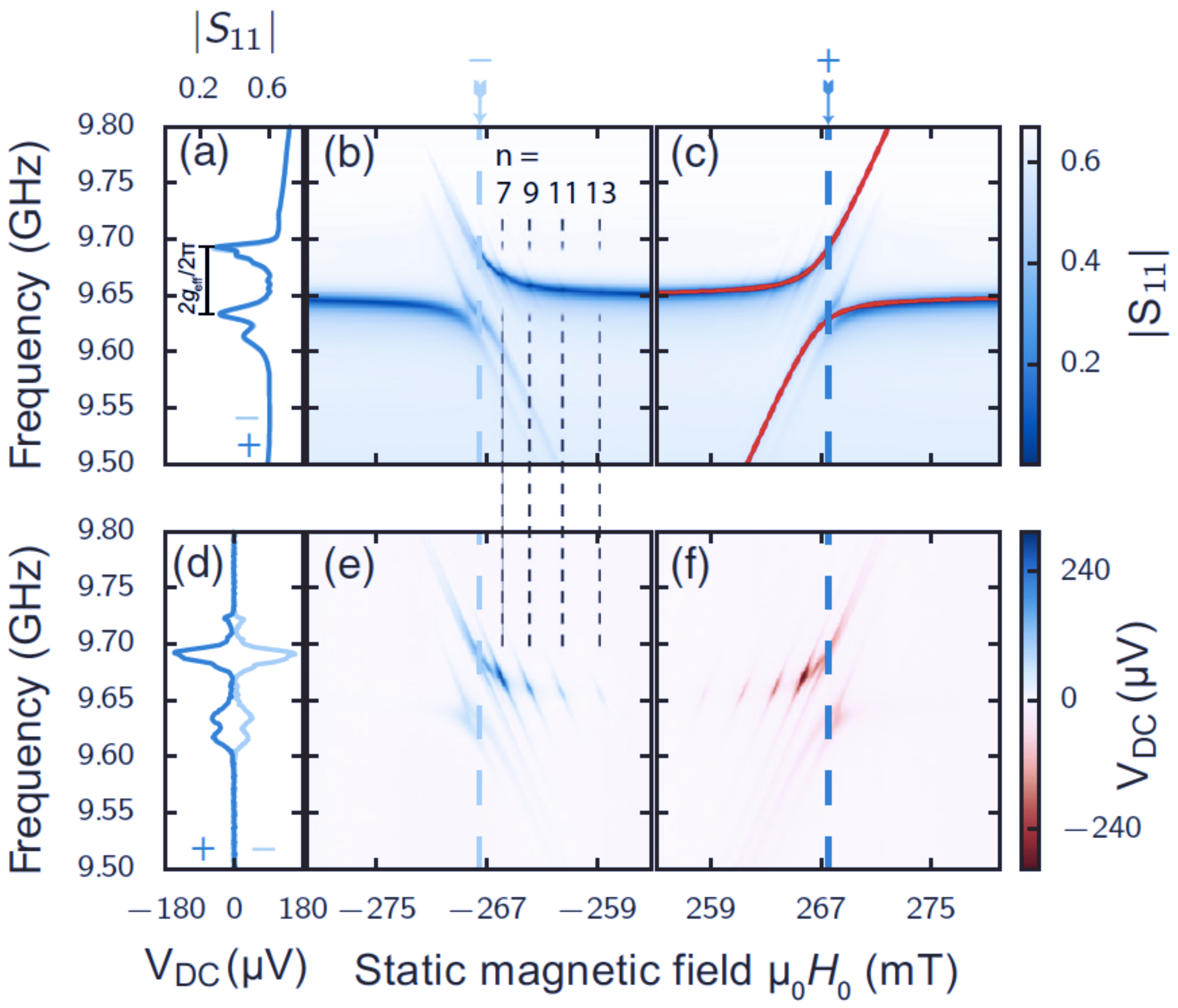}
\caption{\footnotesize{(Color online) (a), (d) Line cuts at $H_0$ = $\pm$267.5 mT show the symmetry under field reversal for $S_{11}$ (a) and $V_{\rm dc}$ (d). (b),(c) Reflection coefficient $S_{11}$ recorded while sweeping the magnetic field and the MW frequency. Strong coupling of the collective spin excitations is indicated by a clear anticrossing, and spin wave modes to the low field side of the main resonance are visible. (e),(f) Simultaneously recorded dc voltage. The strongest signal is from the Kittel mode. Higher magnons modes are visible as well, however, they couple less strongly to the cavity. Adapted from \cite{Maier-Flaig2016}.}}
\label{FigIV5}
\end{figure}

The hybrid magnon-photon modes can be also electrically detected by heavy metal contacts, such as Pt, that convert pumped spin currents into a voltage by the inverse spin Hall effect~\cite{Cao2015,Bai2015,Maier-Flaig2016}. Figs. \ref{FigIV5}b and \ref{FigIV5}c show the amplitude of the reflection coefficient $S_{11}$ recorded on such a device while sweeping the magnetic field and the probe frequency. Strong coupling of the collective spin excitations is indicated by a clear anticrossing, and spin wave modes to the low field side of the main resonance are visible. Figs. \ref{FigIV5}e and \ref{FigIV5}f show the simultaneously recorded dc voltage of CMPs, detected by the spin pumping signal using the Pt detector fabricated on top of the YIG sample. The capability of electrical detection of CMPs has led to the development of cavity spintronics~\cite{Hu2016}, where distant control of the spin current has been demonstrated~\cite{Bai2017}, and the spin current enhancement via an auxiliary spin-wave mode has been achieved~\cite{Xu2020}.

%


Micro-focused Brillouin light scattering (BLS) spectroscopy of spin wave excitations is another tool to access CMPs. These experiments have been carried out on a YIG film coupled to a split-ring MW resonator~\cite{Klingler2016}. Strong coupling with a clear mode anticrossing is observed in the light scattering, which is the first step towards wavelength up-conversion from GHz to THz \cite{Hisatomi2016}. Measurements sensitive to light polarization give insight into the CMP hybridization and the inelastic photon scattering process \cite{Klingler2016}.

Using these experimental techniques, a number of CMP-related interesting phenomena and functionalities have been unearthed, such as magnon dark modes and gradient memory~\cite{Zhang2015dark}, magnon Kerr effect~\cite{Wang2016}, ultrastrong coupling~\cite{Kostylev2016}, cavity-mediated coherent coupling of magnetic moments~\cite{Lambert2016}, cavity-mediated qubit-magnon coupling~\cite{Tabuchi2015}, cavity-mediated remote manipulation of spin current~\cite{Bai2017}, a cavity magnon quintuplet state~\cite{Yao2017}, topological properties and exceptional points~\cite{Zhang2017, Harder2017, Zhang2019}, bistability~\cite{Wang2018}, thermal control ~\cite{Castel2017}, and a nonlinear foldover effect~\cite{Hyde2018}. The research in this direction continues, and more effects will be discovered.

All these effects root on the coherent coupling in hybrid systems, which has potential for both classical~\cite{Maksymov2018, Harder2018b} and quantum information processing~\cite{Lachance-Quirion2019}. A versatile magnon-based quantum information processing platform has taken shape, see Section \ref{SecVI}.

\subsubsection{Dissipative coupling and level attraction}

As reviewed in the previous section, coherent coupling is an active branch of research in the field of cavity magnonics. Historically, level repulsion induced by coherent magnon-photon coupling was first detected by \onlinecite{Artman1953}. They moreover developed a cavity perturbation theory to analyze the coupled system~\cite{Artman1955}. However, unaware of the relevance of such a coupling for studying CMPs in hybrid devices, the magnetism community turned its attention to minimize the cavity perturbation in follow-up experiments, which were often directed towards probing magnons or measuring the magnetic susceptibility of materials. It took more than half a century until the sleeping beauty of coherent coupling was awaken by \onlinecite{Soykal2010}, --- this time with a new perspective of spin-photon entanglement and quantum strong coupling.

\begin{figure}[!t]
\centering
\includegraphics[width=\columnwidth]{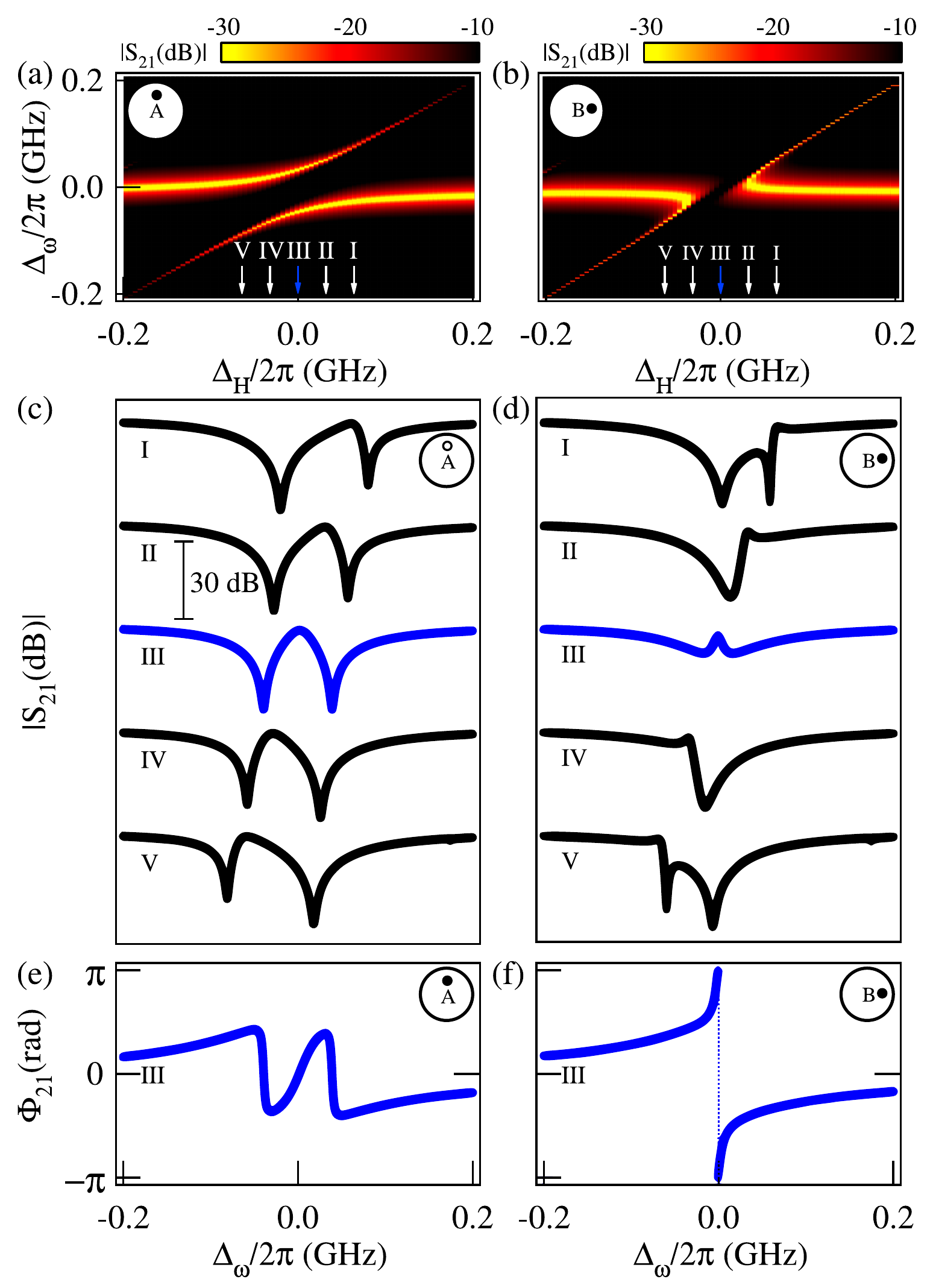}
\caption{\footnotesize{(Color online) Observation of magnon-photon level repulsion and level attraction by setting a YIG sphere in the 1D waveguide depicted in Table \ref{tbl:cavity}. (a) and (b) are the measured dispersions, while (c)-(f) are the measured amplitude and the phase of the transmission coefficient.  Adapted from \cite{Harder2018}}}
\label{FigIV6}
\end{figure}

In contrast, the branch of studying dissipative couplings in cavity magnonics has just started recently. Level attraction caused by dissipative magnon-photon coupling was discovered by \onlinecite{Harder2018} by setting a YIG sphere in the 1D Fabry-Pérot-like cavity (see Table \ref{tbl:cavity}). That cavity, as mentioned in Section \ref{SecIV-A1}, consists of standing wave cavity modes that couple to the outside via travelling waves. Such a cavity exhibits anti-resonances that are characterized by a large external damping rate $\kappa_{\rm ex}$ (see Sec. \ref{secII:io}).
When the YIG sphere is placed on an anti-node of the cavity field for an anti-resonance frequency, i.e. position A  in Fig. \ref{FigIV6}a, the levels repel each other. In Fig. \ref{FigIV6}a, the measured MW transmission amplitude $|S_{21}|$ is plotted as a function of frequency and field detuning $\Delta_\omega = \omega_{\rm D} - \omega_{\rm c}$ and $\Delta_H = \omega_{\rm m}(H)- \omega_{\rm c}$, with $\omega_{\rm m}(H)$ being the field-dependent magnon frequency. It shows a coupling strength of 39 MHz. However, when the YIG sample is placed into a node (position B), the level attraction is observed as shown in Fig. \ref{FigIV6}b, and it can be modelled with a coupling strength of 17 MHz.

The level repulsion between the magnon mode and the cavity anti-resonance is emphasized in Fig. \ref{FigIV6}c.  At $\Delta_H$ = 0,  two sharp dips of equal amplitude are observed. Their splitting and a difference in intensity increases with $|\Delta_{H}|$ as expected for a conventional CMP. In contrast, Fig. \ref{FigIV6}d shows a sharp dip and a relatively broad resonance, that for $\Delta_{H}$ = 0 appear at the same frequency $\omega_{\rm D} = \omega_{\rm m} = \omega_{\rm c}$, i.e. the level attraction is complete. This interpretation is further supported by the phase $\phi_{21}=\arg{S_{21}}$ for $\Delta_{H}$ = 0 in Fig. \ref{FigIV6}e and Fig. \ref{FigIV6}f. The level repulsion is accompanied by two $\pi$-phase shifts, while the single $2\pi$-phase jump in Fig. \ref{FigIV6}f shows that the levels collapsed into one.

Subsequently, \onlinecite{Bhoi2019, Yang2019} reported level attraction by setting a YIG sphere in 2D cavities and \onlinecite{Rao2019a} in a 3D cavity. The common feature of these cavities, as summarized in Table \ref{tbl:LA}, is that they are all similar to waveguides galvanically connected with resonant structures, which support both the standing wave (cavity mode) and the traveling wave propagation. The interference between the standing and traveling waves leads to the cavity anti-resonance, which appears as a dip with a broad background in the transmission spectra of the open cavity. Initially, the coupling between the cavity anti-resonance and the magnon mode was modelled by an effective non-Hermitian term~\cite{Harder2018}, which describes the backaction from the induced RF current impeding the magnetization dynamics, instead of driving it. The model treats the dissipative coupling as a frictional force that couples two harmonic oscillators as depicted in Fig. \ref{FigIV2}.

Searching for the microscopic mechanism of the dissipative coupling, several experiments ~\cite{Wang2019, Yao2019, Rao2020} were performed to investigate the role of travelling waves in different cavities. Theories based on three different approaches are established, all of them consistently attribute the origin of dissipative magnon-photon coupling to traveling wave-induced cooperative external damping: In cavities that support travelling waves, the cavity mode and the magnon mode cooperatively damp to the same traveling waves, leading to an indirect dissipative coupling that causes the level attraction.  Historically, level attraction induced by such an indirect coupling was first observed by Christiaan Huygens in 1673, who found that two pendulum clocks, mounted on the same wall, would eventually swing at the same frequency despite no direct interaction between the clocks. Huygens called the effect ``odd sympathy": The vibration of the wall acts as the common reservoir that correlates the pendulum oscillations, leading to an indirect coupling that ``attracts" the oscillation frequency of the two clocks.

While the traveling wave is the key ingredient of the open cavity magnonic systems in which level attraction has been observed, it is not the only mechanism for inducing dissipative couplings. Theoretically, instead of travelling waves, a damped auxiliary mode has been proposed as an alternative way for mediating the dissipative coupling between two oscillators~\cite{Yu2019}.  The general physical principle of that simple model may be applied to a wide range of coupled physical systems. Furthermore, observing level attraction does not always indicate dissipative coupling. For example, a two-tone driven scheme was proposed by~\onlinecite{Grigoryan2018} to enable level attraction, and it has been realized by \onlinecite{Boventer2019, Boventer2020}. As depicted in Table \ref{tbl:LA}, the key feature of the scheme is that the drive field is split into two paths, one is applied to the cavity input port and the other one is applied through a loop antenna directly to the YIG sample set in the closed 3D cavity.  Another mechanism was proposed by \onlinecite{Proskurin2019}, who showed that magneto-optical coupling, or more specifically, the inverse Faraday effect, may induce the attractive interaction between the magnon and cavity photon modes. This implementation is analogous to the optomechanical approach, where level attraction was realized experimentally in a multimode superconducting MW optomechanical circuit~\cite{Bernier2018}.

Among all these mechanisms, two distinct capabilities stand out for the dissipative coupling mediated by travelling waves: (i) It can be engineered to enable a direction-dependent relative phase between coherent and dissipative magnon-photon couplings, which breaks the time-reversal symmetry for MW propagation. Utilizing the directional interference between coherent and dissipative couplings, nonreciprocal MW transmission has been demonstrated~\cite{Wang2019}. (ii) It sustains the purely dissipative coupling, which enables the realization of anti-parity-time symmetric cavity magnonics. Two types of singularities have been found in such a system: the exceptional points that are square-root singularities appearing in non-Hermitian systems, and an unconventional bound state in the continuum that simultaneously exhibits maximal coherent superposition and slow light capability~\cite{Yang2020}. Moreover, a whole surface of exceptional points has been demonstrated by extending the magnon-polariton system dimensionality into synthetic dimensions given by multiple tuning parameters~\cite{Zhang2019b}.  A theory has further predicted that in systems exhibiting energy level attraction of magnons and cavity photons, parity-time symmetry can also be spontaneously broken, and the magnon and photon can form a high-fidelity Bell state with maximum entanglement in the parity-time symmetry-broken phase~\cite{Yuan2019}. A phase transition to an anti-parity-time-symmetric phase has in turn been demonstrated by using two YIG spheres in a cavity~\cite{Zhao2020}. In general, by utilizing the dissipative coupling, dissipation is no longer a nuisance. On the contrary, it enriches the physics and becomes a resourceful ingredient of open systems. New perspectives for harnessing dissipative couplings, such as dissipation engineering in quantum systems, utilizing the dissipative spin wave bath in cavity spintronics, and developing non-Hermitian metamaterials, have been outlined~\cite{Wang2020dissipative}.

\newpage

\begin{widetext}
\begin{table}[ht!]
\parbox{\textwidth}{\caption{A summary of a few devices and setups used for measuring the level attraction in cavity magnonics. In the experiments listed in the first four rows, the level attraction was due to dissipative magnon-photon coupling mediated by travelling waves, while in the experiment of \onlinecite{Boventer2019}, the level attraction was caused by the interference effect between two driven tones. In all experiments, the intrinsic damping of the standing cavity modes is $\kappa_{\rm in} /2\pi\sim$1--80 MHz. The figures and some of the contents in the table are adapted from \cite{Wang2020dissipative}.} \label{tbl:LA}}
\centering
\begin{tabular}{ccc}
\toprule
Cavity Structure & Key Features & Reference \\
\toprule
1D Fabry-Pérot-like cavity & & \\
\hline
      \begin{minipage}[m]{.45\textwidth}\centering\vspace*{5pt}
      \includegraphics[width=5.5cm]{Table2Fig1}\vspace*{5pt}
      \end{minipage} &
      \begin{minipage}[m]{6cm}
      \begin{itemize}
      \item standing cavity mode and travelling waveguide modes
      \item extrinsic damping:\\ $\kappa_{\rm ex} /2\pi~=~$112~MHz
      \item level attraction
      \end{itemize}
      \end{minipage} &
\cite{Harder2018}
\\
\toprule
  Inverted pattern of the split-ring resonator & & \\
\hline
      \begin{minipage}{.45\textwidth}\centering\vspace*{5pt}
      \includegraphics[width=4.5cm]{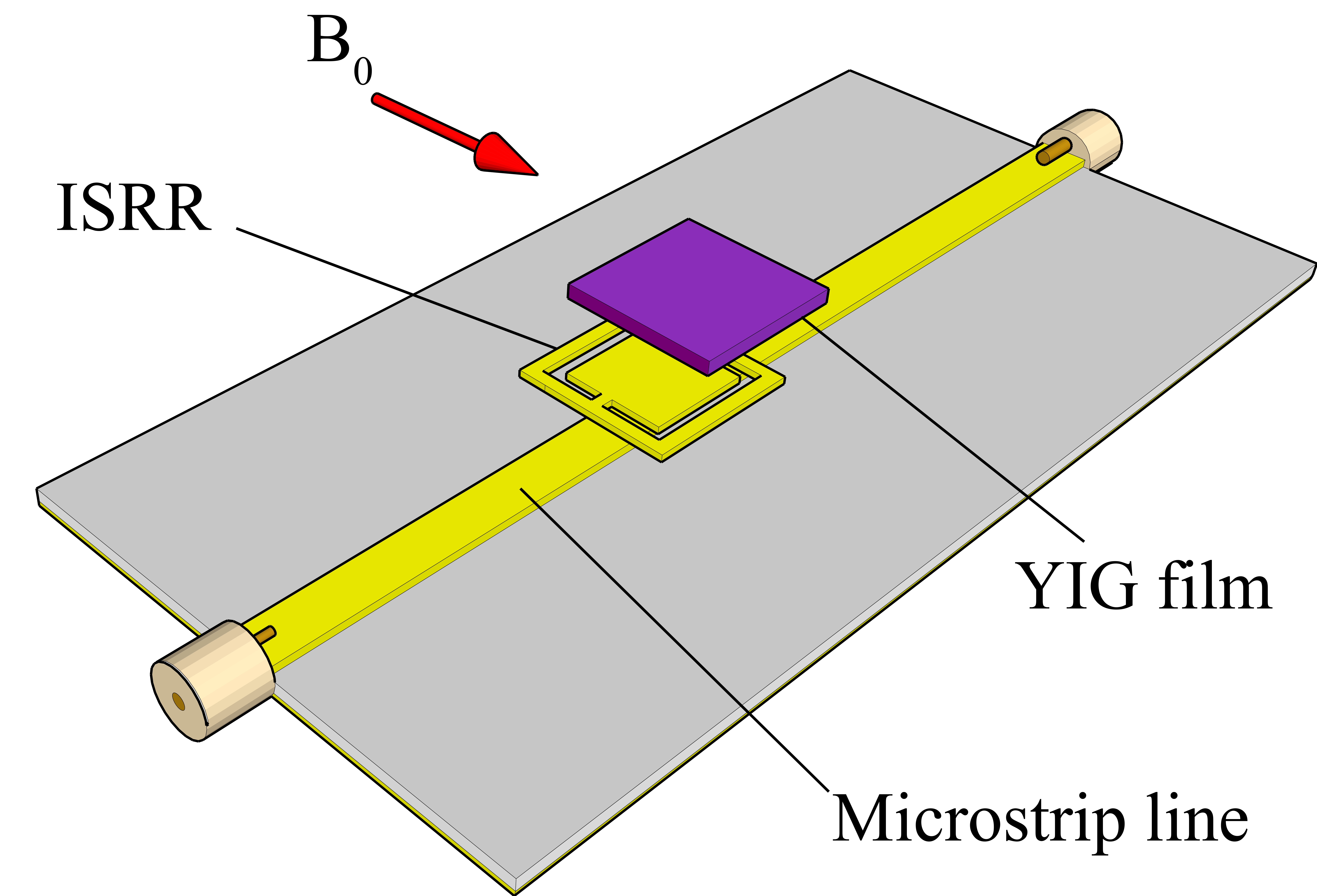}\vspace*{5pt}
      \end{minipage} &
      \begin{minipage}[m]{6cm}
      \begin{itemize}
      \item standing cavity and travelling waveguide modes
      \item level attraction
      \end{itemize}
      \end{minipage} &
      \cite{Bhoi2019}
\\
\toprule
  Cross-line MW circuit & & \\
\hline
      \begin{minipage}{.45\textwidth}\centering\vspace*{5pt}
      \includegraphics[width=5.0cm]{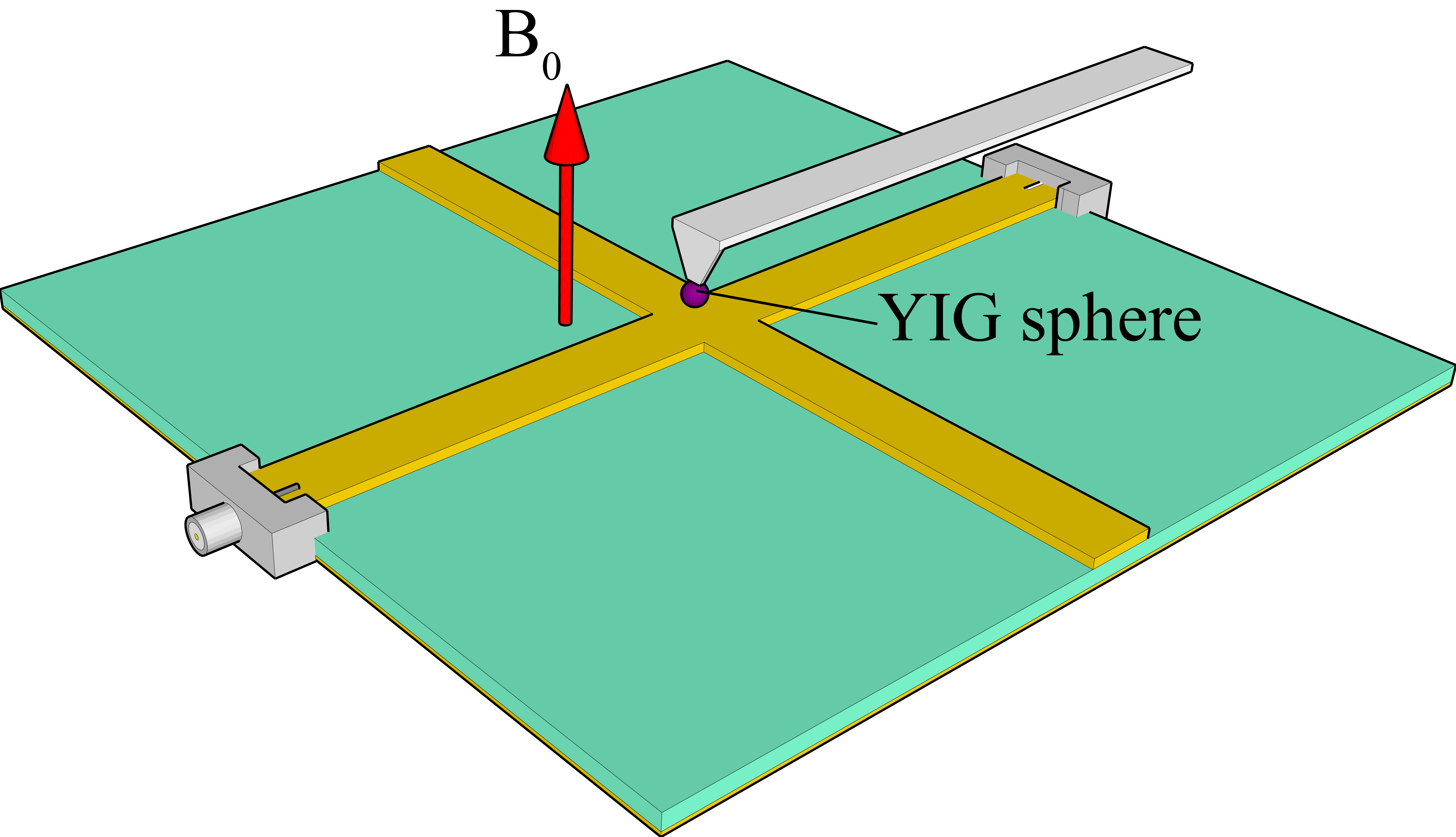}\vspace*{5pt}
      \end{minipage} &
      \begin{minipage}[m]{6cm}
      \begin{itemize}
      \item standing cavity and travelling waveguide modes
      \item extrinsic damping:\\ $\kappa_{\rm ex} /2\pi~=~$880~MHz
      \item level attraction
      \end{itemize}
      \end{minipage} &
      \cite{Yang2019}
\\
\toprule
  Anti-resonance within a 3D cavity & & \\
\hline
      \begin{minipage}{.45\textwidth}\centering\vspace*{5pt}
      \includegraphics[width=4.5cm]{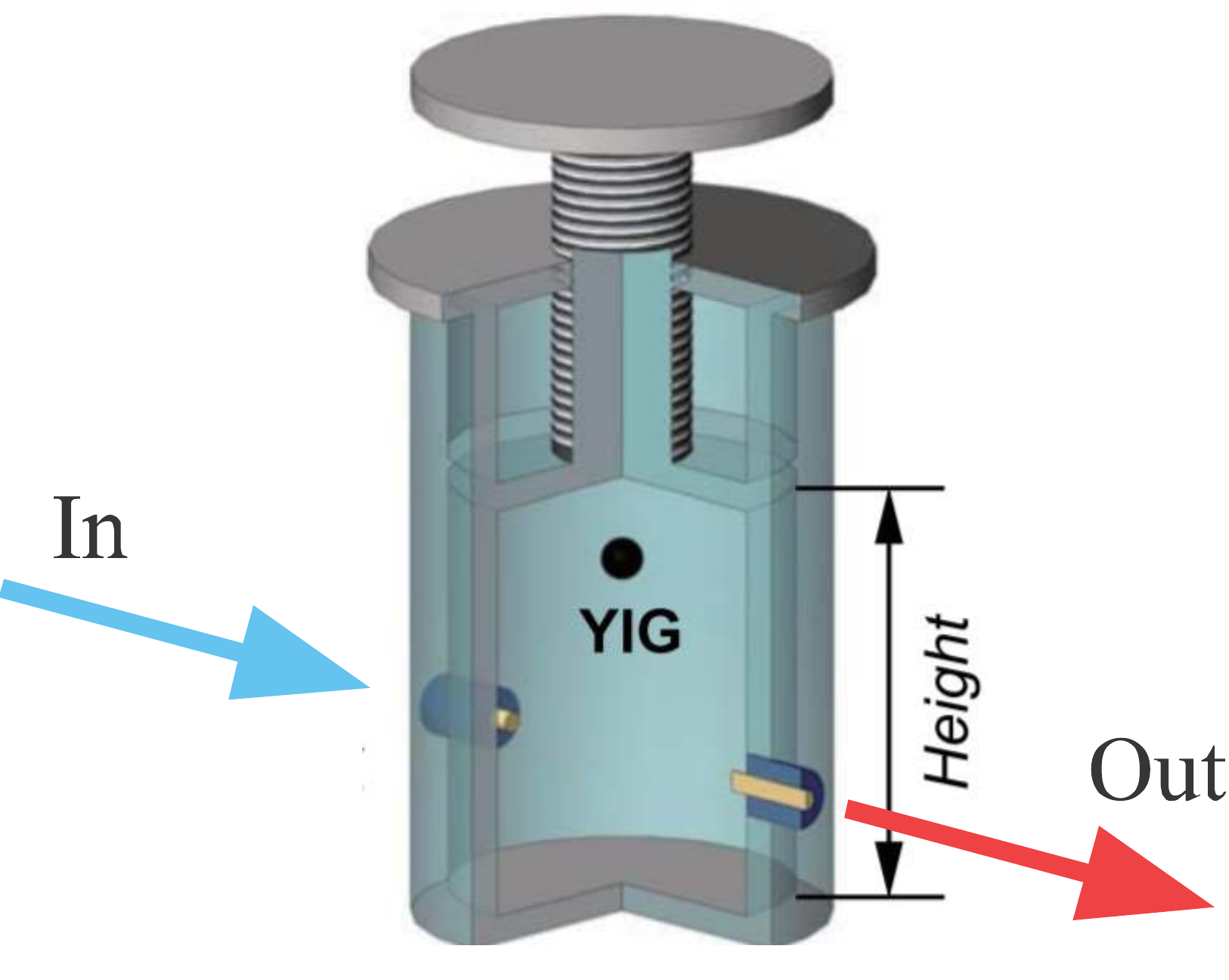}\vspace*{5pt}
      \end{minipage} &
      \begin{minipage}[m]{6cm}
      \begin{itemize}
      \item standing cavity and travelling waveguide modes
      \item extrinsic damping:\\ $\kappa_{\rm ex} /2\pi~=~$14.99~GHz
      \item level attraction
      \end{itemize}
      \end{minipage} &
      \cite{Rao2019a}
\\
\toprule
 3D cavity with two driven tones & & \\
\hline
      \begin{minipage}{.45\textwidth}\centering\vspace*{5pt}
      \includegraphics[width=6.0cm]{Table2Fig5}\vspace*{5pt}
      \end{minipage} &
      \begin{minipage}[m]{6cm}
      \begin{itemize}
      \item standing cavity modes
      \item `cavity' port and `magnon' port
      \item phase shifter
      \item level attraction
      \end{itemize}
      \end{minipage} &
      \cite{Boventer2019}
\\
\hline
\end{tabular}
\end{table}
\newpage
\end{widetext}

\subsection{Nonlinear effects} \label{Sec:IVc}

The physics of coupled driven harmonic oscillators can be explained by classical electrodynamics and linear response to applied MW radiation. However, the dynamics of high-quality magnets can be driven into the  nonlinear regime, causing effects such as Suhl instabilities, see Section \ref{secIIm}. With relative ease, cavity magnonics can  access regimes of nonlinear and quantum dynamics that may be useful for advanced information technology, also discussed in Sec. \ref{SecVI}.

\subsubsection{Instabilities}

It is known for a long time that sufficiently strong MWs drive the magnetization dynamics into the non-linear regime~\cite{Anderson1955, Weiss1959, Gerrits2007, Gui2009}. In the present context, \onlinecite{Wang2018} report a magnon-polariton bistability in a cavity loaded by a YIG sphere in terms of sharp frequency jumps of the resonances that indicate abrupt changes of the amplitudes. A YIG sphere in a Fabry-Pérot-like MW waveguide displays a nonlinear fold-over effect, i.e. a skewed resonance shape as a function of frequency that leads to bistability, a typical signature of non-linear systems~\cite{Hyde2018}. The telltale features are clockwise, counterclockwise, and butterfly-shaped hysteresis loops of the resonance features that depend on the ratio of the magnon and photon components of the magnon polariton excitation.

Since the photon subsystem is linear, fold-over effects must be caused by the nonlinearity of the magnetic subsystem, such as a Kerr nonlinearity~\cite{Wang2018} with positive (negative) coefficient when the static magnetic field is parallel to the [100] ([110]) crystallographic axes of the YIG sphere, respectively~\cite{Zhang2019} (see Sec. \ref{subsec:Beyond-Heisenberg}). These nonlinearity can by captured by modelling the magnon mode by a anharmonic (Duffing) oscillator \cite{Korsch2008} that is coupled to a harmonic oscillator representing the cavity photon mode.

\onlinecite{Makiuchi2021} realized a parametric oscillator in the form of a YIG disk with frequency \(\omega\)  by driving it via a coplanar microwave guide a 2\(\omega\).  The system is a bistable ``parametron", characterized by the phases 0 and \(\pi\) of the magnetic oscillations relative to that of the microwaves. The latter can be read out electrically by the inverse spin Hall effect in Pt contacts. By changing the system parameters, the dynamics can be tuned to form a stable Ising spin system or a randomly fluctuating one with Poissonian statistics. In the latter regime the systems qualifies as a ``probability bit (p-bit)''  in stochastic computing applications \cite{Hayakawa2021}.

\subsubsection{Quantum effects with strong drive}
A number of theoretical proposals explore the quantum nature of magnons in a cavity-magnet architecture. We return in Sec. \ref{SecVI} to quantum effects in the situation when a cavity contains a qubit and a magnet.

\onlinecite{zhang2019quantum} consider the quantum entanglement by the Kerr nonlinearity between the Kittel modes of two YIG spheres in a cavity that is strongly driven by a blue-detuned MW field. This is a strongly driven system, and a large number of magnons is needed, which makes the task of visualizing the quantum effects non-trivial.

\onlinecite{Martinez2019} predicted strong coupling of a MW cavity mode with the gyrotropic motion of a magnetic vortex in sub-\(\mu\)m magnetic disks --- a topologically non-trivial configuration of magnetic structure with potential quantum effects that differ from those associated with the Kittel mode dynamics.

\onlinecite{Elyasi2020} theoretically explored nonlinearities of ferromagnets in MW cavities beyond the macrospin and Duffing approximations. The nonlinearities of a magnet can be interpreted in terms of the Holstein-Primakoff expansion beyond the lowest order term, which introduces interactions between the magnons as discussed in Sec. \ref{secIIm}. The magnon-magnon interactions couple the CMP to energetically degenerate states of backward moving bulk magnons that are pushed in energy by the exchange interaction for sufficiently large wave numbers. The tripartite system under a MW drive and an injection-locking probe can form fixed points that display squeezable quantum fluctuations. They predicted large and distillable quantum entanglement. These quantum resources potentially can be harvested with YIG samples at bath temperatures of around 1 K.

\subsubsection{Nonlinearity induced by microwave feedback}

In all of the experiments reviewed in Section \ref{SecIV-B1}, the measured magnon-photon coupling rate $g_N = g\sqrt{N}$ increases with the spin number $N$, but it is independent of the MW power. This is because in the linear dynamic regime, the excitations in the magnetic subsystem (with $N$ spins) are far from being saturated, so that the number of CMPs $m \ll N$. In such cases, adding photons may increase $m$ but does not change the coupling rate. We note that such a feature is distinctly different from the strong coupling of cavity photons with a single spin, where the single two-level system can be saturated by one photon excitation, so that adding photons enhances the coupling rate.

As reviewed in the two previous subsections, by driving the magnetization dynamics into the nonlinear regime, the coupling features, such as the bistability and the foldover effect, become dependent on the MW driving power. An alternative way to introduce the nonlinearity is to keep the magnetization dynamics in the linear regime ($m \ll N$), but introduce the MW nonlinearity to the coupled system. Such a technique is developed by using the active cavity circuit shown in Table \ref{tbl:cavity}.

\begin{figure}[!t]
\centering
\includegraphics[width=\columnwidth]{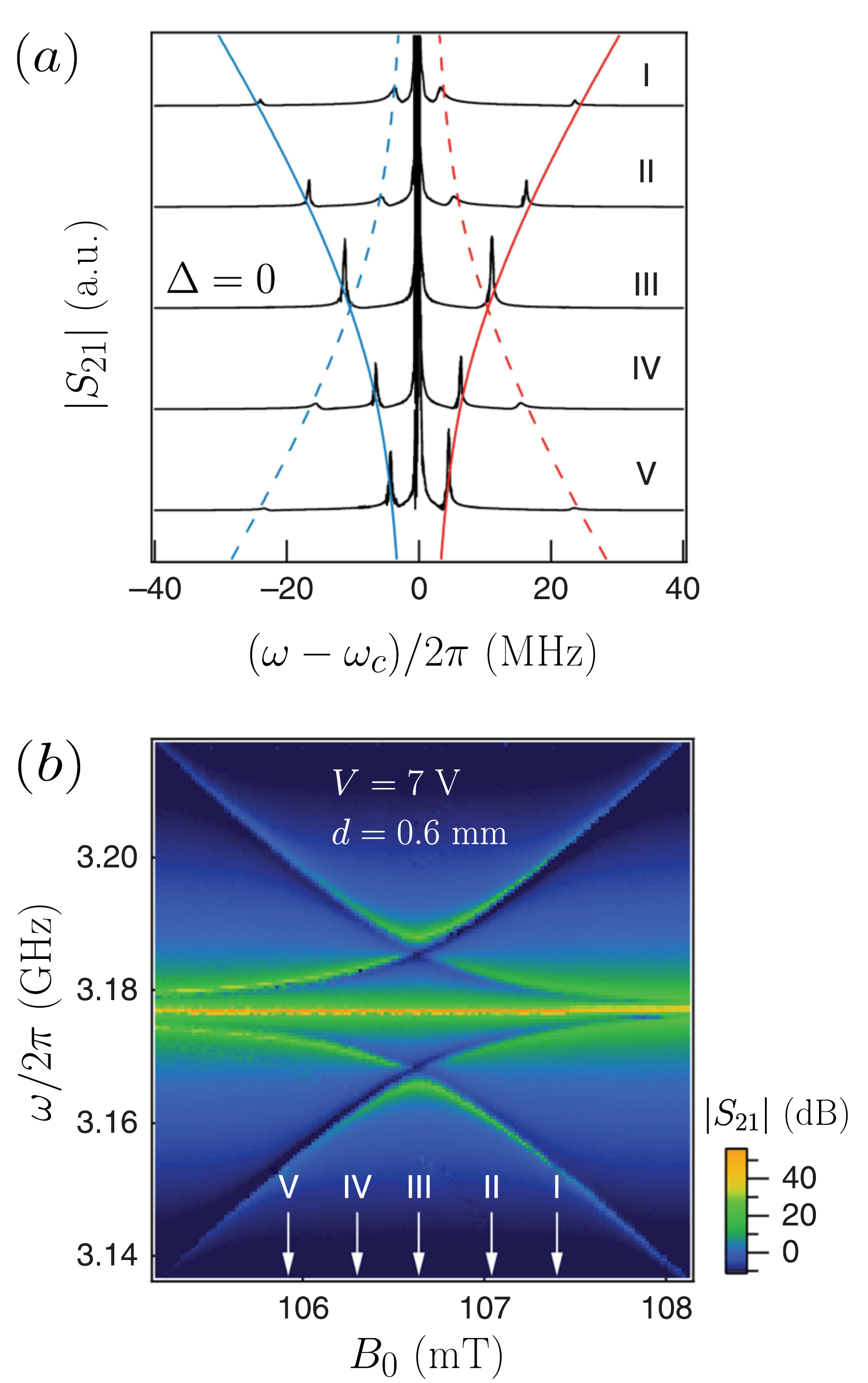}
\caption{(Color online) Cavity magnon quintuplet. Transmission through an active feedback cavity, if driven close to the cavity resonance, shows three peaks as the function of the drive frequency $\omega_{\rm D}$ if the cavity mode is exactly at the resonance with the magnon mode ($\omega_{\rm c} = \omega_{\rm m})$. If the cavity mode is detuned from the magnon mode, the peak splits into five. (a) Dependence of the transmission on the driving frequency for different relations between $\omega_{\rm c}$ and $\omega_{\rm m}$; III corresponds to the resonance between the cavity and the magnon. (b) The same dependence plotted as the function of the drive frequency and of the external magnetic field $\mathbf{B_0}$ which detunes the cavity and the magnons. Adapted from \onlinecite{Yao2017}.}
\label{FigIV7}
\end{figure}

The characteristics of the active cavity circuit have been introduced in Section \ref{SecIV-A1}. It consists of a passive (P) and an active (A) cavity. The A-cavity contains a MW amplifier and acts as a feedback loop that compensates the MW loss of the P-cavity. If one loads a YIG sphere into the cavity circuit, the coherent coupling between the magnons and the MW photons in the P-cavity generates CMPs. But unlike the simple cases reviewed in Section \ref{SecIV-B1}, the $m$-polariton ensemble interacts cooperatively with the $n$ photons fedback from the A-cavity~\cite{Yao2017}. As shown in Fig.~\ref{FigIV7}, instead of the conventional anticrossing between two modes, such a cooperative polariton dynamics leads to five hybridized modes (quintuplet) appearing at  $\omega_{\rm c}$, $\omega_{\rm c}\pm\Omega_+$, and $\omega_{\rm c}\pm\Omega_-$, where
\begin{equation}
\Omega_{\pm} = \sqrt{(\Omega \pm \Delta/2)^2 + 2f^2 g_N^2(\Omega \pm \Delta/2)/\Omega}.
\label{Rabi}
\end{equation}
Here, $\Omega = \sqrt{g_N^2 + (\Delta/2)^2}$ is related to the frequency detuning $\Delta = \omega_{\rm m}-\omega_{\rm c}$, and $f = \sqrt{n/m}$ is the feedback factor that is controlled by the gain of the MW amplifier. At $\Delta$ = 0, Eq. (\ref{Rabi}) reduces to $\Omega_{\pm} = g_N\sqrt{1+2f^2}$, so that the quintuplet reduces to a triplet as shown in Fig.~\ref{FigIV7}. Such a cavity magnon triplet resembles the Mollow triplet  \cite{Mollow1969}, a canonical signature of the light-matter interaction observed in single quantum systems \cite{Xu2007}. It demonstrates that by introducing the MW nonlinearity to the cavity magnonics system, the magnon-photon coupling rate $\Omega_{\pm}$ is controlled by both the spin number and the number of the feedback photons.

\subsection{Two or more magnets in a cavity}

The quantized EM field of cavity photons may coherently interact with spatially separated quantum objects. This allows photon-induced information transfer between distant systems, which is desirable for quantum communication. An indirect interaction between artificial atoms, e.g., superconducting qubits, via a MW resonator~\cite{Majer2007} or a waveguide~\cite{Loo2013}, and in an atomic ensemble through an optical resonator has been reported~\cite{Davis2019}. The coherent coupling of remote paramagnetic spin ensembles, NV centers in diamond via a cavity bus has also been demonstrated~\cite{Astner2017}. If magnon lifetime can be sufficiently improved, combining them with a high-speed intermediary with long coherence length, such as photons, may provide a novel platform for quantum information transfer over macroscopic distances~\cite{Andrich2017,Fukami2021}.

Strong coupling of MW cavity photons with multiple magnets has been realized by \onlinecite{Zhang2015dark}. Two  magnets at the anti-nodes of a cavity mode, Fig.~(\ref{FigIV3}), form a  ``bright" collective mode that precesses in phase with the magnetic field, and a ``dark" mode that does not interact with the cavity mode because the magnetizations of the magnets precess out of phase. The bright mode experiences a Stark frequency shift, while the dark mode is decoupled from the cavity and does not suffer from radiative decay. The coherent long-range coupling of spatially separated magnets via a MW cavity has also been realized in the off-resonant regime~\cite{Lambert2016}, in which magnets are strongly detuned from the cavity modes. The bright  magnon mode is blue shifted when above the main cavity mode, and red shifted otherwise. Note that light-matter interaction has to be nonlinear in order to produce this coupling. The nonlocal coupling between distant magnets in a cavity allows long-range manipulation of spin currents~\cite{Bai2017}. \onlinecite{Zare2018}  interpreted the non-dispersive cavity-photon induced coupling of magnets in terms of a simple molecular model  or ``magnon chemistry''. The dominant mechanism for the coupling is not the coupling of the magnetic field of the cavity to the magnetization, but that of the electric field to the charge polarization in dielectric spheres. The electric component of the cavity mode can dominate the coupling which reaches the ultra-strong coupling regime when the cavity is significantly filled \cite{Zare2018}. A long-range coupling between ferro- and antiferromagnets through a MW cavity has been predicted~\cite{Johansen2018}, which might pave the way toward integrated circuits with new components.

The nature of the interaction mediated by cavity photons among magnets depends on the cavity type and location of the magnets in the cavity. In particular, the cavity can be closed and conserve energy, or dissipative, such as an open waveguide. When the direct coupling of both magnets to the cavity is of the same type, i.e. either coherent to a confined cavity mode, or dissipative to the continuum of modes in an open cavity, the indirect long-range coupling is coherent and levels anti-cross or repel each other. When it is predominantly dissipative, a level-attraction can be expected. A long-range dissipative coupling in a dispersive regime has been proposed~\cite{Grigoryan2019} and observed ~\cite{Xu2019}. The chiral coupling of a chain of magnets with travelling photons in a waveguide that loses energy at the open ends, leads to extended collective magnon modes that are sub-radiant as well as super-radiant edge states with large amplitudes~\cite{Yu2020a, Yu2020b}.

One can think of more complicated ``chemistry'' by putting various coherently coupled objects, not just magnets, to a cavity. In particular, Sec. \ref{SecVI} shows what happens if a magnet and a qubit are coherently coupled inside cavity. \onlinecite{Janssonn2020} considered a magnet and a superconducting sphere and showed that the coherent coupling affects the properties of the superconductor.


\section{Magnons in optical resonators} \label{SecV}

A strong coupling of magnons at GHz frequencies and light at frequencies above 100 THz is difficult to achieve. The large frequency mismatch prohibits an anticrossing and the formation of magnon polaritons at light frequencies. Strong coupling can still be reached in principle when the ``lamp'' shift exceeds the line widths. However, the Zeeman coupling of the spin magnetic moment with the photon magnetic field is suppressed inversely proportional to the detuning. The electric field component, on the the other hand, couples to the spin only via the relatively weak spin-orbit interaction \cite{Borovik-Romanov1982}, see Sec.~\ref{SecIII}. The leading interaction  is a second order process, the electric field-induced two-photon-one-magnon inelastic scattering (see Sec.~\ref{OffresonantMO}). Optimal frequencies are close but below the band gaps in order to resonantly enhance the scattering cross section without significant absorption that would prohibit high photon intensities, which for YIG is in the near infrared. Samples with high dielectric constants act as antenna and confine the photons, which then have ample time to interact with the magnons. YIG spheres and slabs thereby form optical resonators with enhanced magnon-photon interaction. Analogous to optomechanics \cite{Aspelmeyer2014}, the radiation pressure-type interaction while intrinsically weak ($\sim$10\,kHz) is parametrically enhanced by the photon number or optical drive power (see Sec.\,\ref{SecIII}). \onlinecite{Groblacher2009} report strong phonon-photon coupling at high photon intensities. 

The ability to strongly couple magnons to an optical mode would  open up opportunities in quantum technology, such as the optical communication between distant quantum computers with clock frequencies in the MW regime and form an interface between mK and room temperatures \cite{Lambert2019,Lauk2020}. The potential of such transducers are a strong motivation for research on optomechanical systems that reach already efficiencies close to unity \cite{Higginbotham2018,Fan2018}. \onlinecite{Mirhosseini2020} report optical photon emission by superconducting transmon qubits. Repeating such a feat with magnetic systems would offer new functionalities that exploit the intrinsic time-reversal symmetry breaking of the magnetic order, enabling, for example, unidirectional conversion without the need for complex drive schemes \cite{Yu2009, Metelmann2015}. Note that magnetic materials are routinely applied in communication technology as isolators that are transparent only in one direction, at MW \cite{Pozar2004} and optical frequencies \cite{Jalas2013}.

Optical measurements at the single magnon level that complement those in the MW domain \cite{Lachance-Quirion2017} is another challenge. The higher bandwidth of lasers can circumvent measurement problems associated with the magnon dissipation \cite{Lachance-Quirion2019}. Indeed, optomechanics has benefited from the complementary progress on the MW and optical side, culminating in the demonstration of qubit-optical transduction \cite{Higginbotham2018,Mirhosseini2020}. The advantage of magnetic systems is the magnetic field knob that can freeze out background thermal magnons to a large extent even at not so low temperatures.

Optical techniques such as Kerr-Faraday rotation spectroscopy, inelastic light scattering \cite{Demokritov2008,Sebastian2015} and ultrafast pump-and-probe techniques \cite{Walowski2016} are well-established probes of magnetism. Light is much less used to control the magnetic order \cite{Kirilyuk2010}. The ability to manipulate magnetism in the strong optical interaction limit would enable a number of interesting fundamental experiments, such as optical cooling \cite{Sharma2018} or  driving selected magnon modes \cite{Simic2020}.

This Section reviews the progress in enhancing the magnon-optical photon interaction in resonators, starting with the interaction mechanisms, lists the relevant parameters, and introduces proposed optical cavity geometries. Subsequently, we review recent experiments that observe enhanced magnon-photon coupling in YIG optical whispering gallery mode (WGM) resonators.

\subsection{Interaction between magnons and optical photons}\label{sec:OM}

YIG is a wide band-gap electrical insulator with a transparency window in the infrared without measurable absorption, but also with vanishing magnetooptical constants that become significant only by resonant enhancement. Closer to the band gap, both electric dipole and magnetic dipole two-photon transitions become significant \cite{Krinchik1962}. For example, for light at a wavelength of 1.15\,$\mu$m, the ratio between the optical transitions induced by the magnetic and electric field components is 0.06 \cite{Gall1971a}. Electric dipole transitions can be treated in terms of a dielectric tensor $\overleftrightarrow{\varepsilon}{(\mathbf{M})}$ that depends parametrically on the magnetization \textbf{M} \cite{Landau1984, Pershan1966}, see Eq.~(\ref{permitivitty}) and Sec. \ref{SecIII}. This approach captures most magneto-optical phenomena such as the elastic Faraday-Kerr and Cotton-Mouton rotation of the light polarization and the inelastic magnon Brillouin light scattering (BLS) \cite{Wettling1975}. We focus here on the latter process in which a scattered photon suffers a red or blue shift by creating or annihilating a magnon. By the conservation of angular momentum the photon polarization must change during the scattering process as well \cite{Gall1971}, as discussed in Section \ref{SecIII}. Only when rotational symmetry is significantly broken, a net angular momentum of photons and magnons can be transferred to the crystal \cite{Hisatomi2019}.

Historically, magnon BLS is a standard probe of the magnetization dynamics of magnons with small wave vectors \cite{Demokritov2008}, e.g. in identifying a magnon Bose-Einstein condensate \cite{Demokritov2006}, characterizing artificial magnonic crystals \cite{Chumak2009,Sebastian2015,Baba2019}, or probing magnetic textures~\cite{Novosad2002,Schultheiss2019}. By its weak interaction light is relatively non-invasive, i.e. it only weakly perturbs the system under study. Cavity optomagnonics strives to reach a new regime of enhanced magnon-photon coupling that allows manipulating the magnetic system.

The light-magnon interaction in a cubic crystal is governed by Eq.~(\ref{permitivitty}) that leads to the Hamiltonian~(\ref{optomagnonicHam_dispersive}).
In the following we disregard the second order (Cotton-Mouton) term in $\delta\varepsilon_{ij}(\mathbf{M})$ and concentrate on the Faraday part, since the former only renormalizes the interaction parameters in the standard measurement configuration in which incoming and scattered light are normal to the magnetization  \cite{Sharma2017}.  The single adjustable constant $f$ is directly related to the Faraday rotation angle  $\theta_f = \omega f M_s/(2 c n)$ by a magnet with thickness equal to the wavelength.  In the leading order process an input photon in mode $p$ scatters into an output mode $q$  via a magnon in mode $\eta$, governed by the optomagnonic matrix element $G_{pq\eta}$ in the interaction Hamiltonian of Eq.~(\ref{optomagnonicHam_dispersive}),
\begin{equation}
G_{pq\eta} = -i \frac{1}{2\hbar} f M_{\rm s} \sqrt{\frac{4 g \mu_B}{M_{\rm s} V_{\rm m}}}\sqrt{\frac{\hbar \omega_p}{2\varepsilon_0\varepsilon V_p}}\sqrt{\frac{\hbar \omega_q}{2\varepsilon_0\varepsilon V_q}} V_\text{int} .
\end{equation}
The interaction volume $V_\text{int}$ is the overlap integral of the three mode functions in the form of a triple vector product given by
\begin{equation}
V_\text{int} = \int{d\mathbf{r}\,\mathbf{v}_{\eta}(\mathbf{r}) \cdot[\mathbf{u}_q^{*}(\mathbf{r})\times\mathbf{u}_p(\mathbf{r})]},
\label{Vint}
\end{equation}
where $\mathbf{u}_q(\mathbf{r})^{*}$ and $\mathbf{u}_p(\mathbf{r})$ are amplitudes of the optical cavity modes with mode volumes $V_p$ and $V_q$ (see Sec.\ref{secII}), while $\mathbf{v}_{\eta} (\mathbf{r})$ is related to the frequency-normalized Eq. \eqref{eq:m_norm} in Sec. \ref{secIIm},  $\mathbf{w}_{\eta} = \sqrt{{4 g \mu_B}/(M_s V_\textrm{m})} \mathbf{v}_{\eta}$. The effective magnon mode volume here is $V_\textrm{m} = \int{|\mathbf{v}_{\eta}(\mathbf{r})|^2 dV}$, see Eq. \eqref{eq:V_mag_eff}, which is strictly valid only for circularly polarized magnon modes~\cite{Sharma2019}. According to Eq.\,(\ref{Vint}), the TM-TE of TE-TM scattering configuration maximizes the coupling.

In the experiments that follow, the optical mode frequencies and volumes differ only slightly. With $V_p \approx V_q \equiv V_\text{opt}$ and $\omega_p \approx \omega_q \equiv \omega_\text{opt}$, we obtain
\begin{equation}\label{eq:smallVm}
G_{pq\eta} = -i  \theta_f \frac{c}{n} \sqrt{\frac{4 g \mu_B}{M_{\rm s}}} \sqrt{\frac{1}{V_\textrm{m}}} \frac{V_\text{int}}{V_\text{opt}}.
\end{equation}
%
The coupling increases with (i) the material dependent Faraday angle \(\theta_f\), (ii) a geometry with large triple-mode overlap, and (iii) a small magnetic volume.

\subsubsection{Magnon Brillouin light scattering}\label{sec_conv_eff}

\begin{figure}
\includegraphics[width=\columnwidth]{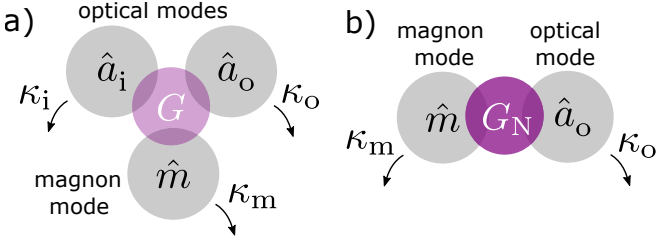}
\caption{(Color online) Limited Hilbert space for resonant magnon light scattering with coupling and dissipation rates (a) for general magnon Brillouin light scattering that is simplified (b) for a strongly driven optical mode \(i\) that can be eliminated in favor of an enhanced coupling rate $G_{N}$.}
\label{fig:V_1}
\end{figure}

Here we address the magnon annihilation rate by anti-Stokes inelastic light scattering at a thermally or MW excited magnet close to resonance as in Fig. \ref{fig:V_1}a. In the interaction  Eq.~(\ref{optomagnonicHam_dispersive}) we focus on three levels, viz. magnon mode with frequency $\omega_\text{m}$ and optical input and output modes $\omega_\text{i}$ and $\omega_\text{o}$ with  $\omega_\text{i}>\omega_\text{o}$,
\begin{eqnarray}
\hat{H}_{\rm opt-m} = \hbar\omega_\text{i}\hat{a}^{\dagger}_\text{i}\hat{a}^{}_\text{i}
& + & \hbar\omega_\text{o}\hat{a}^{\dagger}_\text{o}\hat{a}^{}_\text{o} 
+  \hbar\omega_\text{m}\hat{m}^{\dagger}\hat{m}^{} \nonumber \\
 & + & \hbar G \left( \hat{a}^{\dagger}_\text{o} \hat{a}^{}_\text{i}\hat{m}^{}  + \hat{a}^{\dagger}_\text{i} \hat{a}^{}_\text{o}\hat{m}^{\dagger}\right), \label{eq:H}
\end{eqnarray}%
where  $G_{\text{iom}}^{+} \equiv G$ and $G_{\text{iom}}^{-}=0$.  A proximity laser with frequency $\omega_{\rm D}$, slightly detuned from the input by $\Delta_\text{i}=\omega_\text{i}-\omega_{\rm D}$, drives the system. The input amplitude  $\hat{a}_\text{i}=\langle a_\text{i} \rangle+\delta\hat{a}_\text{i}$ is mainly coherent but has small fluctuations, where $\langle a_\text{i} \rangle=\sqrt{\kappa_\text{i,ex}}\langle a_\text{i,in} \rangle/(\kappa_\text{i}/2 - i \Delta_\text{i})$, $\kappa_\text{i} = \kappa_\text{i,0} + \kappa_\text{i,ex}$ is the total damping rate, and $\langle a_\text{i,in}\rangle$ is the amplitude from the proximity coupling to the external laser. The output mode is not driven, so $\langle a_\text{o}\rangle \sim 0$ (see Sec. \ref{secII:io}). The magnons are populated thermally or driven by MWs to an amplitude $\hat{m}_{\rm in}$.

The linearized Hamiltonian in the rotating frame of the drive frequency is (Fig. \ref{fig:V_1}b)
\begin{align}
\hat{H}'_{\rm opt-m} = \hbar \Delta_\text{i} \delta\hat{a}^{\dagger}_\text{i}\delta\hat{a}^{}_\text{i} 
+ & \hbar \Delta_\text{o} \delta\hat{a}^{\dagger}_\text{o}\delta\hat{a}^{}_\text{o} 
+ \hbar\omega_\text{m}\hat{m}^{\dagger}\hat{m}^{} \nonumber \\
& + \hbar G_{N} \left( \delta\hat{a}^{\dagger}_\text{o} \hat{m}^{}  +  \delta\hat{a}^{}_\text{o}\hat{m}^{\dagger} \right),
\end{align}
with coupling constant $G_{N} = \sqrt{N_{\rm i}}G$, where $N_{\rm i}=|\langle a_\text{i}\rangle|^2$ is the number of photons in the input mode, and the output mode is detuned by $\Delta_{\rm o} = \omega_{\rm o} - \omega_{\rm D}$. The Langevin equations that describe the dynamics of this Hamiltonian have been introduced in Sec. \ref{secII} (see also Sec. \ref{secIII:io}),
\begin{align}
\frac{\partial \hat{m}}{\partial t}  =& -i \omega_\text{m}\hat{m} - \frac{\kappa_\text{m}}{2}\hat{m} + \sqrt{\kappa_{\text{m}\text{,ex}}}\hat{m}_\text{in} - i G_{N} \delta\hat{a}_\text{o},\\
\frac{\partial \delta\hat{a}}{\partial t} =& -i \Delta_\text{o}\delta{\hat{a}}_\text{o} - \frac{\kappa_\text{i}}{2}\delta{\hat{a}}_\text{o} + \sqrt{\kappa_\text{o,ex}}\delta{\hat{a}}_\text{o,in} - i G_{N} \hat{m}.
\end{align}\label{eom}
Their solutions read
\begin{align}
{\hat{m}} =& \frac{\sqrt{\kappa_{\text{m}\text{,ex}}}\hat{m}_\text{in} - i G_{\rm N} \delta\hat{a}_\text{o} }{ \kappa_{\text{m}}/2 - i (\omega - \omega_\text{m})},\\
\delta{\hat{a}}_\text{o} =& \frac{\sqrt{\kappa_{\text{o,ex}}}\delta\hat{a}_\text{o,in} - i G_{\rm N} \hat{m}}{ \kappa_\text{o}/2 - i \left(\omega - (\omega_\text{o}-\omega_{\rm D})\right)}.
\end{align}
where \(\kappa_{\text{m}\text{,ex}}\) is the coupling rate of the magnon to the microwave source. 

Using the input-output relation Eq.~(\ref{losses_cavity_general}) and tuned input frequency $\omega_{\rm D} = \omega_\text{i}$, we get
\begin{equation} \label{output_optical mode}
|\delta\hat{a}_\text{o,out}|^2 = \frac{4  \left( \frac{4 G_{\rm N}^2}{\kappa_\text{o}\kappa_\text{m}} \right) \frac{ \kappa_{\text{m}\text{,ex}}}{\kappa_\text{m}} \frac{ \kappa_{\text{o}\text{,ex}}}{\kappa_\text{o}} }{\left(1+\frac{4 G_{\rm N}^2}{\kappa_\text{o}\kappa_\text{m}} \right)^{2} + \left( \frac{\omega_\text{m} - (\omega_\text{o} - \omega_\text{i})}{\kappa_\text{o}/2} \right)^2} |\hat{m}_\text{in}|^2.
\end{equation}
The scattering is maximized  at the {\it triple resonance condition}  $\omega_\text{m} = \omega_\text{o} - \omega_\text{i}$,
\begin{equation} \label{tripleres}
\max |\delta\hat{a}_\text{o,out}|^2 = \frac{4  C_{\rm OPm} }{\left(1+C_{\rm OPm} \right)^{2}} \frac{ \kappa_{\text{m}\text{,ex}}}{\kappa_\text{m}} \frac{ \kappa_{\text{o}\text{,ex}}}{\kappa_\text{o}} |\hat{m}_\text{in}|^2.
\end{equation}
The magnon  annihilation by anti-Stokes scattering is governed primarily by the factor  $4 C_{\rm OPm} /\left(1+C_{\rm OPm} \right)^{2}$, which is largest when the drive-enhanced optomagnonic cooperativity $C_{\rm OPm} = 4 G_{\rm N}^2 / (\kappa_\text{o} \kappa_\text{m})$ is unity (see Sec. \ref{secIII:io}). The other two factors reflect the impedance matching of magnon and output optical modes and approach unity when the coupling rates to the outside world are much larger than the internal decay rates  $\kappa_{\text{m}\text{,ex}} \gg \kappa_{\text{m}, 0}$ and $\kappa_{\text{o}\text{,ex}} \gg \kappa_{\text{o}, 0}$.  The driven optomagnonic coupling $G_{N}$ is proportional to the optical power stored in the input mode and depends on the power and the proximity impedance matching of the input laser. The single-photon cooperativity $C_{\rm OPm}^{0} =(4G^2) / (\kappa_\text{o} \kappa_\text{m} ) = C_{\rm OPm}/N_\text{i}$ does not depend on the input power. 
The scattering probability Eq. (\ref{output_optical mode}) is proportional to the magnon number that at thermal equilibrium is governed by the Planck distribution function. At low temperatures there are no magnons to annihilate, so anti-Stokes BLS is suppressed. On the other hand, external stimuli such as resonant MWs, can strongly enhance the magnon number and BLS cross section. The correlation between MW absorption and BLS spectra helps in to assign magnon modes as discussed in Sec. \ref{SecVHigherorder}.

\subsection{Optical cavity designs}
\subsubsection{Materials}
The materials parameters that determine the optomagnonic coupling are the refractive index, the Faraday angle, and the saturation magnetization. For the cooperativities we also need the damping rates of the optical and magnon modes. YIG is currently the best available material since it has been optimized for commercial applications at both MW and optical frequencies for MW generation or filtering and optical isolators or circulators.  The Faraday rotation angle of undoped YIG at wavelength 1.5\,$\mu$m is $\sim$4~rad/cm$^{-1}$  per unit of thickness and the Gilbert damping \(\alpha \approx 10^{-4}-10^{-5}\), see Sec. \ref{sec:magnon_dissipation}. The figure of merit  for optical isolators is the Faraday rotation divided by the optical loss \cite{Stadler2014}. Doping can increase this number at the cost of increased damping \cite{Wood1967}. This is not an issue in static optical isolators, but the trade-off from doping often leads to a reduced cooperativity. 

Van der Waals ferromagnetic and antiferromagnetic materials \cite{Huang2017, Gong2017} show sizable magneto-optic activity for only a few monolayers.  \onlinecite{Huang2017} report a Kerr rotation angle of 0.005\, rad for a monolayer of CrI$_3$ (but not the optical absorption). With $\sim$1\,nm thickness this corresponds to $\approx$50,000~rad/cm$^{-1}$. The low magnetic damping in vanadium tetracyanoethylene \cite{Zhu2016} make such organic-based ferromagnets also attractive for optomagnonics \cite{Liu2018}.

\subsubsection{Optical resonators}

The cavities that confine light with wave length in \(\mathrm{\mu}\)m are quite different from those of microwaves with cm wave length. They range from simple Fabry-P\'erot resonators, to complex photonic crystal devices. Light can be trapped simply by a material with high dielectric constant with the experimental challenge to couple such a resonator in a controlled manner to the external laser input and output. The key geometric parameters are the overlap of the magnon and photon modes as well as the volume of the magnon mode. In order to meet the resonance condition the TM-TE photon mode splitting should be comparable to the magnon mode frequency that can be fine-tuned by a magnetic field. The optical losses by absorption and disorder scattering by the magnet and at the proximity coupling node should be minimized as well. 

\begin{figure}%
\includegraphics[width=\columnwidth]{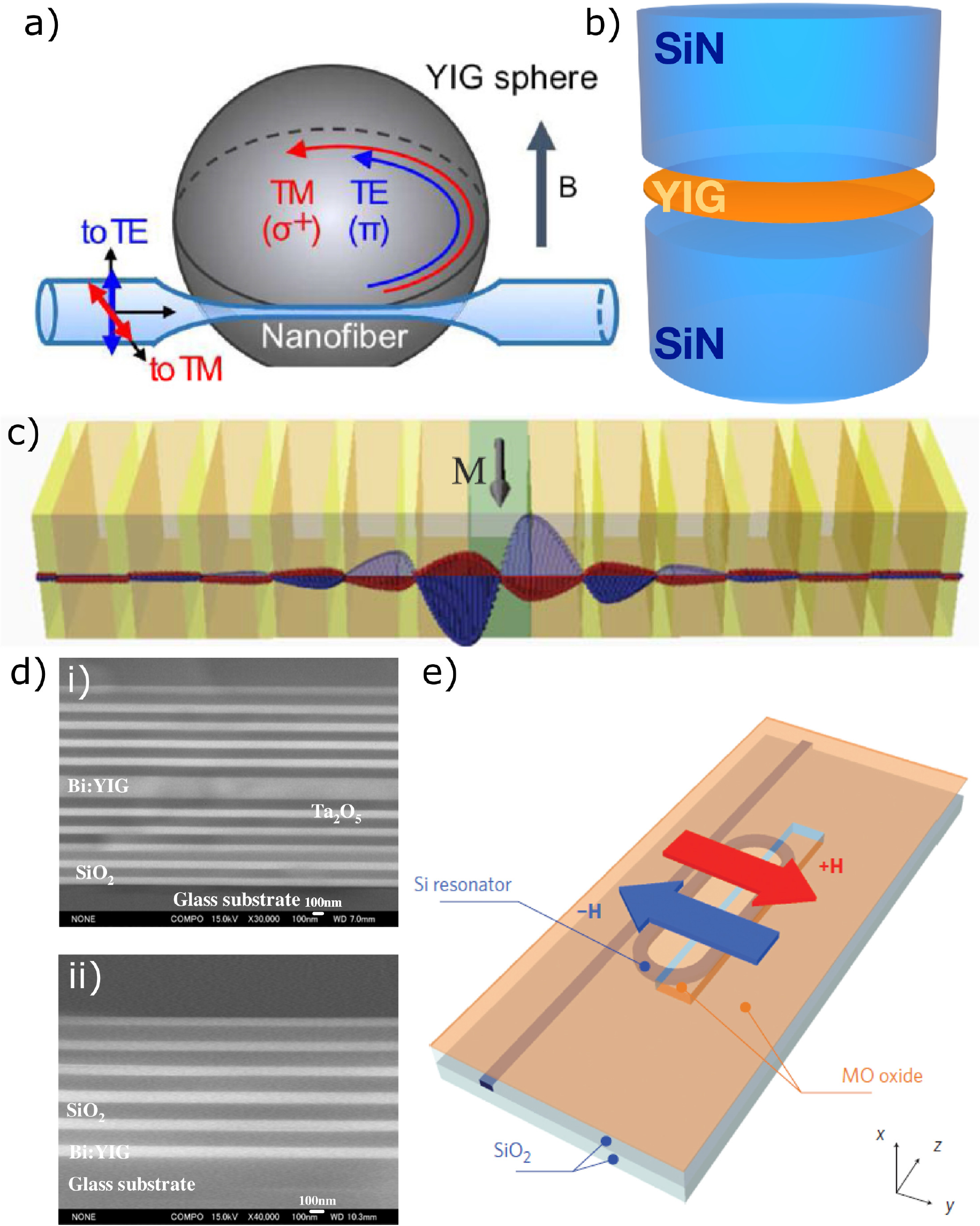}%
\caption{(Color online) Optical cavities with embedded magnetic elements. (a) A solid sphere of YIG with optical whispering gallery modes~\cite{Osada2016}. (b) The reduced mode volume in cylindrical YIG samples can achieve higher magneto-optical coupling and support different magnetic textures \cite{Graf2018}. (c) Photonic crystals can confine the optical field \cite{Pantazopoulos2019}, also in narrow-band optical isolators \cite{Inoue2006} (d). (e)  Compact optical isolator with magnetic elements based on a silicon photonic ``racetrack'' resonator \cite{Bi2011} .}%
\label{MOcavities}%
\end{figure}

Whispering gallery mode (WGM) resonators are frequently used to enhance optical interactions with phonons, and they turned out to be very useful in optomagnonics as discussed below. Since WGMs are confined to the sample boundary, the surface roughness must be suppressed. Mechanically polished YIG spheres with sub-mm diameters are commercially available for microwave applications. The high refractive index $n_{\rm YIG}\approx2.2$ and transparency in the infrared enable these spheres to support well-defined optical WGMs as well.  

Optical resonators with embedded magnetic elements enhance the static Faraday effect, for application in compact optical isolators \cite{Stadler2014}. Photonic crystal devices may increase the Faraday rotation by a factor of 4 \cite{Inoue2006}. A single magnetic layer sandwiched between two non-magnetic Bragg reflectors \cite{Takayama2000} or magnetic multilayers  \cite{Fedyanin2004} generate localized optical Tamm states with enhanced amplitudes \cite{Goto2008} that also show enhanced magnetic second harmonic generation \cite{Fedyanin2002}. \onlinecite{Wang2005} proposed  optical isolators based on defects in two-dimensional photonic crystals. High-quality silicon photonic resonators with embedded YIG elements  \cite{Bi2011} raise the hope for monolithic integration of optical isolators into photonic integrated circuits \cite{Dai2012}.

Optical isolators are a useful reference point for the design of future integrated magneto-optical cavities, but while they enhance the static Faraday rotation this is not necessarily the case for the interaction of light with the {\it dynamical} magnetization. Even with annealing-induced-recrystallization, the quality of sputter-deposited YIG has a Gilbert damping ten times larger than that of liquid phase epitaxial-grown YIG \cite{Hauser2016}, which is not tolerable for cavity optomagnonic purposes. 

Theoretical studies can guide cavity design for the optimization of the magneto-optical coupling \cite{AlmpanisBook}. For example, compared to a sphere, in a disk with thickness $\sim\lambda$ the total mode volume of a WGM is reduced \cite{Graf2018}. Defects in photonic crystal structures made from or containing magnetic material can act as a small mode volume resonators. This has been considered for 1D planar structures consisting of dielectric mirrors with an embedded magnetic layer \cite{Pantazopoulos2019}, or multilayers of magnetic and non-magnetic dielectrics that confine both the magnon mode and the optical mode to the same volume \cite{Pantazopoulos2018}. 1D photonic crystals formed by equidistant holes in a dielectric beam \cite{Graf2021} may also localize the optical and magnetic modes to the same volume.

The magneto-optical interaction can also be enhanced for a given sample by selecting magnon modes that maximize the triple-mode overlap. In magnetic spheres, also  magnons  form WGMs at the equator \cite{Sharma2018}. The Damon-Eshbach surface modes are chiral, which increases the asymmetry between Stokes and anti-Stokes scattering. However, they are so localized that the overlap with the optical WGMs is still small. Almost perfect overlap can be achieved by choosing modes with slightly smaller orbital angular momentum, however~\cite{Sharma2019}.  The wave vector of the (close to) surface magnons is large, such that angular momentum can be conserved the photons are back- rather than forward scattered. These surface modes can be efficiently actuated only by microwaves with matched wave length, which are not easily generated. Alternatively, two-beam stimulated Raman scattering process can selectively populate magnetic surfaces modes with large momenta \cite{Simic2020}.  \onlinecite{Graf2018} suggest to couple to a gyrotropic mode \cite{Thiele1973}  of a magnetic vortex \cite{Shinjo2000} in a thin magnetic disk  to enhance the magneto-optical coupling. The gyroptropic mode frequencies are typically much smaller that those of magnetostatic modes, which may enhance, e.g.,  magnetoelastic phenomena. Indeed, magnons and photons may couple via an intermediate mechanical mode \cite{Losby2015}, exploiting the high sensitivity and large Q-factors of optomechanical systems.  This kind of coupling has already been exploited in measurements of single spins in nitrogen vacancy centers of diamond \cite{Arcizet2011}, and as a possible route to highly efficient MW optical transduction \cite{Rudd2019}.

\subsection{Whispering gallery modes} \label{sec:VC}

The magneto-optical coupling in WGM resonators has been the subject of several studies.  The selection rules for Brillouin light scattering are understood in terms of a number of conservation rules and the overlap integral in Eq.~(\ref{Vint}), and the coupling rates to different modes can be calculated \cite{Sharma2017}. Here we discuss the observations of BLS by WGMs in YIG spheres that are in general well understood.

\subsubsection{Optical WGMs}

A dielectric sphere with high refractive index supports WGMs, i.e. light modes that cannot escape due to total internal reflection at the dielectric/air boundary \cite{Ilchenko2006}. These modes can have $Q$-factors as high as 10$^8$ \cite{Armani2003} and strongly enhance interaction effects in non-linear optics \cite{Braginsky1989}, optomechanics \cite{Schliesser2008}, and biosensing \cite{Vollmer2008}.

\begin{figure}%
\includegraphics[width=\columnwidth]{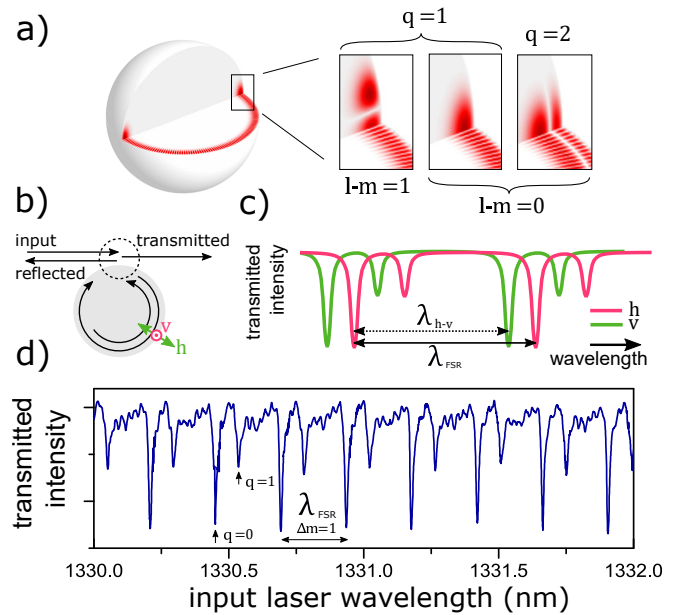}%
\caption{(Color online) (a) The polar, azimuthal, and radial indices $\{l,m,q\}$ of the whispering gallery modes in (YIG) spheres count the nodes in the mode amplitudes. (b) Outline of the measurement in the WGM plane. The optical modes are populated by an input laser, while detecting the polarization and frequency of transmitted or reflected light. The (quasi-) TE and TM modes have linear E-field polarization perpendicular (h) and parallel (v) to the WGM plane, respectively. (c) Schematic  mode structure in the transmission spectrum. The free spectral range $\lambda_\textrm{FSR}$ and TE-TM splitting $\lambda_\textrm{TE-TM}$ are indicated. (d) The dips in the measured transmitted intensity with the same polarization (TE) and frequency as the input identify the resonant WGMs. Panel (c) is adapted from \cite{Haigh2016}, (d) is adapted from \cite{Haigh2015a}, (a) and (b) (J. A. Haigh) were not previously published.}%
\label{WGMs}%
\end{figure}

The modes of a dielectric resonator are solutions of the Helmholtz equation [Eq. (\ref{eq:Helmholtz})],  see Sec.\,\ref{secII}. For a rotationally symmetric spheroid, the  indices $\{l,m,q\}$ count the nodes in the mode amplitude in the polar, azimuthal, and radial directions, respectively, as illustrated by Fig.\,\ref{WGMs}a. Modes with linear polarization parallel and perpendicular to the WGM plane (see Fig.\,\ref{WGMs}b) have the same nodal structure, but their frequencies differ by birefringence,  i.e.  the different boundary conditions for the electric field. Since the interface is curved we label them ``quasi"-TM (h) and TE (v) in Fig.\,\ref{WGMs}b,c.

Experimentally, the WGMs can be probed by evanescent coupling to an external optical mode with matched energy and wave vector \cite{Gorodetsky1999}. A tapered glass fiber can be used \cite{Osada2016}, but the wave-vector matching is poor. Impedance mismatch can be minimized by a proximity material with refractive index close to that of YIG  $(n_{\textrm{YIG}}\approx2.2)$ such as high refractive index silicon nitride waveguides \cite{Zhang2016}. Precise wave-matching can be achieved only with prism couplers \cite{Haigh2015} that are available with high-refractive index materials such as silicon and rutile and show excellent selectivity of the optical WGMs of attached YIG spheres.

In the configuration of  Fig.\,\ref{WGMs}b the injection of photons into the sample reduces the transmission of the input light as a function of wavelength as sketched in Fig.\,\ref{WGMs}c. The observations in Fig.\,\ref{WGMs}d show a periodic feature of two dips that can be assigned to the modes with $q=1$ and $q=2$ \cite{Haigh2016}.  The difference in wavelength between neighboring main resonances is the so-called free spectral range, and depends on the sphere radius. The $Q$-factors corresponding of the line widths of these optical modes can reach $10^7$ when the surface is polished and cleaned \cite{Zhang2016}. At frequencies optimized for BLS, the remaining losses are consistent with the corresponding YIG absorption coefficient of $\sim$0.1\,cm$^{-1}$ \cite{Wood1967}.

The TE-TM splitting by geometrical birefringence is  $\lambda_{\textrm{TE-TM}}=\lambda_{\textrm{FSR}} \sqrt{1-1/n_{\textrm{YIG}}^2 } \sim 0.9\lambda_{\textrm{FSR}}$, where $\lambda_\textrm{FSR}$ is the free spectral range. This implies that the closest spacing \( \lambda_{\textrm{TE-TM}}-\lambda_{\textrm{FSR}} \) between the dominant TE and TM modes belong to the same radial, but different azimuthal mode indices, as shown schematically in Fig.\,\ref{WGMs}c. In a 1\,mm sphere they are split by 7\,GHz, which is conveniently close to typical magnon frequencies. The selection rules, based on the symmetry of the Faraday effect, dictate that only modes of opposite polarization can cause Brillouin light scattering. The magnon mode frequencies can be tuned by external magnetic fields to match a triple resonance condition that maximizes the BLS scattering probability.

\subsubsection{Uniform magnon mode}

We first consider the Brillouin light scattering of WGMs by the Kittel (uniform) mode. The WGMs are not simply plane waves (see Fig.\,\ref{WGMs}a). The Kittel mode has zero orbital momentum and spin angular momentum \(S=1\). This momentum must be transferred between the optical modes with a splitting that matches the FMR frequency. As discussed above, this can be achieved easily in 1\,mm-scale YIG spheres, tuning the Kittel mode by a magnetic field to match the closest-spaced TM and TE WGMs, whose difference in azimuthal mode index accounts for the change in spin angular momentum. This creates an asymmetry in the Stokes/anti-Stokes scattering, since  the ordering of the input and output mode fixes the change in azimuthal index that can match the spin angular momentum transfer for either magnon creation or annihilation, but not both. These selection rules are implicit in the matrix elements Eqs. (\ref{G_pqalpha}) and  (\ref{G+pqalpha}). The prefactors in Stokes and anti-Stokes matrix elements are also not the same since in general the specific WGMs and also the Cotton-Mouton effect as mentioned in Sec. \ref{SecIII} should be taken into account.

\begin{figure}%
\includegraphics[width=8.3cm]{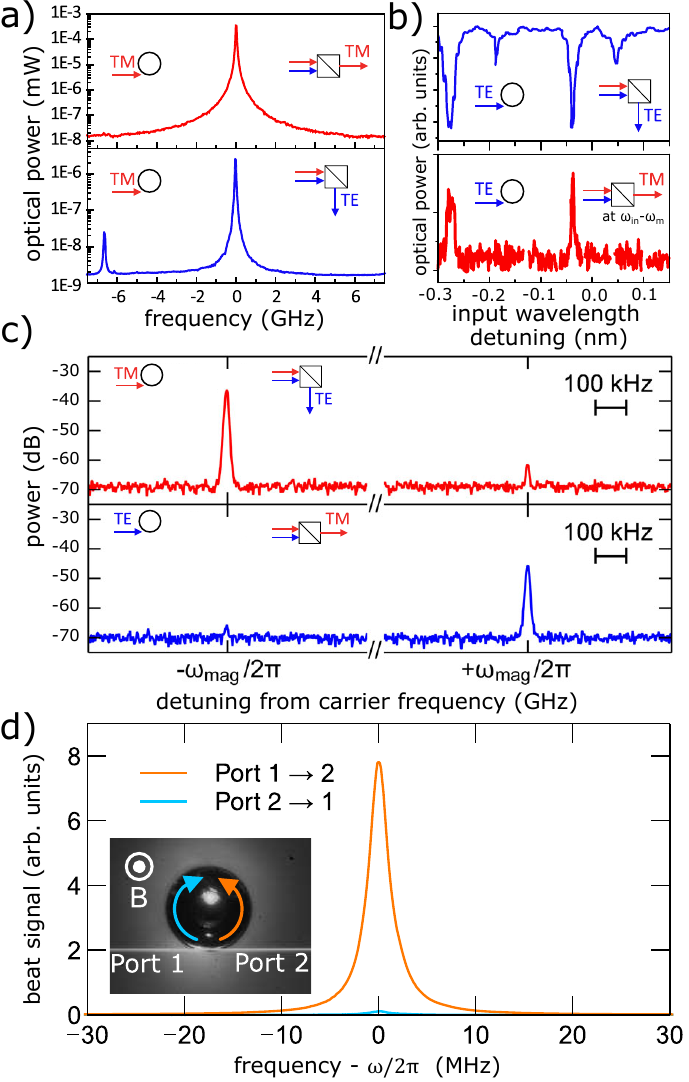}%
\caption{(Color online) Magnon BLS in WGMs for TM/TE input laser polarizations (indicated in inset by the red/blue arrow into the WGM), and forward scattered light polarization (indicated in the inset by red/blue arrows leaving the polarizing beam splitter). (a) Polarization and scattering selectivity, reproduced from \onlinecite{Zhang2016}. The magnon scattering process only occurs for cross-polarized input/output fields. For TM input polarization, only the Stokes process is observed. (b) Magnons scatter light only when resonant with the WGMs (from \onlinecite{Haigh2016}). The upper panel shows a WGM resonance in the elastic transmission spectrum while the lower panel is the scattered light intensity, close to  the anti-Stokes frequency $\omega_\text{in} -\omega_{\rm m}$.  (c) The Stokes/anti-Stokes process is highly selective and controlled by the input polarization (from \onlinecite{Osada2016}). (d) The scattering is non-reciprocal: The anti-Stokes peak indicating magnon annihilation for fixed magnetization direction (up) and TM input mode is only observed for one direction of the WGM circulation (from \onlinecite{Osada2016}).}%
\label{BLS}%
\end{figure}

The asymmetry can also be interpreted in terms of a  circular component of the evanescent optical polarization in the curved geometry of the TM WGMs \cite{Zhang2016, Osada2016}. The associated effective optical {\it spin-orbit} coupling \cite{Onoda2004,Bliokh2015} changes sign with the photon field rotation. The difference in the angular moments between the TM and TE modes breaks their degeneracy. The integral over the mode volume leads to relative shifts between the mode families that comes down to the geometric birefringence inferred above.

Resonant magnon Brillouin light scattering in magnetic spheres was first observed by \onlinecite{Zhang2016} and \onlinecite{Osada2016}, followed by the demonstration of the triple resonance condition by \onlinecite{Haigh2016}. In these experiments, a WGM of certain polarization was pumped by an input laser, while recording the output power spectrum by a optical heterodyne measurement with a fast photodiode and a MW spectrum analyzer or a Fabry-P\'erot etalon filtering spectrometer.  The BLS intensity scales with the number of magnons created by MW drives.

Figure \,\ref{BLS}a demonstrates that the polarization governs the scattering. For TM input polarization, scattered light is only observed in the output channel with TE polarization. Next, the input laser must be tuned to the WGM frequencies (see Fig.\,\ref{BLS}b). Thirdly, the direction of the energy flow, from absorption (anti-Stokes) to emission (Stokes) of magnons, is controlled by the input polarization (for constant magnetization direction), as seen in Fig.\,\ref{BLS}c, which shows for TE input (upper panel) only a red-shifted Stokes line, while for TM input  (lower panel) a blue-shifted anti-Stokes line.

As discussed above, the asymmetry of the WGM modes with respect to the light polarization explains the strong sideband selectivity. The azimuthal mode index \(m\) of the WGM must change by one in the scattering process. When the magnon is tuned to a $\Delta m =-1$ transition, the $\Delta m =1$ transition is off-resonant due to the geometrical birefringence and vice versa.

Figure \,\ref{BLS}d shows Brillouin light scattering spectra for input pump from two ports that couple to WGMs with opposite circulation. The large non-reciprocity implies suppression of competing side bands in the transduction of MW to optical photons. 

The key observations from these experiments are that the scattering is (i)~single sideband, (ii)~non-reciprocal, (iii)~depends on the input polarization, and (iv)~can be controlled by MWs. All these observations agree with the theoretical description.

\begin{figure}%
\includegraphics[width=\columnwidth]{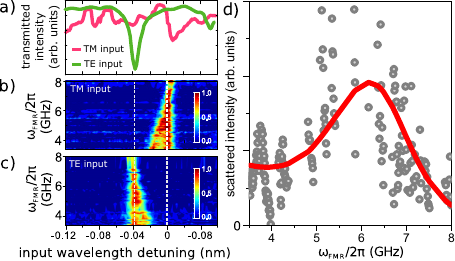}%
\caption{(Color online) Triple resonance condition for magnon scattering. (a) (Elastic) transmission spectra of TE and TM polarized light relative to the TM mode frequency identify the neighboring WGMs. (b,c). Color map of the amplitude of the magnon BLS signal, normalized to the peak value for each magnon frequency, for TM (b) and TE (c) input polarization. (d)~Peak scattering amplitude as a function of magnon frequency. The maximum agrees with the triple resonance at which the magnon frequency matches the TE-TM splitting. Reproduced from \onlinecite{Haigh2016}. }%
\label{TRC}%
\end{figure}

The tunability of the magnon mode with applied magnetic field allows a precise mapping of the triple-resonance condition, as shown in Fig.\,\ref{TRC}. When detuned, the BLS broadens into two peaks as a function of input laser wavelength, as seen in the color plots Fig.\,\ref{TRC}b,c for $\omega_{\rm m}/2\pi \approx4$\,GHz. These correspond to the input and output optical frequencies close to resonance with the TM and TE modes. These correspond to the peak in $G_N$ associated with resonantly driving the input mode, and the minimum in the denominator of Eq.~(\ref{output_optical mode}), respectively. With increasing magnon frequency, the condition for resonance with the output mode shifts by the same amount until the two peaks coalesce at the triple resonance point. While this is not so clear from the color plots because each horizontal spectrum is separately normalized to a maximum amplitude of unity to emphasize the off-resonant behavior, but it is emphasized in Fig.\,\ref{TRC}d, which shows the expected maximum  at the triple resonance. The red curve is a plot of Eq.~(\ref{output_optical mode}) with an optical damping rate $\sim1$\,GHz.

The non-reciprocal nature, tunability, and cavity enhancement of the magnon-photon coupling at optical frequencies can be used to distill the single photon coupling rate in Eq.~(\ref{eq:H}).   The measured value $G/2\pi \approx $~5.4\,Hz \cite{Osada2016} agrees with that calculated from the model parameters. The coupling $G_{N}$ for manageable MR drive powers is still many orders of magnitude smaller than the combined damping rates, so the present experiments are still far from the strong coupling regime. 

\subsubsection{Higher-order magnon modes \label{SecVHigherorder}}

Following the experimental discovery of BLS by the uniform Kittel mode, \onlinecite{Sharma2017} theoretically considered the general problem of BLS of WGMs in a magnetic sphere by ``Walker" magnon modes close to the Kittel mode and Damon-Eshbach surface modes. \onlinecite{Sharma2019} found an almost perfect overlap with the optical WGMs not for the magnetostatic Damon-Eshbach but for dipolar-exchange surface magnon modes close to the equator, with a single-photon coupling rate enhanced by two orders of magnitudes in a backscattering configuration \cite{Sharma2019}. 

The selection rules for magnon modes other than the Kittel mode, including the Cotton-Mouton effect, are found by working out Eqs.~(\ref{G_pqalpha}),  (\ref{G+pqalpha}) in a cylindrical basis \cite{Sharma2017}. The resulting Clebsch–Gordan coefficients do not vanish when (i) the polar and radial mode indices  $\{l,q\}$ of a WGM do not change in the scattering process, (ii) the magnon amplitude does not have a node at the equator and (iii) the dynamic magnetization rotates by $2\pi$ with respect to the photon propagation direction around the sphere. This leads to optical coupling of a $\{l,m,q\}$ magnon only when $l$ is odd and $m=\pm1$, depending on the WGM circulation and magnetic field direction, see also \cite{Osada2018a}.

\onlinecite{Osada2018} and \onlinecite{Haigh2018} observes many low-frequency magneto-static Walker modes in the BLS spectra. The results confirm the selection rules and demonstrate that the coupling rate increase for higher order modes by an order of magnitude which raises the hope for significantly larger coupling of the surface modes \cite{Sharma2019}. The magnon modes in BLS can be indexed with high confidence by  comparison with MW absorption spectra and mode-selective MW excitation  \cite{Haigh2018}. \onlinecite{Gloppe2019} carried out a detailed tomography of the low-frequency modes in  magnetic spheres. 

\subsection{Experiments in other cavities}\label{other_cavities}

\onlinecite{Zhu2020} recently demonstrated a YIG waveguide-based Fabry-P\'erot cavity with a $\sim50$-fold enhancement in the optomagnonic coupling rate over the WGM devices. A rib waveguide with polished end-facets with reflective coatings increases the quality factor of the optical modes to $Q\approx2\times10^5$. 

\onlinecite{Haigh2020} report a sub-picoliter optical mode volume for a YIG film in a  laterally confined Fabry-Perot cavity formed by two dielectric mirrors. Cavities of this type can have mode volumes as small as 1\,fL \cite{Dolan2010}, which would yield coupling rates in the MHz range. Low-impedance MW resonators \cite{MckenzieSell2019} will be required to couple MWs efficiently into such small volumes.

\subsection{Applications}

An important milestone for cavity magnonics would be an efficient conversion between MW and optical photons. 
However, interesting and potentially applicable effects can be expected even for smaller coupling rates. For example, \onlinecite{Bittencourt2019} show that protocols for heralding magnon Fock states can work for cooperativities as small as $\sim 10^{-2}$. In the following we address other examples.

\subsubsection{Photon transducer}

As discussed in Sec. \ref{sec:OM}, the interaction between optical photons and magnons benefits from small magnetic volumes, see Eq. \eqref{eq:smallVm}. \onlinecite{Kusminskiy2016} estimate that for a YIG optical cavity with a mode volume of the order the optical wavelength cube $\lambda^3 \approx 1$\,$\mu$m$^3$, the single photon coupling rate would be 0.1\,MHz. For an optical dissipation rate $\kappa_{\rm i}/2\pi =1$ GHz and an input power of 100\,mW  this leads to $G_{N}/2\pi \approx2$\,GHz which is larger than the damping and would allow efficient transduction between MW and optical photon via magnons. On the other hand, the resonant coupling between a MW cavity photon and a magnon, Eq. (\ref{resonant_coupling_arb_mode}), is proportional to $\sqrt{V_{\rm m}}$.  The efficiency of optomagnonic transducers between MW and IR photons is therefore largest at an intermediate magnetic volume $V_{\rm m}$.

The transduction efficiency in terms of the cooperativity for an optomagnonic transducer at the triple resonance point is given by \cite{Zhu2020} 
\begin{equation}
\xi = \frac{4  C_{\rm{OPm}} C_{\rm{MWm}}}{\left(1+C_{\rm{OPm}} +C_{\rm{MWm}} \right)^{2}} \frac{ \kappa_{\text{MW}\text{,ex}}}{\kappa_\text{MW}} \frac{ \kappa_{\text{o}\text{,ex}}}{\kappa_\text{o}}\,, 
\end{equation}
where $C_{\rm{OPm}}$ ($C_{\rm{MWm}}$) is the cooperativity for the magnon mode coupling to optical (MW) photons and $\kappa_{\text{o}\text{,ex}}/\kappa_\text{o}$ ($\kappa_{\text{MW}\text{,ex}}/\kappa_\text{MW}$) the optical (MW) ratio of external coupling rates to total losses (see for comparison Eq. \eqref{tripleres}, where $\max |\delta\hat{a}_\text{o,out}|^2 /|\hat{m}_\text{in}|^2$ gives the magnon to optical photon conversion efficiency). Note that an efficiency approaching one can be achieved when both cooperativities are equal and large, and the losses are dominated by the external coupling rates.

In the YIG waveguide-based cavity realized by \onlinecite{Zhu2020}, a stripline underneath the YIG film acts as a MW source. The YIG thickness modulation confines the magnon modes to the rib, increasing the overlap between magnons, MW and optical photon modes. The result is a conversion efficiency \(\xi\) between MW and light that is strongly enhanced compared to that in YIG spheres. \onlinecite{Zhu2020} report $\xi = 5\times10^{-7}$  at the triple resonance condition with room for further improvements.

\subsubsection{Inverse Faraday effect vs. stimulated Raman scattering}

The Faraday effect is rotation of the linear polarization plane of light passing through a material with magnetization parallel to the wave vector. Since action implies reaction, the magnetization is affected by this process as well. A light beam generates an effective magnetic field that interacts with the magnetization in the {\it inverse} Faraday effect that was predicted by \onlinecite{Pitayevsky1960} and discussed in textbooks, \onlinecite{Landau1984}.  Ultrafast, high intensity laser pulses can induce magnetization dynamics by this effect \cite{Kimel2005} and even switch the magnetization \cite{Hadri2017}.  Longer pulses at these intensities may destroy the samples, however. The use of optical cavities to enhance the optomagnonic interaction might allow controlled driving of the magnetization dynamics \cite{Simic2020} under continuous wave conditions at much lower input powers~\cite{Zhu2021}. 

\onlinecite{Zhu2021} experimentally observed a stimulated Raman scattering process as proposed by \onlinecite{Simic2020} in a rib like cavity \cite{Zhu2020}. Two slightly detuned input lasers with TM and TE polarization excite magnons resonating with the frequency difference that are detected by their microwave stray fields. In contrast to the surface magnons with large wave numbers addressed by \onlinecite{Simic2020}, the lasers are co-propagating and the magnon wave numbers are small. The effect can be understood in terms of the Hamiltonian introduced in Sec.~\ref{sec_conv_eff}, in which both optical modes ($a_\text{i}$ and $a_\text{o}$) are coherently driven at frequencies $w_\text{D,i}$ and $w_{D,\text{o}}$, respectively, $\hat{a}_\text{i}\approx\langle a_\text{i} \rangle e^{i w_\text{D,i} t}$ and $\hat{a}_\text{o}\approx\langle a_\text{o} \rangle e^{i w_\text{D,o} t}$. The coupling term in Eq. (\ref{eq:H}) becomes
\begin{eqnarray}
\hbar \langle a_\text{i} \rangle \langle a_\text{o} \rangle  G \left( e^{i (w_\text{D,i} - w_\text{D,o})t} \hat{m}^{}  +  e^{-i (w_\text{D,i} - w_{D,\text{o}})t} \hat{m}^{\dagger}\right).\qquad \label{inverse_faraday_drive}
\end{eqnarray}%
This coherent wave field drives a magnon mode, see Eq.~(\ref{eq:H_driving}), with amplitude proportional to $\sqrt{n_\text{i} n_\text{o}} G$, at frequency $\omega_\text{D,i} - \omega_{D,\text{o}}$, which becomes resonant when $\omega_\text{D,i}=\omega_\text{i}$, $\omega_\text{D,o}=\omega_\text{o}$, and $\omega_\text{i} - \omega_{\text{o}}=\omega_m$.  

Strictly speaking, the stimulated Raman scattering is not the same as the inverse Faraday effect. \onlinecite{Zhu2021}  observe the Stokes scattering process of optical magnon creation, but the process could be used as well to annihilated them (see next section). A similar effect has been used to show the bidirectional nature of magnon-photon scattering for MW-optical conversion \cite{Hisatomi2019} in experiments, be it without an optical cavity. 

\subsubsection{Magnon cooling}

The non-reciprocal magnon-photon coupling allows manipulation of magnon modes with light \cite{Sharma2018}.  Anti-Stokes (Stokes) scattering removes (adds) a magnon from a selected magnon mode, which can be interpreted as selective cooling (heating), respectively. Analogous phonon-photon scattering processes have been used with much success in optomechanics \cite{Aspelmeyer2014}. 

The optically-induced magnon annihilation rate under the triple resonance condition can be estimated from Eq.\,(\ref{output_optical mode}) as an effective optically-induced damping,
\begin{equation}
\Gamma_\text{opt} = \frac{4G_{N}^2}{\kappa_\text{o}}.
\end{equation}
Unlike the intrinsic magnetic damping that forces equilibrium with a thermal phonon bath, the optomagnonic damping strives to bring the magnon mode into equilibrium with the optical mode, which is at high frequencies (\(\sim 2000\)~K) and therefore not thermally excited. We can estimate a steady-state temperature under resonant illumination  by comparing the rates of absorption $\kappa_{\rm m} \overline{n}_{\rm m} (n_\text{th}+1)$ and injection $\kappa_{\rm m} (\overline{n}_{\rm m}+1) n_\text{th}$ of magnons by the thermal bath, where $\overline{n}_{\rm m}$ is the population of the magnon mode, and $n_\text{th}=1/(\exp{(\hbar \omega_\text{m}/(k_BT))}-1)$ is the number of magnons at thermal equilibrium. Considering the additional absorption in the presence of the optical fields $\Gamma_\text{opt}\overline{n}_{\rm m}$, and equating the rates of absorption and emission, we obtain the equilibrium number of magnons,
\begin{equation}
\overline{n}_{\rm m} = n_\text{th}\frac{\kappa_{\rm m}}{\kappa_{\rm m} + \Gamma_\text{opt} } = \frac{n_\text{th}}{ 1 + C_{\rm OPm} }. \label{estimate}
\end{equation}
Therefore,  significant cooling $\overline{n}_{\rm m} \ll n_\text{th}$ requires a large optomagnonic cooperativity $C_{\rm OPm}$. The above estimate holds for coupling rates smaller than the optical damping rate $G_N <\kappa_{\rm o}$, which implies that scattered photons are efficiently dissipated. If this is not the case, the number of magnons may become very small,  which requires a quantum mechanical treatment \cite{Bittencourt2019}, e.g. by the Langevin equations (Sec.\,\ref{eom}) that describe the thermal fluctuations in terms of the inputs $m_\text{in}$ and $\delta a_{\rm o,\text{in}}$. The result,
\begin{equation}
\overline{n}_{\rm m} = n_\text{th}\frac{1}{1+C_{\rm OPm}}\left( 1 + \frac{\kappa_\text{o}}{\kappa_\text{m}(1+C_{\rm OPm})} \right),
\end{equation}
reduces to Eq. (\ref{estimate}) when  $G_N <\kappa_{\rm o}$. Similar expressions have been used in optomechanics \cite{Galland2014}. 

Preparing a mode with only a few number of magnons is a prerequisite for quantum manipulation \cite{Bittencourt2019}, see also Sec. \ref{SecVI}. The dynamical cooling discussed here should be combined with conventional refrigeration of the lattice and by ``freezing" the magnons out by applying a large magnetic field. The thermal energy at 100\,mK corresponds to $\sim$2\,GHz. It is now routinely possible to make optical measurements at these temperatures \cite{Higginbotham2018,Mirhosseini2020}.

\subsubsection{Nonlinear effects}

We often treat the magnetization dynamics by assuming small-amplitude oscillations or a small number of magnons. This is equivalent to the lowest order terms in the Holstein-Primakoff expansion discussed in Sec. \ref{secIIm}. The magnon system is then equivalent to an ensemble of classical  harmonic oscillators. However, the spin system is inherently nonlinear. When the modulus of a spin variable is constant, the dynamics is restricted to stay on the Bloch sphere. The nonlinear regime is easily reached by MW drives, see  \ref{Sec:IVc}, but in principle also in magneto-optical devices under a strong optical drive \cite{Kusminskiy2016}. The dynamics of a macrospin $\mathbf{S}$ is governed by an optically-induced effective magnetic field. The optically modified damping can become negative, leading to period doubling and ultimately chaotic dynamics, which is much more difficult to envision in optomechanical systems. 

\section{Quantum magnonics}\label{SecVI}

The dynamics of the magnetic order behaves like a collection of non-interacting harmonic oscillators provided that the magnon occupation numbers are much smaller than the total number of spins~(Sec. \ref{secIIm}). At low temperatures and weak excitation the system response is linear. However, non-linearities are essential for phenomena such as the creation and observation of non-classical states~\cite{Haroche2006}. In cavity magnonics non-linearities arise when a magnetostatic mode couples to the electromagnetic field, giving rise to radiation pressure~\cite{Kusminskiy2016}, or to a phonon mode \cite{Zhang2016}, but they are weak.  To date, only magnons coupled to an intrinsically non-linear quantum system such as a SQUID enable genuine quantum magnonics~\cite{Tabuchi2015,Tabuchi2016}. In this section, we address first the theory of a specific realization of quantum magnonics based on superconducting qubits as the nonlinear element and subsequently discuss experimental results.

\subsection{Theory}

\subsubsection{Origin of the coupling}

The coherent interaction between magnetostatic modes in a magnetically-ordered system and superconducting circuits requires two key ingredients. The first ingredient is the coherent coupling between magnetostatic modes and MW cavity modes through the magnetic-dipole interaction discussed in Sec.~\ref{SecIII}. The second ingredient is the electric-dipole coupling of superconducting qubits to the cavity modes through the electric-dipole interaction employed in conventional circuit QED. These interactions enable control of an effective cavity-mediated interaction between these two very different macroscopic modes~\cite{Tabuchi2015,Tabuchi2016}.

\subsubsection{Superconducting qubits}

The to date arguably most advances qubits are based on superconducting circuits using the Josephson effect~\cite{Devoret2013}, whose dissipationless nonlinearity provides long-lived and tunable effective two-level systems~\cite{Makhlin2001}. The ``transmon'' regime~\cite{Koch2007} of the Cooper-pair box~\cite{Shnirman1997,Nakamura1999} is particularly relevant due to its simplicity and insensitivity to charge noise. The transmon qubit is well described by the Duffing oscillator Hamiltonian
\begin{equation}
\hat{H}_\mathrm{q}=\hbar\left(\omega_\mathrm{q}-\frac{K_\mathrm{q}}{2}\right)\hat q^\dagger\hat q+\hbar\frac{K_\mathrm{q}}{2}\left(\hat q^\dagger\hat q\right)^2,
\label{eq:Duffing}
\end{equation}
where $\omega_\mathrm{q}$ ($\omega_\mathrm{q}+K_\mathrm{q}$) is the angular frequency of the transition between the ground state $|g\rangle$ (first excited state $|e\rangle$) and the first excited state $|e\rangle$ (second excited state $|f\rangle$). In Eq.~\eqref{eq:Duffing}, the ladder operator $\hat q$ ($\hat q^\dagger$) annihilates (creates) an excitation in the circuit. The anharmonicity $K_\mathrm{q}$ is negative in the transmon regime and parameterizes the difference between the angular frequencies of the first and second transitions. In the transmon regime of the Cooper-pair box the anharmonicity $|K_\mathrm{q}/(2\pi)|\approx 0.1-1 $~GHz is much larger than the intrinsic line width \(\kappa/(2\pi) \approx \)$1$~MHz. When the bandwidth of the control pulses is smaller than the anharmonicity, the transmon becomes a pseudo-spin system with  Hamiltonian~\cite{Koch2007,Schreier2008}
\begin{equation}
\hat{H}_\mathrm{q}=\frac{1}{2}\hbar\omega_\mathrm{q}\hat\sigma_z,
\end{equation}
where $\hat\sigma_z=|e\rangle\langle e|-|g\rangle\langle g|$.

Superconducting qubits can have a large electric dipole moment resulting in coupling strengths to ac electric fields with the frequency of a few hundreds of MHz in coplanar waveguide resonators~\cite{Wallraff2004} and 3D MW cavities~\cite{Paik2011}. As long as the coupling between  qubit and cavity mode is not ultrastrong, the Jaynes-Cummings coupling is valid, see Eq. \eqref{HopfieldHam}, 
\begin{equation}
\hat{H}_\textrm{q-c}=\hbar g_\textrm{q-c}\left(\hat a\hat q^\dagger + \hat a^\dagger\hat q \right),
\label{eq:qubit_cavity_coupling}
\end{equation}
where $g_\textrm{q-c}$ is the electric-dipole coupling strength~\cite{Blais2004}. 

\subsubsection{Resonant interaction}

The magnon-photon coupling [Eq.~(\ref{Cavity_magnon_coupling_RWA})] and that between the qubit and the same cavity mode~[Eq.~\eqref{eq:qubit_cavity_coupling}] lead to a cavity-mediated magnon-qubit interaction. Detuning the cavity mode from both subsystems leads to the ``beam-splitter'' interaction,
\begin{equation}
\hat{H}_\textrm{q-m}^\mathrm{res.}=\hbar g_\textrm{q-m}\left(\hat q\hat m^\dagger + \hat q^\dagger\hat m \right) \ ,
\label{eq:qubit_magnon_coupling}
\end{equation}
where $g_\textrm{q-m}$ is the qubit-magnon coupling strength, i.e. the effective interaction between the magnetostatic mode and the qubit~\cite{Tabuchi2015,Tabuchi2016}. This description is valid when $\left|\omega_i-\omega_\mathrm{c}\right|\gg\left|g_{i\textrm{-c}}\right|$ with $i=\mathrm{q,m}$  and $\left|\omega_\mathrm{q}-\omega_\mathrm{m}\right|\ll\left|g_{i\textrm{-c}}\right|$~\cite{Tabuchi2016}. Physically, the modes exchange energy at a rate $2g_\textrm{q-m}$ through virtual photons in the cavity mode.
In this regime, to second order in perturbation theory one obtains
\begin{equation}
g_\textrm{q-m}\approx\frac{g_\textrm{q-c}g_\textrm{m-c}}{\omega_\mathrm{q,m}-\omega_\mathrm{c}},
\label{eq:qubit_magnon_coupling_strength}
\end{equation}
where $\omega_\mathrm{q}=\omega_\mathrm{m}\equiv\omega_\mathrm{q,m}$ is the angular frequency of the qubit and the magnetostatic mode~\cite{Tabuchi2015,Tabuchi2016}. In the presence of multiple cavity modes, Eq.~\eqref{eq:qubit_magnon_coupling_strength} should be summed over all relevant modes. Since the electric- and magnetic-dipole interactions are coherent, the contributions from the different cavity modes can interfere constructively or destructively. Careful MW engineering can therefore maximize $g_\textrm{q-m}$ in a multimode MW cavity.

Strong coupling requires $\left|g_\textrm{q-m}\right|\gg\kappa_\mathrm{q},\kappa_\mathrm{m}$, where $\kappa_\mathrm{q,m}$ are the qubit and magnon relaxation rates or line widths. $\kappa_\mathrm{q}$ is related to the qubit coherence time $T_2^*$ as $\kappa_\mathrm{q}=2/T_2^*$. The strong coupling regime enables the exchange of quanta between both modes at an angular frequency $2g_\textrm{q-m}$, which may generate nonclassical, e.g. Fock, states with negative values of Wigner function, in harmonic oscillator systems~\cite{Haroche2006,Hofheinz2008,Hofheinz2009}.  The qubit-magnon coupling strength can be maximized by balancing the system such that \(g_\textrm{q-c} \approx g_\textrm{m-c}\).

\subsubsection{Dispersive qubit-magnon interaction}
The resonant interaction between the magnetostatic mode and the qubit is suppressed when $|\Delta_\textrm{q-m}|\equiv|\omega_\mathrm{q}-\omega_\mathrm{m}|\gg |g_\textrm{q-m}|$. The interaction Hamiltonian then becomes
\begin{equation}
\hat{H}_\textrm{q-m}^\mathrm{disp.}=2\hbar\chi_\textrm{q-m}\hat q^\dagger\hat q\hat m^\dagger\hat m,
\label{eq:qubit_magnon_dispersive}
\end{equation}
where $\chi_\textrm{q-m}$ is the \textit{dispersive coupling strength}~\cite{Tabuchi2015}. The Hamiltonian given by Eq.~\eqref{eq:qubit_magnon_dispersive} describes a shift of the angular frequency of one subsystem by $2\chi_\textrm{q-m}$ for every excitation in the other system. For a transmon,
\begin{equation}
\chi_\textrm{q-m}\approx\frac{K_\mathrm{q}g_\textrm{q-m}^2}{\Delta_\textrm{q-m}\left(\Delta_\textrm{q-m}+K_\mathrm{q}\right)},
\label{eq:qubit_magnon_dispersive_strength}
\end{equation}
provided that $\left|\Delta_\textrm{q-m}\right|,\left|\Delta_\textrm{q-m}+K_\mathrm{q}\right|\gg g_\textrm{q-m}$~\cite{Koch2007}. Equation~\eqref{eq:qubit_magnon_dispersive_strength} is valid both outside and inside the {\em straddling regime}, i.e.  $\omega_\mathrm{m} \in [\omega_{\rm q},\omega_{\rm q}+K_{\rm q}]$. The dispersive shift is positive and larger inside than outside the straddling regime for the same detuning. Neglecting the second excited state of the transmon by letting $K_\mathrm{q}\rightarrow\infty$ in Eq.~\eqref{eq:qubit_magnon_dispersive_strength} leads to $\chi_\textrm{q-m}\approx g_\textrm{q-m}^2/\Delta_\textrm{q-m}$.

When  $\left|2\chi_\textrm{q-m}\right|\gg\kappa_\mathrm{q},\kappa_\mathrm{m}$, we enter the strongly dispersive regime that  allows resolving single quanta of  excitation~\cite{Gambetta2006,Schuster2007,Lachance-Quirion2017,Sletten2019,Arrangoiz-Arriola2019} and preparing quantum states~\cite{Leghtas2013,Vlastakis2013} in the linear system.

\subsubsection{Other qubit-mediated interactions}

The coupling between magnetostatic modes and a superconducting qubit can lead to an even richer set of interactions. For example, driving the system at the angular frequency $\omega_\mathrm{D}=\left(\omega_\mathrm{q}+\omega_\mathrm{m}\right)/2$  leads to parametric coupling described by the Hamiltonian
\begin{equation}
\hat{H}_\textrm{q-m}^\mathrm{param.}=\hbar\tilde g_\textrm{q-m}\left(\hat q\hat m + \hat q^\dagger\hat m^\dagger \right),
\label{eq:qubit_magnon_coupling_parametric}
\end{equation}
where $\tilde g_\textrm{q-m}$ is the effective parametric qubit-magnon coupling strength that depends on the drive power~\cite{Tabuchi2015}. Hereby one achieves a dynamically-tunable coupling strength, which is a useful resource for quantum state transfer~\cite{Satzinger2018}.

Another non-linearity is the ``cross-Kerr'' dispersive interaction between magnon and cavity modes mediated by the qubit with Hamiltonian
\begin{equation}
\hat{H}_\mathrm{m-c}^\mathrm{Kerr}=2\hbar\chi_\mathrm{m-c}\hat a^\dagger\hat a\hat m^\dagger\hat m,
\label{eq:cross-Kerr}
\end{equation}
where $\chi_\mathrm{m-c}$ is the cross-Kerr coupling strength~\cite{Nigg2012}.  Equation~\eqref{eq:cross-Kerr} leads, for example, to a frequency shift of the cavity mode depending on the magnon number~\cite{Haigh2015}. In quantum magnonics, this interaction can be useful, for example, for the detection of magnons~\cite{Helmer2009} and the preparation of quantum states~\cite{Holland2015}.

Finally, the nonlinearity of the qubit leads to a ``self-Kerr" interaction between magnetostatic modes that modifies the magnon Hamiltonian as
\begin{equation}
\hat{H}_\mathrm{m}=\hbar\left(\omega_\mathrm{m}-\frac{K_\mathrm{m}}{2}\right)\hat m^\dagger\hat m+\hbar\frac{K_\mathrm{m}}{2}\left(\hat m^\dagger\hat m\right)^2,
\label{eq:self-Kerr}
\end{equation}
where $K_\mathrm{m}$ is the qubit-induced anharmonicity of the self-Kerr coefficient, see Eq.~\eqref{eq:HKerr}. The induced nonlinearity with an amplitude $\left|K_\mathrm{m}\right|/2\pi\sim10^5$~Hz is much larger than the intrinsic nonlinearity of magnons in millimeter-sized YIG samples ~\cite{Haigh2015,Zhang2019}.

\subsection{Experiments}

\subsubsection{Resonant interaction}

\begin{figure*}[t]\begin{center}
\includegraphics[scale=1]{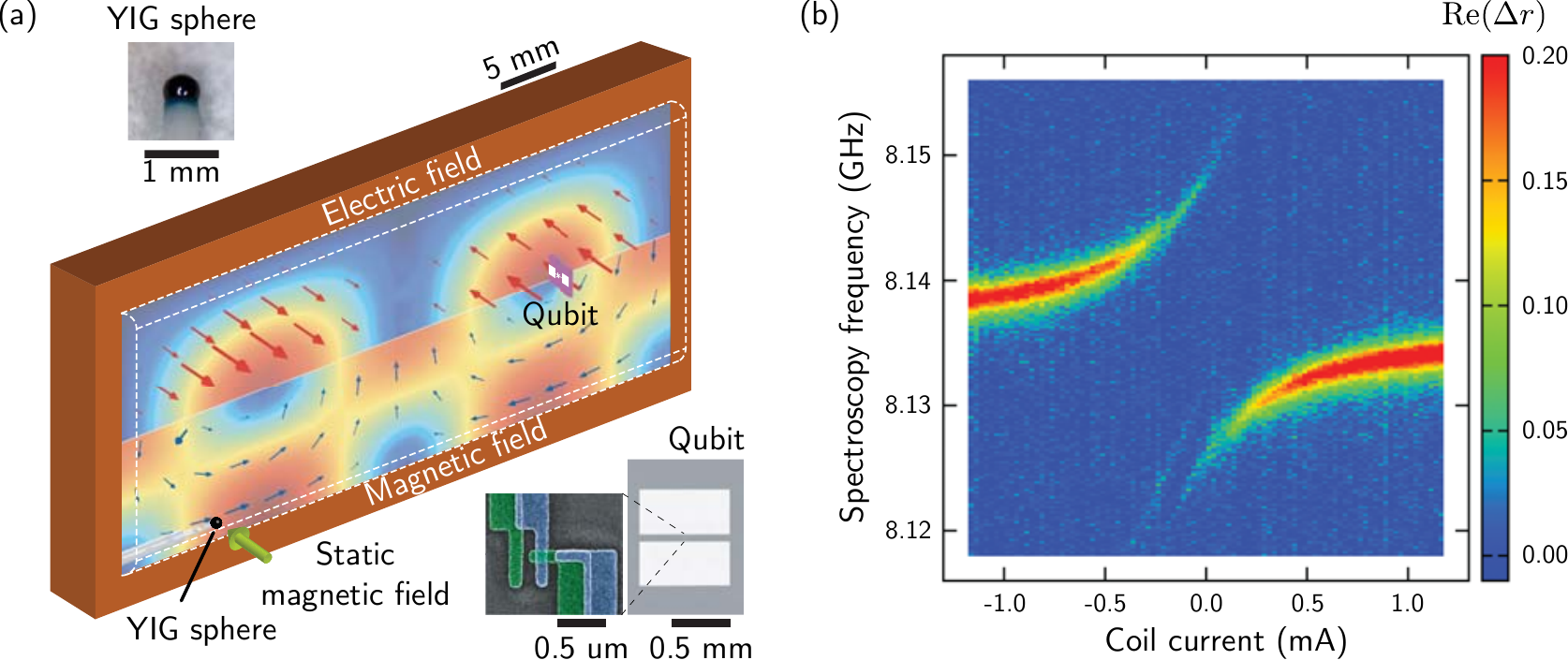}
\caption{
(Color online) (a)~Schematic of a hybrid device with strong coupling between a magnet and a qubit. A transmon-type superconducting qubit and a ferrimagnetic YIG sphere of a diameter of $0.5$~mm are  placed inside a 3D MW cavity near the antinode of the electric and magnetic fields of the TE$_{102}$ cavity mode, respectively. A static magnetic field saturates the magnetization.
(b)~Qubit spectrum $\mathrm{Re} (\Delta r)$ measured as a function of the current in a superconducting coil that tunes the static magnetic field at the magnetic sphere. The avoided crossing is the signature of a strong coherent interaction between the qubit and the Kittel mode with a coupling strength $g_\textrm{q-m}/2\pi=10.0$~MHz.
Adapted from~\onlinecite{Tabuchi2015}.
\label{fig:Quantum_magnonics_resonant}}
\end{center}\end{figure*}

\onlinecite{Tabuchi2015} discovered the resonant interaction between a magnon mode and a superconducting qubit in the strong coupling regime via the microwave modes of a 3D MW copper cavity as shown in Fig.~\ref{fig:Quantum_magnonics_resonant}. A spherical YIG crystal and a transmon-type superconducting qubit are placed inside the cavity near to antinodes of the, respectively, magnetic and electric fields of the TE$_{102}$ mode at $\omega_\mathrm{c}/2\pi=8.488$~GHz (Fig.~\ref{fig:Quantum_magnonics_resonant}a). The TE$_{103}$ cavity mode at $10.461$~GHz is used to read out the qubit state. At temperatures of about $10$~mK in a dilution refrigerator all relevant modes are close to their (vacuum) ground state.

The YIG sphere with a diameter of $0.5$~mm is magnetized to saturation by a pair of permanent magnets placed outside the cavity that generate a magnetic field of $\sim0.29$~T at the YIG sphere. The Kittel and TE$_{102}$ cavity modes are coupled through a magnetic-dipole interaction of coupling strength $g_\mathrm{m-c}/2\pi=21.0$~MHz. The transmon-type superconducting qubit has a resonant frequency of $\omega_\mathrm{q}/2\pi=8.136$~GHz with an anharmonicity $K_\mathrm{q}/2\pi=-0.158$~GHz. The $0.7$~mm-long dipole antenna of the qubit leads to an electric-dipole interaction with the TE$_{102}$ cavity mode with coupling strength $g_\mathrm{q-c}/2\pi=121$~MHz, a typical value for circuit QED in 3D cavities~\cite{Paik2011}.

The absorption spectrum of the qubit measured through two-tone spectroscopy probes the coupling between the Kittel magnon mode and the qubit. The reflection coefficient $r$ of the probe tone, close to resonance with the TE$_{103}$ cavity mode, is measured as a function of the frequency close to resonance with the qubit. The dispersive interaction between the qubit and the  TE$_{103}$ cavity mode causes changes in the reflection coefficient $r$ when the
spectroscopy tone is absorbed by the qubit~\cite{Schuster2005}. The qubit spectrum measured as a function of coil current, and thereby magnon frequency, shows an avoided crossing, the hallmark of a strong coherent interaction~(Fig.~\ref{fig:Quantum_magnonics_resonant}b). Indeed, the qubit-magnon coupling strength $g_\textrm{q-m}/2\pi=10.0$~MHz is larger than the line widths of the qubit and the Kittel mode, $\kappa_\mathrm{q}/2\pi=1.2$~MHz and $\kappa_\mathrm{m}/2\pi=1.3$~MHz, respectively. Furthermore, this value agrees well with the value of $11.8$~MHz calculated with Eq.~\eqref{eq:qubit_magnon_coupling_strength} when considering only the TE$_{102}$ cavity mode.

A few follow-up experiments corroborated this first demonstration of strong qubit-magnon coupling. First, realigning the YIG sphere reduces the coupling to higher-index magnetostatic modes~\cite{Tabuchi2016}. Secondly, \onlinecite{Lachance-Quirion2017,Lachance-Quirion2020} employed a device identical to that of \onlinecite{Tabuchi2015}, but with a qubit of resonance frequency $\omega_\mathrm{q}/2\pi=7.991$~GHz, i.e. a larger detuning with respect to the TE$_{102}$ cavity mode. A three-dimensional MW cavity made out of both copper and aluminum also indicates strong coupling \cite{Wang2020}, but it is currently unclear how much that design reduces internal losses.

\subsubsection{Dispersive interaction}

The dispersive regime of quantum magnonics was first accessed by~\onlinecite{Lachance-Quirion2017}. The dispersive interaction between the qubit and the Kittel mode was monitored by the qubit absorption spectrum in the presence of a pump tone close to resonance with the Kittel mode that injects an average number of magnons $\overline{n}_\mathrm{m}$ into the Kittel mode~\cite{Rezende1969}.

\begin{figure*}[t]\begin{center}
\includegraphics[scale=1]{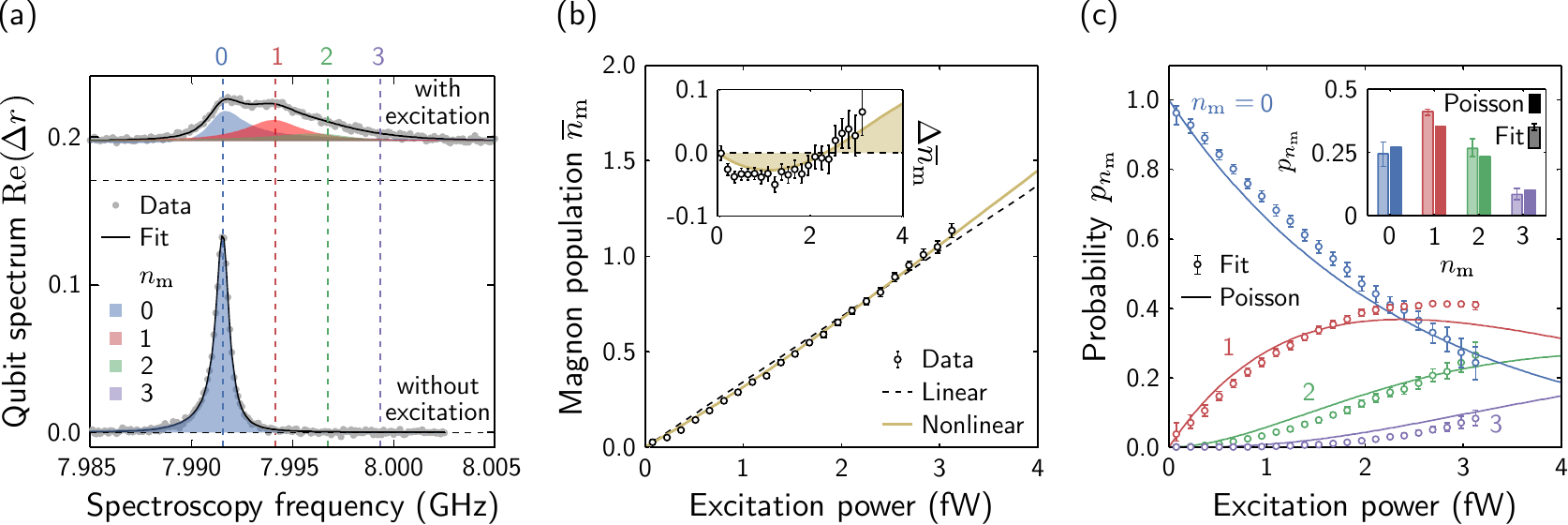}
\caption{
  (Color online) (a)~Qubit spectrum measured without (bottom) and with (top) MW excitation close to the ferromagnetic resonance of the YIG sphere. The solid black lines are fits to the data. A shift per excited magnon of $2\chi_\textrm{q-m}/2\pi=3.0$~MHz is observed, demonstrating the strong dispersive ``cross-Kerr'' interaction Eq.~\eqref{eq:cross-Kerr}. The components of the spectrum contributed by the different magnon Fock states $|n_\mathrm{m}\rangle$ are indicated by the color-coded shaded areas generated by the fit (black lines). The integer vertical dashed lines indicate the frequencies of the qubit coupled to the Kittel mode in the magnon Fock states $|n_\mathrm{m}\rangle$.
(b)~Magnon population $\overline{n}_\mathrm{m}$ as a function the excitation power. The black dashed line indicates a linear fit to the data. The solid gold line is a numerical fit with a ``self-Kerr" interaction $K_\mathrm{m}/2\pi=0.2$~MHz for the Kittel mode. The inset shows the difference $\Delta\overline{n}_\mathrm{m}$ between the data and the nonlinear fit from the linear fit.
(c)~Probability distributions $p_{n_\mathrm{m}}$ of the first four magnon Fock states as a function of the excitation power. The solid lines show the Poisson distributions based on the magnon populations shown in (b). The inset shows the probability distribution for the highest excitation power.
Adapted from~\onlinecite{Lachance-Quirion2017}.
\label{fig:Quantum_magnonics_dispersive}}
\end{center}\end{figure*}

According to Eq.~\eqref{eq:qubit_magnon_dispersive}, the qubit-magnon dispersive interaction shifts the qubit frequency by $2\chi_\textrm{q-m}$ for each injected magnon~\cite{Gambetta2006}. The observed shift per magnon of $2\chi_\textrm{q-m}/2\pi=3.0$~MHz is larger than the line widths of the qubit and the Kittel mode of respectively $0.78$~MHz and $1.3$~MHz, i.e.\ the experiment reached the strong dispersive regime of quantum magnonics~\cite{Lachance-Quirion2017}. Figure~\ref{fig:Quantum_magnonics_dispersive}a shows individually resolved magnon Fock states $|n_\mathrm{m}\rangle$ in the qubit spectrum.

Both the average number of magnons $\overline{n}_\mathrm{m}$ (Fig.~\ref{fig:Quantum_magnonics_dispersive}b) and the probability $p_{n_\mathrm{m}}$ of having $n_\mathrm{m}$ magnons (Fig.~\ref{fig:Quantum_magnonics_dispersive}c) were obtained by fitting an analytical model to the data~\cite{Gambetta2006}. The magnon population $\overline{n}_\mathrm{m}$ in the absence of a pump confirms that the Kittel mode is well thermalized with a population below $0.01$~magnons at $T\sim10$~mK. The magnon probabilities are Poissonian distributed, as expected for a linear system such as the Kittel mode~\cite{Rezende1969}.

The first experimental demonstration of a strong dispersive interaction in quantum magnonics was achieved in the straddling regime with $\omega_\mathrm{m} \in [\omega_{\rm q},\omega_{\rm q}+K_{\rm q}]$ \cite{Koch2007,Lachance-Quirion2017}. A large dispersive shift can also be obtained by tuning the angular frequency of the Kittel mode $\omega_\mathrm{m}$ close to  $\omega_\mathrm{q}+K_\mathrm{q}$ of the second qubit transition~\cite{Lachance-Quirion2019, Lachance-Quirion2020,Wolski2020}, which also greatly limits the self-Kerr nonlinearity of the Kittel mode~\cite{Juliusson2016,Lachance-Quirion2017}. In these papers, Ramsey interferometry characterizes the strong dispersive interaction better than standard two-tone spectroscopy by avoiding the broadening of the qubit absorption spectrum from both the probe and spectroscopy MW tones.

\subsubsection{Other qubit-mediated interactions}

\onlinecite{Tabuchi2015} demonstrated the parametric coupling described by Eq.~\eqref{eq:qubit_magnon_coupling_parametric}. Here, the Kittel mode was detuned from the qubit by $\Delta_\textrm{q-m}/2\pi=-274$~MHz, with modulus much larger than the coupling strength $g_\textrm{q-m}/2\pi=10$~MHz. A large detuning suppresses the static coupling of Eq.~\eqref{eq:qubit_magnon_coupling}. However, driving the hybrid system at an angular frequency $\omega_\mathrm{D}$ close to the average angular frequency $\left(\omega_\mathrm{q}+\omega_\mathrm{m}\right)/2$ leads to an avoided crossing in the spectrum of the Kittel mode, i.e., strong coherent coupling. The parametric coupling strength increases linearly with the drive power up to $\tilde g_\textrm{q-m}/2\pi=3.4$~MHz~\cite{Tabuchi2015} and generates a time-controlled interaction between the fixed-frequency transmon qubit and the Kittel mode, whose frequency in the current implementations can only be changed on a timescale much longer than the lifetimes.

\onlinecite{Lachance-Quirion2017} observed the qubit-mediated self-Kerr interaction of the Kittel mode as described by Eq.~\eqref{eq:self-Kerr} in terms of a  nonlinear scaling of the magnon population $\overline{n}_\mathrm{m}$ as a function of the pump power close to resonance with the Kittel mode (Fig.~\ref{fig:Quantum_magnonics_dispersive}b). The observed self-Kerr coefficient $K_\mathrm{m}/2\pi=0.2$~MHz was smaller than the line width $\kappa_\mathrm{m}/2\pi=1.3$~MHz of the mode, therefore keeping the Kittel mode in the linear regime at the level of a single magnon. Because the self-Kerr interaction depends strongly on the frequency of the Kittel mode relative to the frequencies of the first two qubit transitions~\cite{Juliusson2016,Lachance-Quirion2017}, it can readily be controlled with a static magnetic field.

\subsection{Applications and challenges}

\subsubsection{Quantum sensors}

Quantum magnonics can be applied to quantum sensing. The engineering of a strong coherent interaction between magnets and superconducting qubits allows applying the tools developed in quantum technologies~\cite{Degen2017} to, for example, sensing of magnons. The strong dispersive regime of quantum magnonics was used to entangle the Kittel mode of a millimeter-sized YIG sphere with a superconducting qubit~\cite{Lachance-Quirion2020}. The high-fidelity single-shot readout of the qubit state allows detection of a single magnon with a quantum efficiency reaching $\sim70\%$. The protocol can be made quantum non-demolition (QND) by replacing the non-QND high-power qubit readout technique~\cite{Reed2010} by a dispersive readout technique~\cite{Walter2017}. The demonstration of the single-magnon detector, the equivalent  of the single-photon detector to magnonics, paves the way, for example, to the heralded generation of single magnons.


Alternatively, a steady-state magnon population can be detected with a sensitivity of approximately $10^{-3}$~magnons$/\sqrt{\mathrm{Hz}}$ through Ramsey interferometry of a qubit that is dispersively coupled to a magnon mode~\cite{Wolski2020}. In the strong dispersive regime, the qubit is sensitive to the magnon population through dissipation by the magnons, in stark contrast to the entanglement-based method of \onlinecite{Lachance-Quirion2020}. Such a sensing method could be useful to characterize weak magnon-creation processes.

All protocols of quantum sensing rely on the coherence of the qubit~\cite{Degen2017}. The performance of single-magnon detectors can be improved via the qubit coherence time of presently $T_2^*\sim1~\mu$s ~\cite{Lachance-Quirion2020},  limited by relaxation through the cavity modes and dephasing from a finite thermal population of the same modes~\cite{Tabuchi2015,Tabuchi2016,Lachance-Quirion2017,Lachance-Quirion2020}. Both contributions can be suppressed by smaller internal losses of the cavity that are of the order of $1$~MHz in 3D MW cavities made out of copper~\cite{Tabuchi2015,Tabuchi2016,Lachance-Quirion2017,Lachance-Quirion2020}. Superconducting MW cavities with lower internal losses that still allow saturation of the magnetic order, would greatly improve quantum sensing of magnons. Finally, the $\sqrt{N}$-enhancement of the magnetic-dipole interaction between magnetostatic and cavity modes can be harnessed in quantum magnonics to improve the detection of static or MW magnetic fields close to the FMR frequency~\cite{Crescini2020, Crescini2020a}.

\subsubsection{Quantum transducers}

Quantum magnonics may lead to a bidirectional MW-to-optical quantum transducer for MW-only superconducting circuits~\cite{Hisatomi2016,Haigh2016,Lachance-Quirion2019}. We discussed the perspectives to achieve strong coupling between magnons and an optical cavity in Section \ref{SecV}. Here we address the MW part, which is, at the time of this review, significantly more advanced. Quantum information transfer from a superconducting qubit and optical light via a magnonic  transducer requires faithful  encoding of an arbitrary quantum state of the qubit into a nonclassical state of magnons. This can be achieved by employing both the resonant and dispersive regimes of the strong coherent coupling of the fundamental excitations of a magnet and a superconducting qubit.

In the resonant regime, the beam-splitter interaction Eq~\eqref{eq:qubit_cavity_coupling} can be used to transfer an excitation in the qubit to a single magnon in the Kittel mode~\cite{Meekhof1996,Hofheinz2008,Hofheinz2009,Satzinger2018,Chu2018} by dynamical control of either the detuning or the coupling strength~\cite{Hofheinz2009}. In quantum magnonics this can be achieved by two methods. First, the parametric coupling described by the Hamiltonian of Eq.~\eqref{eq:qubit_magnon_coupling_parametric} can be used to obtain a tunable coupling strength between the magnetostatic mode and the qubit. Secondly, the detuning between both systems can be tuned dynamically either by changing the frequency of a flux-tunable qubit~\cite{Hofheinz2008,Hofheinz2009,Satzinger2018,Chu2018} or a fixed-frequency qubit through a time-controlled ac-Stark shift~\cite{Chu2017}.

In the dispersive regime, the interaction described by the Hamiltonian of Eq.~\eqref{eq:qubit_magnon_dispersive} can be used to encode arbitrary qubit states into a nonclassical state of magnons~\cite{Leghtas2013,Vlastakis2013}. Such schemes have the advantage of working with qubits and magnon modes of fixed frequency that are coupled through a static dispersive strong  interaction. However, the encoding schemes based on such a dispersive interaction are inherently slower than those based on a resonant interaction. Both approaches require larger coupling strengths than demonstrated to date, as well as longer qubit coherence times and magnon lifetimes. The qubit-magnon coupling strength can be increased by careful quantum engineering, for example, by increasing the magnetic-dipole coupling strengths between the Kittel and cavity modes without increasing losses. This can be achieved by increasing the spatial overlap, characterized by a filling factor $\eta$ between the Kittel and the cavity modes since $g_\mathrm{m-c}\propto\sqrt{\eta}$. Increasing the lifetime of magnons, beyond the current $\sim100$~ns, requires understanding and control of the magnon decay through a bath of two-level systems of unknown microscopic origin~\cite{Tabuchi2014,Pfirrmann2019,Kosen2019}.



\section{Challenges} \label{SecVIII}
The past ten years witnessed systematic studies of the coherent magnon-photon interaction in cavities and resonators over a wide frequency spectrum in small structures and devices,  with contributions from optics, magnetism, acoustics, MW technology, and spintronics~\cite{Li2020,Awschalom2021}. Our review only provides a somewhat subjective snapshot of the state of the art of this rapidly progressing field. In the following, we sketch some challenges for the next decade.
\begin{itemize}

\item {\em Materials}. YIG is the material of choice because of its high Curie temperature and superior optical, magnetic, and mechanical quality even for thin films~\cite{Schmidt2020} that facilitates strong coupling of macroscopic samples at room temperature. Nevertheless, the search for alternative materials continues. Rare-earth iron garnets with open 4f shells show large magneto-optical constants~\cite{Dionne2009} and  thin film perpendicular magnetization~\cite{Avci2021}, be it at the cost of increased Gilbert damping. Antiferromagnetic insulators grow with high crystal perfection, but their resonance frequencies are usually in the THz regime for which high-quality cavities still have to be developed~\cite{Bialek2020}. Mono- or multilayers of two-dimensional van der Waals magnets are a new and  promising class of materials with dimensionally enhanced magnon-photon interactions~\cite{Mandal2020}.

\item {\em Nanostructures}. We expect increased activity in the cavity magnonics of nanostructures, because smaller sized magnets facilitate the coherent control of the order parameter. YIG is a difficult material to pattern at the nanoscale without sacrificing its magnetic quality, but progress is being made~\cite{Schmidt2020}. We have seen that the coupling with light is strongly enhanced with decreasing volume of the magnet since the overlap integral of the magnon-photon matrix elements is proportional to $V^{-1/2}$. Nanoscale periodic structures in the form of optomagnonic crystals can be a promising approach to this end~\cite{Graf2021}. The long-term goal is to develop a superior transducer between MW and light for classical and quantum information exchange applications. The decrease in the coupling to MWs with the number of spins can on the other hand be compensated by cavity design, which leads to strong magnon-photon coupling in conventional metallic magnets~\cite{Hou2019,Li2019}. 
  
\item {\em Nonlinearities}. The ease by which magnets can be driven into the nonlinear regime by MWs is an important advantage over other systems~\cite{Bertotti2009}. Non-linearities cause chaotic dynamics, instabilities of the Kittel mode, allow parametric excitation, and may lead to magnon condensation with associated spin superfluidity~\cite{Sonin2020}. Theory predicts that the increased coupling should lead to complex non-linear behavior beyond the Duffing model~\cite{Kusminskiy2016,Elyasi2020}. We expect more experimental emphasis on such nonlinearities in high-quality MW cavities.
  
\item {\em Multiple loads}. The coherent coupling of various systems in MW cavities has been a main effort of cavity QED research and the coupling of magnets with superconducting qubits has been a milestone of cavity magnonics. The coherent coupling between magnetic systems is of great interest as well, since the emergence and control of dark and bright (super- and subluminescent) states can be applied to  memories~\cite{Zhang2016,Yu2020a}. The MW photons couple magnets to form exotic ``magnon molecules''\cite{Zare2018, Yu2020a}.

\item {\em Hybrid systems}. 
We reviewed only the physics of spin in photon cavities and resonators. However, the confinement of any wave leads to strong modulation of its density of states and the interaction with spins inside. The strong coupling between a Kittel mode and the spin waves in a proximity film leads to remote coherent dynamic coupling between different magnets~\cite{Chen2019}. The elastic and magnetic collective modes in a magnet are coupled, holding the promise of combining the best features of optomechanics and optomagnonics~\cite{Zhang2016,Costache2019}, with applications to thermometry \cite{Potts2020}. The strong coupling between magnons and phonons would allow for pumping of a phonon spin current into a phononic cavity. A material with high acoustic quality allows for coherent coupling of spins over macroscopic distances at room temperature~\cite{An2020}. 
  
\item {\em Chirality}. A unique feature of magnetic order is its broken time reversal symmetry. The well-known chirality of (Damon-Eshbach) surface spin waves follows from the topology of magnetic half space~\cite{Yamamoto2019}. The non-chiral magnons in thin films can be excited unidirectionally by chiral magnetodipolar stray fields ~\cite{Chen2019,YuLiu2019b,Yu2019c} or at  chirality lines of MW modes in cavities and wave guides~\cite{Yu2020a}. Analogous effects exist for spin waves coupled to surface phonons~\cite{Yamamoto2020} and plasmons~\cite{Oue2020}. 
  
\item {\em Techniques}. A challenge that accompanies the trends above is the design of MW cavities to smaller sizes and higher frequencies without reduced quality. The efficiency of the proximity coupling of laser light to magnets by tapered fibers or prisms could be improved. Advanced magnetometry with NV centers in diamond provides spatiotemporal images of the stray fields of  and thermal \cite{Du2017} and coherent magnons \cite{Bertelli2020,Zhou2020,Bertelli2021} with valuable information of the coupling process.
    
\item {\em Quantum magnonics}. The observation of macroscopic quantum effects in magnonics remains a or the major challenge. Quantum effects can be unequivocally observed only in the non-linear regime and the relatively large damping of even YIG has to be overcome. As reported in Sec. \ref{SecVI},  quantum effects are observed by coherent coupling to the genuine quantum state in superconducting qubits that gives access to its nonlinear dynamics. Quantum effects in purely magnetic systems require cooling to low temperatures in order to suppress dephasing by phonons. In the strong coupling regime, quantum effects are observable by either sufficiently fast measurements or studies of the magnetic noise properties. The prize would be the predicted magnon squeezing of the massive and distillable entanglement of the magnon-photon system~\cite{Elyasi2020}. Another route for creating nonclassical magnetic states, such as magnon Fock states ~\cite{Bittencourt2019} and spin cat states~\cite{Sharma2021} are protocols involving heralding, where the nonlinearity is provided by the projective measurement.
\end{itemize}

\section*{Acknowledgments}

B.Z.R. would like to acknowledge Amir Eskandari-asl, S. M.-Reza Taheri, M.-F. Miri, F. Pirmoradian and S. Hamed Aboutalebi for helpful discussions. This work was supported by Institute for Research in Fundamental Sciences (IPM), Iran's National Elites Foundation and Iran Science Elites Federation. S.V.K. acknowledges support from the Max Planck Gesellschaft through an Independent Max Planck Research Group. J.A.H. would like to acknowledge Andrew Ramsay. This work was supported by the European Union's Horizon 2020 research and innovation programme under grant agreement No 732894 (FET Proactive HOT). K.U., D.L.-Q., and Y.N. would like to acknowledge Yutaka Tabuchi, Arnaud Gloppe, Alto Osada, Ryusuke Hisatomi, and Samuel Piotr Wolski. This work is partly supported by Japan Society for the Promotion of Science (JSPS) KAKENHI (18F18015), JST Exploratory Research for Advanced Technology (ERATO) (JPMJER1601), and FRQNT Postdoctoral Fellowships. D.L.-Q. is an international research fellow of JSPS. C.-M.H. would like to acknowledge the financial support of NSERC Discovery Grants and NSERC Discovery Accelerator Supplements. C.-M.H. thanks all his colleagues from the Dynamic Spintronics group at University of Manitoba for contributions and discussions. H.X.T. acknowledges support from National Science Foundation (EFMA-1741666) and a Packard Fellowship in Science and Engineering. This work was supported by JSPS KAKENHI (Grant No. 19H006450). G.E.W.B. and Y.M.B. thank their colleagues Sanchar Sharma, Mehrdad Elyasi, and Tao Yu and acknowledge support of Netherlands Organisation for Scientific Research (NWO).

\bibliography{Refs}

\end{document}